\pdfoutput=1

\documentclass[journal,onecolumn]{IEEEtran}
\ifCLASSINFOpdf
   \usepackage[pdftex]{graphicx}
   \graphicspath{{../pdf/}{../jpeg/}}
   \DeclareGraphicsExtensions{.pdf,.jpeg,.png}
\else
   \usepackage[dvips]{graphicx}
   \graphicspath{{../eps/}}
   \DeclareGraphicsExtensions{.eps}
\fi

\usepackage{amsmath}
\usepackage{amsthm}
\usepackage{amsfonts}
\usepackage{amssymb}
\usepackage{dsfont}
\usepackage{cancel}

\usepackage{color}
\definecolor{red}{rgb}{1,0,0}

\begin{document}

%
\title{A computational model describing the interplay of basal ganglia and subcortical background oscillations during working memory processes}
%
%
%
\author{Utku \c{C}elikok, Eva M. Navarro-L\'{o}pez and~Neslihan Serap \c{S}eng\"{o}r 
       %
\thanks{U. \c{C}elikok is with the Faculty of Engineering and Natural Sciences, Istanbul Bilgi University, 34060, Ey\"up, Istanbul, Turkey and with the Department of Electronics and Telecommunication, Istanbul Technical University, Maslak, 34469, Sariyer, Istanbul, Turkey.
\texttt{\small celikok.utku@gmail.com}} 
\thanks{E.M. Navarro-L\'opez is with the School of Computer Science, The University of Manchester, Oxford Road, Kilburn Building, Manchester M13 9PL, UK. 
\texttt{\small eva.navarro@manchester.ac.uk}} 
\thanks{N.S. \c{S}eng\"or is with the Department of Electronics and Telecommunication, Istanbul Technical University, Maslak, 34469, Sariyer, Istanbul, Turkey.
\texttt{\small sengorn@itu.edu.tr}}

}
\maketitle



\begin{abstract}
Working memory is responsible for the temporary manipulation and storage of information to support reasoning, learning and comprehension in the human brain.\ Background oscillations from subcortical structures may drive a gating or switching mechanism during working memory computations, and different frequency bands may be associated with different processes while working memory tasks are performed.\ There are three well-known relationships between working memory processes and specific frequency bands of subcortical oscillations, namely: the storage of new information which correlates positively with beta/gamma-frequency band oscillations, the maintenance of information while ignoring irrelevant stimulation which is directly linked to theta-frequency band oscillations, and the clearance of memory which is associated with alpha-frequency band oscillations.\ Although these relationships between working memory processes and subcortical background oscillations have been observed, a full explanation of these phenomena is still needed.\ This paper will aid understanding of the working memory's operation and phase switching  by proposing a novel and biophysical realistic mathematical-computational framework which unifies the generation of subcortical background oscillations, the role of basal ganglia-thalamo-cortical circuits and the influence of dopamine in the selection of working memory operations and phases: this has never been attempted before.
 
\end{abstract}


%
\IEEEpeerreviewmaketitle


\section{Introduction}

Working memory facilitates the maintenance and processing of information required to learn, reason, comprehend, compare and build a cause-effect relationship in complex cognitive processes. It is interpreted as the scratchpad of the brain for momentary recall and processing of information. It is claimed that working memory is a kind of short-term memory that acts as an interface operating between the long-term memory and actions to support thinking processes \cite{ref1}. The most outstanding feature of working memory is the fact that it is not a passive system where information is only stored. It is, indeed, an active system where information is processed. 

Models for memory have evolved over the years from a single unitary memory system to short-term, long-term memory, long-short-term memory, and finally to working memory \cite{ref2}. In these works, neuropsychological data have shown that it is not necessary for information to pass through short-term memory to gain access to long-term memory. It has been observed that, information can still be transferred to the long-term memory in patients with short-term memory impairments. The working memory model by Baddeley \cite{ref1} indicates a clear distinction between short-term memory and working memory, and currently, Baddeley's model is accepted as the most plausible model for working memory. It includes an executive control mechanism (central executive) that interacts with short-term memory for auditory-verbal (phonological loop), visual-spatial (visuo-spatial sketchpad) information and an episodic buffer which is in charge of binding diverse types of information \cite{ref4}. As Baddeley pointed out: working memory is a temporary storage system under attentional control that underpins our capacity for complex thought \cite{ref3}. 

Baddeley's working memory model provides a suitable mechanism to describe working memory functions. However, the central executive subunit in it is an artificial definition since it is not clear how this subunit selects the information to be maintained or ignored. Basal ganglia loops can be considered as a gating mechanism in action selection, as well as for information selection in cognitive tasks. Consequently, they might be interpreted as a candidate for the central executive subunit. The basal ganglia are a collection of subcortical structures -- relatively large in primates, particularly in humans -- that are highly interconnected with the cortex. This suggests that basal ganglia play a key role in planning, cognition and reinforcement learning. Neurophysiological and anatomical studies have shown that inputs coming from different parts of the cortex are directed to the thalamus through the basal ganglia circuits and, finally, projected back to the cortex from which the circuit is originated \cite{ref5,ref6}.

As working memory is thought to be a system operating in real time with limited capacity, incoming information is needed to be loaded and maintained while irrelevant task distractors are ignored. Furthermore, as its capacity is typically thought to be limited, the information has to be cleared away once a task has been completed \cite{ref3,ref4}. Experimental studies have revealed that the basal ganglia exert an inhibitory influence on a number of cortical areas through the thalamus, and this inhibition allows information to be loaded. Information loading starts as soon as the sensory information is registered in the posterior cortex. The basal ganglia project to the thalamus, which is mostly inhibited by pre-synaptic activation from subcortical areas. The external stimulus ``releases the brakes'' and excites the thalamus, resulting in neuronal activation. Recurrent activity, caused by the activation of the thalamus, opens the gates for the prefrontal cortex and allows working memory to load  information. Therefore, the basal ganglia can be interpreted to have a ``switching behaviour'' which is modulated by dopamine \cite{ref7}, and the thalamus can be thought of as a translator of the message received from the basal ganglia and projected back to the cortex areas related to a specific working memory task \cite{ref8}. Recently, we have proposed a multi-level hybrid automaton model for the switching mechanism of the basal ganglia and the thalamus linked to some working memory processes \cite{eva:2015}.

In this paper, we propose a mathematical-computational model for the most relevant working memory processes which includes the switching control of the basal ganglia, the thalamus and the key role of dopamine in basal ganglia's selection mechanisms. Our work focuses on obtaining a model where the prefrontal cortex activation corresponds to distinct working memory processes due to the  incoming activity from the thalamus. For this purpose, the dopamine-dependent activity of the thalamus due to opposing effects of the direct and indirect pathways is modelled. We associate three different pathways within the basal ganglia with different behavioural patterns of the thalamus. First, the direct pathway of the basal ganglia allows incoming information to be loaded in working memory, which matches with beta-gamma-frequency band activity present in the thalamus. Second, the activation of the indirect pathway of the basal ganglia is related to theta-frequency band activity in the thalamus, corresponding to the maintenance of information. Finally, the clearance of information corresponds to the balance between the direct and the indirect pathways, giving rise to alpha-frequency band activity in the thalamus. Our study is driven by experimental observations: the storage of new information in working memory correlates positively with oscillations in the beta-gamma-frequency band ($13-30$Hz and $30-120$Hz) \cite{ref9,ref10};  theta-frequency band oscillations ($4-8$Hz) \cite{ref11} are directly linked to the maintenance of information while ignoring irrelevant stimulation, and finally, the clearance of memory is associated with alpha-frequency band ($8-13$Hz)  oscillations \cite{ref12}.

The aim is to show that our working memory network model is able to generate subcortical background oscillations which correspond to oscillations observed on human electrophysiology during working memory tasks. In particular, our model will effectively model how these oscillations are generated in the thalamus due to the interaction of the different pathways in the basal ganglia and the regulation of dopamine.  Our model will show the onset of different behaviours during working memory tasks according to the different frequencies of the oscillations generated in the thalamus, driven by the basal ganglia's pathways. The importance of oscillatory activity patterns to control brain states and switching in memory phases has been already established \cite{Schmidt:2013, Torres:2014}. We associate different behaviours in the prefrontal cortex (working memory) with specific incoming oscillatory patterns from the thalamus, that are associated with specific pathways in the basal ganglia. As far as we are aware, this is the first time that a biologically plausible spiking neural network of this type has been proposed. The results obtained might be a stepping stone to better understand the mechanisms behind the different frequency-band oscillations observed during working memory tasks. The main contribution of this work is to propose a modelling framework to clarify the role of the basal ganglia in the generation of the subcortical oscillations associated with different working memory processes.


This paper is organised as follows. Section II gives some background information on the basal ganglia and the different  basal ganglia's pathways, linked to relevant subcortical oscillations. Section III summarises the already-existing computational models that have been a source of inspiration for our work. Section IV presents our mathematical-computational model. In order to clarify how the simulation results have been obtained and to ensure the reproducibility of our results, all the mathematical details and parameter values are explicitly presented for all the types of single neurons considered, as well as the connection dynamics in each population of neurons and between different populations. Section V presents the simulation results that reveal the emergence of beta-gamma band activity due to the dominance of the direct pathway, and the onset of theta band activity due to the dominance of the indirect pathway of the basal ganglia. The simulation results are explained by considering the delayed-match-to-sample (DMS) task, which will help understand the neuronal activity during working memory tasks. Conclusions are given in the last section.  

\section{Basal Ganglia and Working Memory}

The basal ganglia are subcortical structures located in the human forebrain and buried deep into the brain. It is well known that the basal ganglia receive massive projections from the posterior cortex. This suggests that they have a role in planning and cognition. The basal ganglia project mostly into the thalamic nuclei, which in turn project to the prefrontal cortex \cite{ref13}, completing the cortex-basal ganglia-thalamus loop, which has been studied thoroughly since the work of Alexander et al. \cite{ref6}. Under resting conditions, the network output to the thalamus is inhibitory. The basal ganglia's circuit works by modulating this inhibition. One of basal ganglia's major roles is to integrate associative and limbic information for the production of context-dependent behaviours \cite{ref14}. Besides its importance as a switching control mechanism for several functions, the basal ganglia are linked to neurological disorders such as Parkinson's disease, Huntington's disease and schizophrenia. 

The prefrontal cortex and the basal ganglia interact with each other. This interaction is key in working memory processes and is based on disinhibition or what is usually referred to as ``releasing the brakes'' in the cortex \cite{ref15}. While the cortex provides robust maintenance of task-dependent information, the basal ganglia perform a gating-like function for appropriate context, which allows working memory representations to be updated at appropriate time. The neurotransmitter dopamine helps the basal ganglia perform this gating mechanism by strengthening the efficacy of cortical inputs.

Two main aspects of the basal ganglia are especially relevant to the model proposed in this work: the modulating role of dopamine and  the dominance of inhibition. On the one hand, the activation of different basal ganglia's pathways -- namely, direct, indirect and hyperdirect -- is modulated by dopamine. That is, the selection of an action depends on the level of dopamine or on the striatal neuron's receptor-type receiving the dopamine. On the other hand, unlike other circuits in the brain, the basal ganglia are mostly dominated by inhibitory projections. Inhibition facilitates the decision of which units are needed to be silent and provides a winner-takes-all mechanism. Dopamine decreases some units' activity and due to this, units which are disinhibited become active. The release of dopamine in the striatum makes medium spiny neurons (MSNs) bistable.\ Depending on the dopamine level and the receptor type, MSNs may stay silent or get activated.
 
\begin{figure}[!ht]
\centering
\includegraphics[width=5.5in]{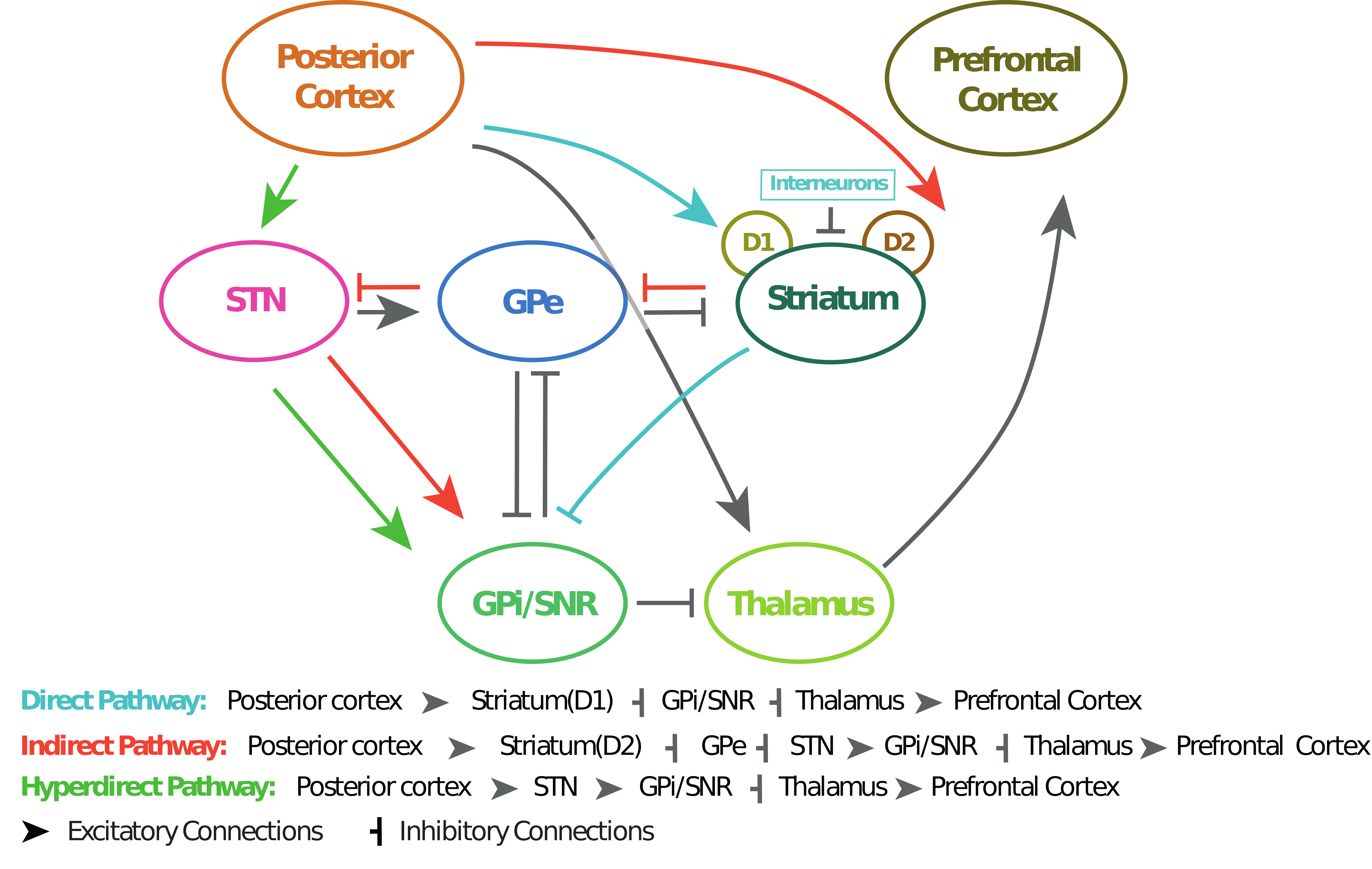}
\caption{Connection graph of the neuronal structures considered in the working memory model. Our model consists of two different cortical structures, the basal ganglia nuclei and the thalamus. The proposed model integrates the direct, indirect and hyperdirect pathways of the basal ganglia, in addtition to some internal regulatory connections.}
\label{fig:connection diagram1}
\end{figure}

The neural populations considered in our model correspond to units of the posterior cortex, prefrontal cortex, striatum, internal and external segments of globus pallidus (GPi and GPe), subthalamic nucleus (STN) and thalamus.\ In the GPi's population, we integrate the activity of the substantia nigra pars reticulata (SNr).

There are two main routes that connect the striatum, the input nuclei of the basal ganglia to the GPi and the output nuclei of the basal ganglia: the direct pathway and the indirect pathway. The direct pathway integrates inhibitory projections from the striatum $D_1$-receptor-type neurons to the GPi. The indirect pathway goes through the striatum $D_2$-receptor type neurons, the GPe, the STN and the GPi \cite{ref16}. These two pathways have opposite effects on the thalamic activity since their different striatum populations have different chemical and physiological properties. Most importantly, they express different types of dopamine receptors \cite{ref17}. As a result, dopamine has opposite effects on them: it excites $D_1$-type neurons but inhibits $D_2$-type neurons. While dopamine enhances the activity of  the thalamus via $D_1$-receptor-type neurons in the direct pathway, it reduces the activity of $D_2$-receptor type neurons, which provokes the reduction of the activity of the thalamic nuclei in the indirect pathway. Depending on the level of dopamine, the striatum triggers different pathways.

 There is also a hyperdirect pathway which bypasses the striatum and connects the cortex to the GPi via the STN. The hyperdirect pathway is critical for suppressing erroneous movements.\ When it fails, patients are unable to inhibit unwanted motor patterns and cause involuntary movements, which results in a condition known as hemiballismus \cite{ref44}. The hyperdirect pathway message can be translated as ``hold your horses'' as it is suggested in \cite{ref21}. The main connections are shown in Figure \ref{fig:connection diagram1}. 

Now, we relate the activity of the direct and indirect pathways of the basal ganglia to some key working memory processes. Although we do not associate the hyperdirect pathway of the basal ganglia with a specific operation in the working memory directly, we consider the connections which form the hyperdirect pathway as important internal regulatory mechanisms within the proposed network model. The interpretations that we make sets the basis for the mathematical-computational model of the working memory network that we propose in this work. We consider three main operations in the working memory: loading, maintenance and clearance of information. These operations are related to basal ganglia's pathways and subcortical background oscillations in the following way.

\begin{itemize}

\item \textbf{Loading of Information through the Direct Pathway}. In the direct pathway, cortical cells project excitatory inputs to the striatum and by the overabundance of dopamine, $D_1$-receptor-type neurons get activated while the activity of $D_2$-receptor type neurons is reduced. Active $D_1$-type neurons, in turn, project inhibitory activation onto the neurons of the GPi/SNr. The GPi/SNr neurons project directly onto the thalamus. The striatal inhibition of the GPi is coupled with the disinhibition of the thalamus. The result is a reduction of the inhibition of the thalamus via the striatum. The thalamus projects excitatory glutamatergic neurons to the cortex itself. The direct pathway, therefore, results in a net excitation of the cortex by the thalamus. When the direct pathway is  more active, it transmits a ``Go'' signal to the corresponding region. This signal temporarily disables the recurrent connections in the prefrontal cortex and interrupts the currently maintained representation \cite{ref5,ref6}. In our model, the update or loading of information during working memory tasks is associated to the dominance of the direct pathway activity which is triggered by an external input received in the posterior cortex. If this information is needed to be loaded, the substantia nigra pars compacta (SNc) releases dopamine to the striatum. This release enhances the activation of $D_1$-receptor-type neurons in the striatum. Consequently, the direct pathway becomes dominant over the indirect pathway. The activation of $D_1$-receptor-type neurons in the striatum triggers the beta-gamma-frequency band oscillations in the thalamus. This allows the prefrontal cortex to have a stable active state which corresponds to the loading of information in working memory \cite{ref22}. This scenario is represented in Figure \ref{fig:load}.

\begin{figure}[!ht]
\centering
\includegraphics[width=5.5in]{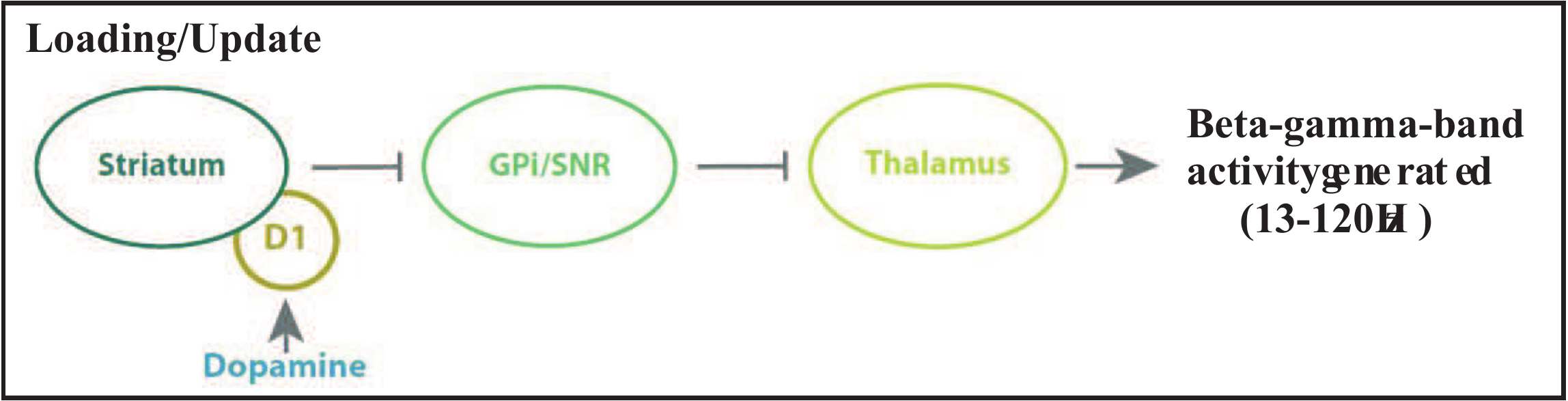}
\caption{Loading of information via the direct pathway of the basal ganglia.\ Cortical excitation through the thalamus `opens the gates' and allows new information to be loaded.\ Information is stored temporarily thanks to the steady and high frequency activity of cortical pyramidal neurons that is supported by the thalamic excitation. The arrow $\longrightarrow$ corresponds to excitatory connections, and $\dashv$ corresponds to inhibitory connections.}
\label{fig:load}
\end{figure}

\item \textbf{Maintenance of Information through the Indirect Pathway}. In our model, the maintenance of information starts at the $D_2$-receptor type neurons in the striatum through the indirect pathway. Once they are stimulated by the cortex and with dopamine depletion, striatal $D_2$-type neurons project inhibitory axons onto the neurons of the GPe, which tonically inhibits the STN. This inhibition of the GPe by the striatum results in the net reduction of the inhibition on the STN. The STN, in turn, projects excitatory inputs to the GPi, which inhibits the thalamus. The result is the inhibition of the thalamus and, therefore, a decrease of the stimulation of the prefrontal cortex by the thalamus. This scenario gives rise to  thalamic theta-frequency band activity. 

\begin{figure}[!ht]
\centering
\includegraphics[width=6.3in]{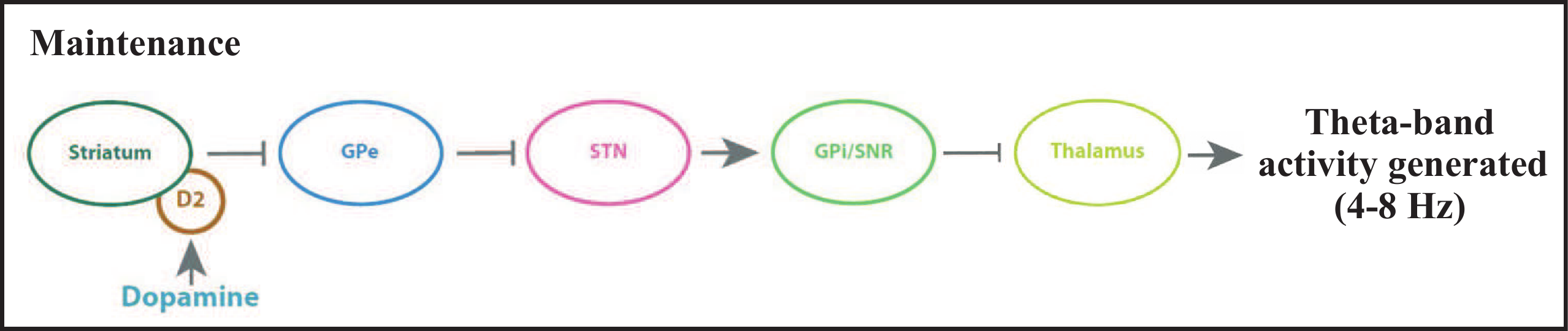}
\caption{Maintenance of  information is achieved by a depletion of the dopaminergic effect on $D_2$-receptor type neurons via the indirect pathway of the basal ganglia.\ Dopamine depletion results in the excitation of the GPi, which  in turn, inhibits the thalamus. The thalamus presents theta-frequency band oscillations and is incapable of causing a persistent spiking activity in the cortical neurons. The arrow $\longrightarrow$ corresponds to excitatory connections, and $\dashv$ corresponds to inhibitory connections.}
\label{fig:maintain}
\end{figure}

In this case, the activity of the thalamus is insufficient to change the ongoing activity in the prefrontal cortex. Consequently, the prefrontal cortex exhibits robust active maintenance by its recurrent connections as given in Figure \ref{fig:maintain}. Moreover, if there is no information to maintain  in  the working memory, the indirect pathway prevents new representations from being loaded \cite{ref22}. The indirect pathway encodes a ``No-Go'' signal that protects prefrontal cortex representations by preventing new contents from being loaded \cite{ref5,ref6}.

\item \textbf{Clearance of Information through the Direct and Indirect Pathways}. In working memory, clearance of information is achieved by updating the maintained information or by a temporal decay of existing information. No specific pathway of the basal ganglia is used for clearance. In the model proposed here, this will correspond to the resting state of the basal ganglia when there is no dopamine release or depletion, which is shown in Figure \ref{fig:clear}. In this case, the alpha-frequency band activity in the thalamus will cause the information to be cleared. There are some EEG studies showing that the thalamus at resting state presents alpha-frequency band activity \cite{ref23, ref24}. By integrating  these studies with the role of the basal ganglia in working memory, clearance should be a result of the balance between the direct and indirect pathways, with the basal ganglia in a resting state \cite{ref25,ref26}.

\begin{figure}[!ht]
\centering
\includegraphics[width=6.3in]{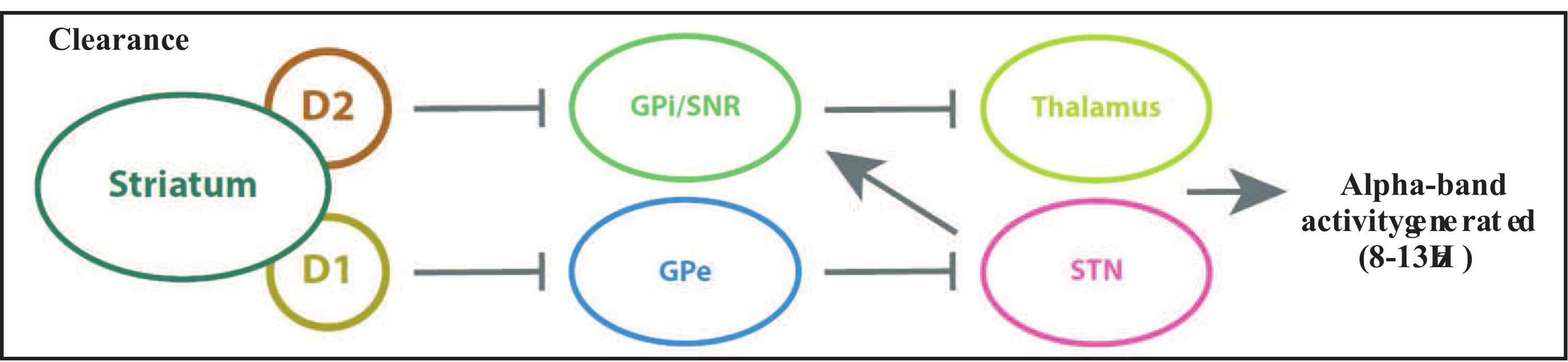}
\caption{Clearance of information by a resting state of the basal ganglia and its effect on the direct and indirect pathways.\ Basal ganglia does not employ a specific pathway to convey the information. In other words, there is no dopaminergic regulation over the system. The arrow $\longrightarrow$ corresponds to excitatory connections, and $\dashv$ corresponds to inhibitory connections.}
\label{fig:clear}
\end{figure}

\end{itemize}

\section{Relevant Computational Approaches}

In this section, three different computational models that have inspired our work will be explained. The first model is the work of Frank et al. \cite{ref22}, which provides a general and well-defined basal ganglia's architecture. In our work, we followed the connectivity of the basal ganglia's model given in \cite{ref22}. We will also integrate the striatum model proposed by Humphries et al.  \cite{ref27}. This model is of special interest for our study since it includes the dopamine regulation and its effect on the basal ganglia circuit. We will also consider the novel approach of \cite{ref28} for the interaction between the basal ganglia and the prefrontal cortex. In this paper, the thalamus is considered as the source of background oscillations during working memory tasks.

The model presented in \cite{ref22} may be classified as a connectionist model according to its neural plausibility and population architecture. It presents the basal ganglia circuitry as a mechanism that enables cortex functions to take place at appropriate time. In other words, the basal ganglia decide `when to do' and the cortex knows `what to do'. Here, a computational framework is proposed to explain how these subsystems work in harmony and enable working memory to process information in real time. They use the 12-AX cognitive task to test the robust maintenance mechanism of the cortex and the gating-like function of the basal ganglia. The model operates according to point neurons  and rate-coded output activations. Different parts of the cortex and the basal ganglia are modelled with a hierarchically-organised stripe-like architecture, where every stripe corresponds to an individually-updated region. Even though simulations start with predetermined parameters, the error-driven and associative learning mechanism enables the model to be self-governing. As a result of this study, the prefrontal cortex maintenance mechanism boosts activation due to its excitatory ion channels which are able to keep the activation without any stimulation. Furthermore, the basal ganglia are shown to be able to switch on/off the maintenance mechanism.

Another model that has motivated our work is the one proposed in \cite{ref27}. This paper focuses on the striatal microcircuit and its role in signal processing and action selection. In this paper, biologically plausible models of MSNs and striatal interneurons are given. The proposed neuron models in \cite{ref27} are able to reproduce the effect of dopamine on striatal neurons with $D_1$ and $D_2$-receptor type. As in \cite{ref22}, the model gives rate-coded output activations. During action selection, the model enhances the difference between the cortical inputs to allow the basal ganglia to select between competing informations. Another important aspect of this model is that it considers gap junctions in the fast-spiking interneurons in the striatum. In our model, we use the dynamics of the MSNs and fast-spiking interneurons as given in \cite{ref27}. We prefer to use these neuron models as they have a well-defined dopamine sensitivity, which is crucial for our model. 

The last model that has inspired our work is the one proposed in  \cite{ref28}. In this paper, subcortical background activation is considered as an input responsible to drive different processes during working memory tasks. Neurons are modelled as quadratic integrate-and-fire point neurons and the cortical population consists of only excitatory neurons. Different frequencies due to background activity are observed as the main gating mechanism for working memory computations. Such a flexible gating mechanism allows information to be selectively loaded, maintained, blocked, or cleared. Different working memory processes are identified as resting states or active states. Moreover, in \cite{ref28}, a bihemispheric spiking network is designed to explain the lateralisation of alpha frequency. It is also shown that relevant and irrelevant information are segregated in different hemispheres according to different background input frequencies.

Inspired by these works, the model we propose here is unique in the sense that: 1) it provides a computational framework that links the role of dopamine to working memory processes, and additionally, 2) it explains the onset of subcortical background oscillations from the activity of the direct and indirect pathways within the basal ganglia circuit.

\section{A Computational Model of the Basal Ganglia-thalamo-cortical network for the Analysis of Working Memory}    

In this section, we propose a mathematical-computational model to reproduce the relationships between the basal ganglia's pathways and the working memory processes explained in previous sections. As mentioned above, we consider the following neural populations: posterior cortex, striatum, GPe, GPi, STN, thalamus and prefrontal cortex. The activity of the GPi  and the SNr populations are integrated in the model as these two structures are considered to form the output of the basal ganglia's circuit \cite{ref6}. Each individual neuron will be modelled as a spiking point-like neuron of the threshold-firing type. That is: they include an auxiliary reset condition for the generation of the action potentials or spikes. All the neurons in the model are connected according to a random graph and the probabilities both in local connections within a population, and the interconnections between neurons of different populations are chosen in accordance to neurophysiological studies in the literature. However, some of the connection probabilities are chosen to be higher to allow the pre-synaptic population to drive the post-synaptic population more accurately. We also make inhibitory synaptic couplings stronger than excitatory ones to allow inhibitory neurons to adequately regulate excitatory population behaviour. 

In the rest of this section, we will describe the model by giving details of each neural population and their interconnections. All the parameters used will be given to provide necessary information to make our results reproducible.

\subsection{Model of the Posterior Cortex} \label{model_pctx}

The first neural structure to consider is the posterior cortex, where sensory information -- represented by an external input -- arrives.\ The external input represents the information that would be loaded or  maintained for future use during working memory tasks, or cleared if the information is already in the working memory.\ The model of the posterior cortex consists of two neural populations: excitatory (glutamatergic neurons with NMDA receptors) and inhibitory (GABAergic).\ We choose the ratio of excitatory to inhibitory neurons to be 4:1.\ This ratio is determined by the anatomy of a mammalian cortex \cite{ref29}.\ The excitatory NMDA receptors for glutamate stay inactive at their membrane potentials.\ Ion channels are mostly blocked and prevent positively charged ions from flowing in.\ An incoming signal from a pre-synaptic unit releases the inhibition on the ion channels, which excite the post-synaptic neurons.\ In contrast, GABAergic neurons use inhibitory neurotransmitters for communication.\ They show fast responses to inputs.\ Fast-spiking activity allow them to have a sufficient effect on NMDA receptors.\ They work as a regulatory unit within the population. 

In our model, excitatory neurons of the posterior/prefrontal cortex show regular-spiking activity as most of the neurons in the cortex have regular-spiking behaviour.\ When they receive a prolonged stimulus,  the neurons produce some spikes with short inter-spike period \cite{ref30}.\ During the following stimulation, these neurons show a reduction in the firing frequency of their spike response \cite{ref31}.\ This is called the spike-frequency adaptation.\ These neurons have large spike-after-hyperpolarisations.\ Consequently, even if the injected current increases, a fast-spiking behaviour cannot be observed.

For excitatory neurons, inspired by \cite{ref33},  dynamics for the membrane potential ($v_{{NMDA}_{pCtx}}$) and the recovery current ($u_{{NMDA}_{pCtx}}$) for  pyramidal neurons with NMDA-type receptors are described by the following equations. In the dynamical system below, $v_{{NMDA}_{pCtx}}$ and $u_{{NMDA}_{pCtx}}$ are vectors that contain the membrane potential and the recovery currents for all the neurons within the population of excitatory NMDA-type neurons within the posterior cortex:

\begin{equation} \label{eq:NMDApCtx} 
\begin{split}
C\frac{dv_{{NMDA}_{pCtx}}(t)}{dt} & = k \bigl[v_{{NMDA}_{pCtx}}(t)-v_r\bigr]\bigl[v_{{NMDA}_{pCtx}}(t)-v_t\bigr]-u_{{NMDA}_{pCtx}}(t)+I_{{NMDA}_{pCtx}}(t),\\
\frac{du_{{NMDA}_{pCtx}}(t)}{dt} & = a\biggl\{b\bigl[v_{{NMDA}_{pCtx}}(t)-v_r\bigr]-u_{{NMDA}_{pCtx}}(t)\biggr\},
\end{split}
\end{equation}
where $v_{{NMDA}_{pCtx}}  \in \mathbb{R}^{N_{{NMDA}_{pCtx}}}$, $u_{{NMDA}_{pCtx}}  \in \mathbb{R}^{N_{{NMDA}_{pCtx}}}$ with $N_{{NMDA}_{pCtx}}$ the number of NMDA-type neurons within the posterior cortex.\ $C$ is the membrane capacitance, $v_r \in \mathbb{R}^{N_{{NMDA}_{pCtx}}}$ the resting membrane potential for each neuron of the population and $v_t \in \mathbb{R}^{N_{{NMDA}_{pCtx}}}$ the instantaneous threshold potential for each neuron of the population.\ All the elements of vectors $v_r$ and $v_t$ are the same.\  $I_{{NMDA}_{pCtx}}(t) \in \mathbb{R}^{N_{{NMDA}_{pCtx}}}$ is the total synaptic current flowing into every excitatory neuron at time $t$.\ Parameter $a$ is the recovery time constant; $k$ and $b$ are derived from the single-neuron frequency-current curve ($f-I$ curve) by measuring the instantaneous firing-rate versus the net synaptic current.\ We also applied frequency analysis to single neuron models to obtain real-like electrophysiological behaviour of the firing activity within the neurons of the corresponding region of the brain.

\begin{figure}[!ht]
\centering
\includegraphics[width=4.5in]{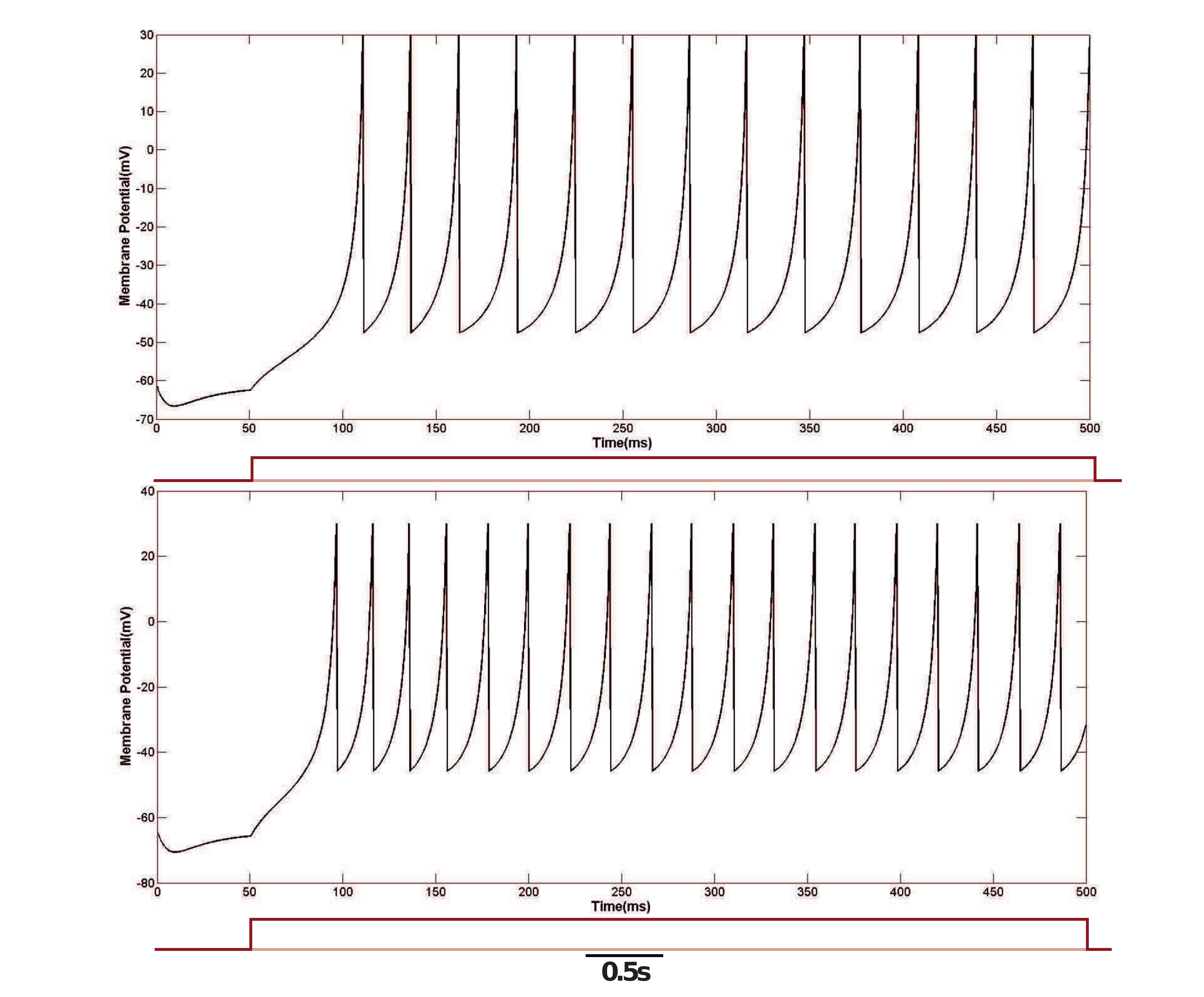}
\caption{Membrane potential of a single NMDA-receptor-type neuron in the posterior cortex. In the figure at the top, a relatively low DC current is applied to an NMDA-type neuron. After a short adaptation period, the NMDA-type neuron reduces its activity. A higher input is considered in the response shown in the figure at the bottom. An NMDA-type neuron never gets too excited even if the injected current is amplified.}
\label{fig:sNMDA}
\end{figure}

We consider the following spike-generation conditions and reset of every element $i$ of vectors $v_{{NMDA}_{pCtx}}$ and $u_{{NMDA}_{pCtx}}$ at $t_{{peak}_{{NMDA}_{pCtx}}}$:

\begin{align}\label{eq:NMDApCtx_reset1}
\text{for all } i, \, { if~~} v_{{NMDA}_{pCtx}}(i) \geq v_{{peak}_{{NMDA}_{pCtx}}} \text{ then} \begin{cases} v_{{NMDA}_{pCtx}}(i)\leftarrow c_{{NMDA}_{pCtx}}\\ u_{{NMDA}_{pCtx}}(i) \leftarrow u_{{NMDA}_{pCtx}}(i)+d_{{NMDA}_{pCtx}} \end{cases} 
\end{align}
with $v_{{peak}_{{NMDA}_{pCtx}}}$ the spike cut-off value, and $c_{{NMDA}_{pCtx}}$ the voltage reset value -- that is, the value of the membrane potential immediately after the neuron fires.\ The parameter $d_{{NMDA}_{pCtx}}$ is tuned to achieve the desired rate of spiking output.\ Responses for a single pyramidal neuron with NMDA-type receptors in the posterior cortex are given in Figure \ref{fig:sNMDA}.

We will use similar models to equations \eqref{eq:NMDApCtx}-\eqref{eq:NMDApCtx_reset1} to describe the expected dynamics of GABAergic neurons of the posterior cortex, medium spiny neurons ($D_1$ and $D_2$-receptor types), fast-spiking interneurons of the striatum, the thalamus, the GPi, the GPe, the STN and the prefrontal cortex.

\begin{figure}
\centering
\includegraphics[width=5.5in]{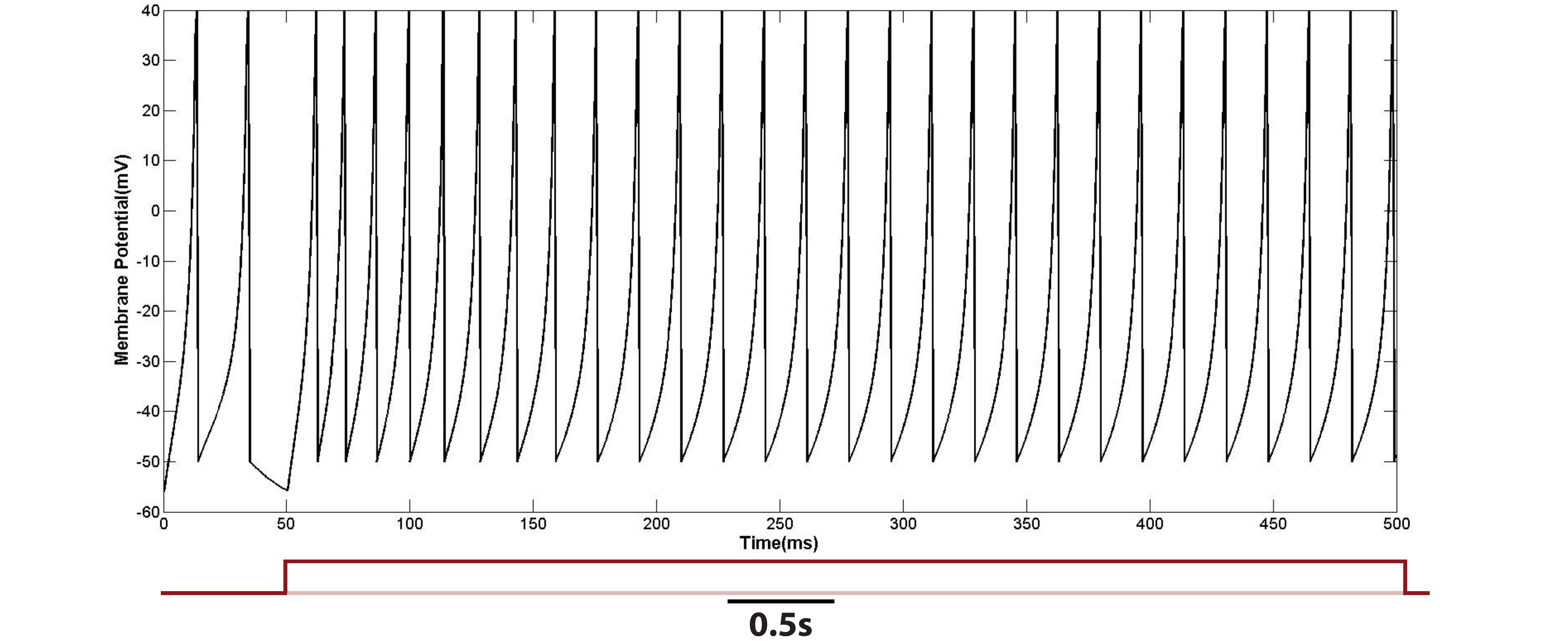}
\caption{Membrane potential of a single GABAergic neuron in the posterior cortex. Even though a low DC current is injected to a GABAergic neuron, it may show high firing activity. As it is observed in NMDA-type neuron dynamics, a low-threshold spiking GABAergic neuron shows spike-frequency adaptation.}
\label{fig:sGABA}
\end{figure}

Inhibitory GABAergic neurons of the posterior cortex generate low-threshold spikes (LTS).\ These neurons can produce high-frequency trains of action potentials \cite{ref30}.\ Similarly to regular-spiking neurons of the cortex, LTS neurons exhibit spike-frequency adaptation.\ They have a low spike threshold, which enables them to generate spikes with high frequencies, but they may also generate instability.\ That is, depending on the membrane potential and the current input, they may also exhibit regular, irregular, and burst firing \cite{ref32}.\ LTS neurons have higher depolarised resting potentials and lower input resistances than regular-spiking neurons \cite{ref33}.\ Inhibition by LTS neurons prevent cortical neural populations from getting overexcited.

The dynamical evolution of the membrane potential and the recovery current for GABAergic inhibitory neurons of the posterior cortex will follow similar equations to \eqref{eq:NMDApCtx}-\eqref{eq:NMDApCtx_reset1}, but substituting the following variables and parameters $v_{{NMDA}_{pCtx}}$, $u_{{NMDA}_{pCtx}}$, $I_{{NMDA}_{pCtx}}$, $v_{{peak}_{{NMDA}_{pCtx}}}$, $c_{{NMDA}_{pCtx}}$ and $d_{{NMDA}_{pCtx}}$ by $v_{{GABA}_{pCtx}}  \in \mathbb{R}^{N_{{GABA}_{pCtx}}}$, $u_{{GABA}_{pCtx}}  \in \mathbb{R}^{N_{{GABA}_{pCtx}}}$, $I_{{GABA}_{pCtx}}  \in \mathbb{R}^{N_{{GABA}_{pCtx}}}$, $v_{{peak}_{{GABA}_{pCtx}}}$, $c_{{GABA}_{pCtx}}$ and $d_{{GABA}_{pCtx}}$, respectively.\ The constant $N_{{GABA}_{pCtx}}$ is the number of GABAergic-type neurons within the posterior cortex.\ The parameters $C$, $k$, $a$, $v_r$, $v_t$, $b$, $c$ and $d$ are different for inhibitory and excitatory neurons.\ Here, the current  $I_{{GABA}_{pCtx}}$ models the total synaptic input current for inhibitory neurons in the posterior cortex.\ The membrane potential behaviour for a single  GABAergic neuron is given in Figure \ref{fig:sGABA}.\ The parameters used for NMDA-type and GABAergic neurons are given in Table \ref{table:pCtx}.

\begin{table}[ht]
\caption{Posterior Cortex Parameters}
\centering
\begin{tabular}{c c c c}
\hline\hline
Parameter & Description & NMDA Receptor & GABA Receptor \\ [0.5ex]
\hline 
$N$ &  Number of neurons & 80 & 20\\
$C$ &  Capacitance & 50pF & 25pF\\
$v_{r}$ & Reset potential for each neuron of the population & -55mV & -56mV\\
$v_{t}$ & Instantaneous threshold potential for each neuron of the population & -45mV & -42mV\\
$k$ & Izhikevich's parameter & 0.2 & 0.2\\
$a$ & '' & 0.13 & 0.03\\
$b$ & '' & -1.5 & 2\\
$c$ & '' & -50 & -50\\
$d$ & '' & 60 & 2\\
$v_{peak}$ & Spike threshold & 30mV & 40mV\\
$\eta_{ext}$ & External input frequency & 40 & -\\ 
$\eta_{background}$ & Background input frequency & 5 & 5\\
$\tau$ & Decaying-effect constant in dynamical connections & 14000 & 5000\\[1ex]
\hline
\end{tabular}
\label{table:pCtx}
\end{table}

Now, we will look more closely at the synaptic currents $I_{{NMDA}_{pCtx}}(t)$ and $I_{{GABA}_{pCtx}}(t)$, which are given in  equations \eqref{eq:INMDApCtx} and \eqref{eq:IGABApCtx}, respectively:

\begin{equation}\label{eq:INMDApCtx}
I_{NMDA_{pCtx}} =I_{Local}-I_{pCtx_{GABA}}+I_{ext}+I_{{back}_{NMDA_{pCtx}}},
\end{equation}

\begin{equation} \label{eq:IGABApCtx}
I_{{GABA}_{pCtx}}=I_{pCtx_{NMDA}}-I^{*}_{Local}+I_{{back}_{GABA_{pCtx}}}.
\end{equation} 

\begin{figure}[!ht]
\centering
\includegraphics[width=3.0in]{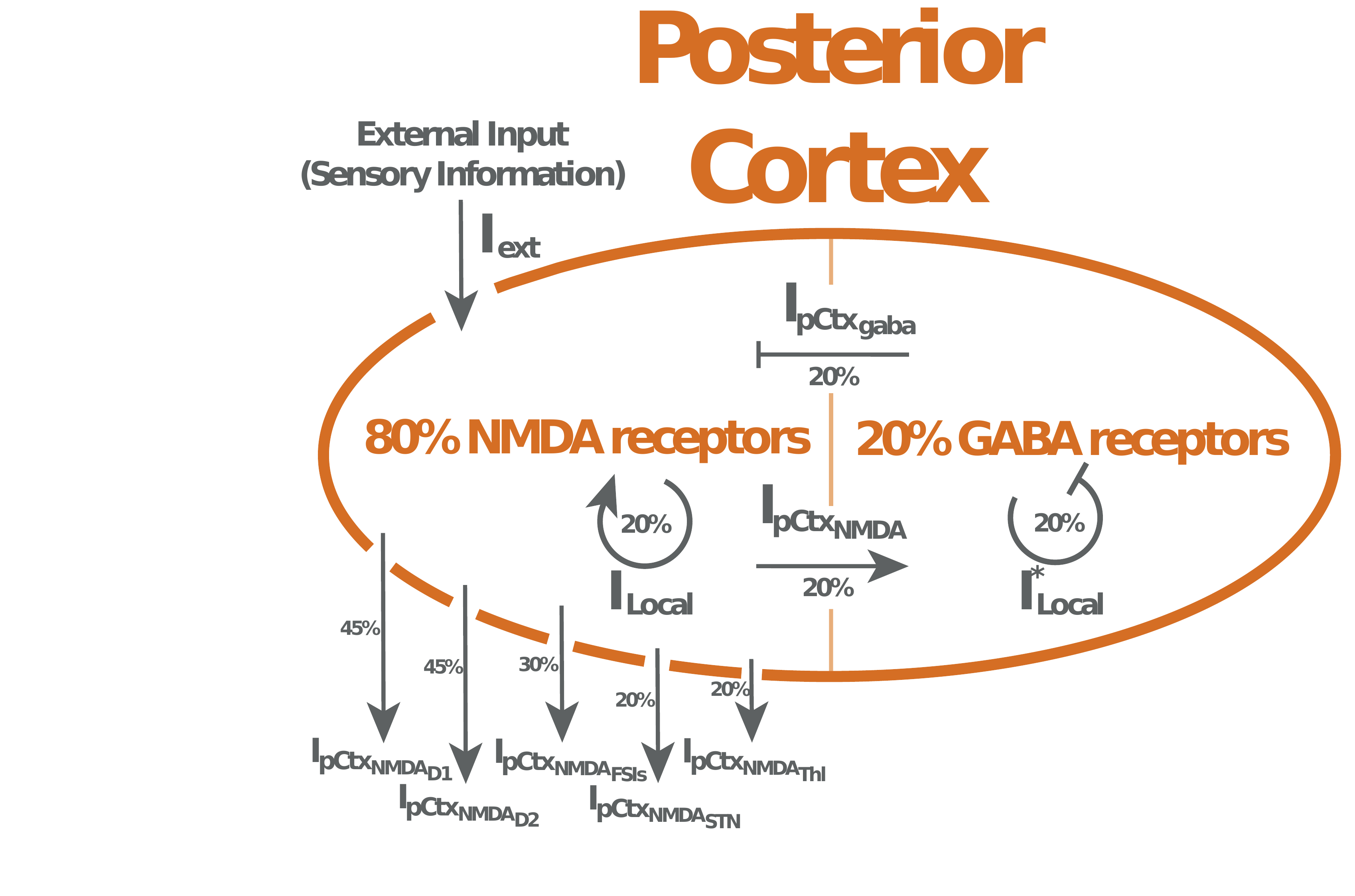}
\caption{The posterior cortex consists of excitatory (glutamatergic with NMDA receptors) and inhibitory (GABAergic) populations of neurons.\ Glutamatergic neurons show regular-spiking patterns, while GABAergic neurons have a low-threshold spiking activity.\ Excitatory neurons receive signals from recurrent connections which come again from excitatory and inhibitory neurons of the posterior cortex.\ The external stimulus is sent to excitatory neurons.\ Inhibitory neurons receive signals from local connections and excitatory neurons.\ Inhibitory neurons do not receive/send any signal from/out of the posterior cortex.\ Excitatory neurons send signals to five different targets, namely: medium spiny neurons ($D_1$ and $D_2$), fast-spiking interneurons in striatum (FSIs), the subthalamic nucleus (STN, hyperdirect pathway input) and the thalamus.\ The percentage values indicate the connection probabilities of considered connections.\ For the posterior cortex, all the internal neurons are connected to each other with a probability of $20\%$.\ The arrow $\longrightarrow$ corresponds to excitatory connections, and $\dashv$ corresponds to inhibitory connections.}
\label{fig:pctx}
\end{figure}

In order to explain these synaptic currents, we will describe how the internal and external connections of the posterior cortex are modelled, following the summary of connections given in Figure \ref{fig:pctx}. 

Figure \ref{fig:pctx} shows the connections of the posterior cortex with their connection probabilities.\ In this section, we will give details for the pre-synaptic inputs for the excitatory NMDA-receptor-type neurons.\ That is, we will explain how we compute $I_{NMDA_{pCtx}}$.\ Input $I_{{GABA}_{pCtx}}$ is calculated in a similar way to $I_{NMDA_{pCtx}}$.

In equation \eqref{eq:INMDApCtx}, $I_{ext}$ represents the external incoming sensory information which will be loaded, maintained or ignored during working memory tasks, according to a subcortical decision.\ The external incoming sensory stimulus is only applied to the posterior cortex, specifically, to the NMDA-receptor-type population.\ The input triggers the activation of the posterior cortex.\ Striatal neurons receive this cortical activity and if the direct pathway in the basal ganglia is dominant, this stimulus may cause an activation in the prefrontal cortex. If the indirect pathway is dominant, the update of the external stimulus is not considered.\ 

The input $I_{{back}_{NMDA_{pCtx}}}$ in equation \eqref{eq:INMDApCtx} represents the random background activity of the MDA-receptor-type population in the posterior cortex. A background input is applied to all the populations of neurons, and is calculated separately for each of them.\ There are two main reasons to use a background input.\ First, the populations in the model do not only receive signals from the populations considered in the basal ganglia's circuit, but they also receive signals from other brain areas, and this activity is represented by the background input.\ Furthermore, the background input provides the possibility to reach the threshold needed for the activation of populations and to allow them to have a random firing activity in the absence of any other input.\ The background and the external inputs are modelled as random values generated with a Poisson probability distribution.\ However, there are two differences between them.\ The external stimulus, which corresponds to sensory information, is a temporary signal. Consequently, it has to be applied for a short-time interval.\ Therefore, in the simulations, this stimulus is only injected between 25-75ms.\ On the other hand, the background input is applied during the total duration of the simulation.\ Another difference between these two inputs is their frequency.\ Since the external input represents sensory stimuli, it is sensible and efficient to start activity in NMDA-type neurons.\ For this reason, the external input $I_{ext}$ is modelled as an input with a higher frequency than the background input.

Poisson probability distributions provide a good description for irregularities in neurons' spike times.\ This formulation generates the probability density of a single-spike train.\ With a Poisson probability  distribution, each spike is generated independently from other spikes with an instantaneous firing rate, and these generated spikes are used as an input.\ Thus, the background and the external inputs can be thought as the firing rate of a pre-synaptic population.\ In the simulations, both inputs are calculated using the $poissrnd$ command from the Statistics Toolbox of MATLAB.\ This function returns the value of a number of spikes which will be used as the background or the external input for each time step and each neuron.\ The parameter $\eta$ describes the firing rate for the background and the external inputs.\ The parameter values are given in the parameter tables for each population as $\eta_{ext}$ and $\eta_{background}$.\ Both  $I_{{back}_{NMDA_{pCtx}}}$ and $I_{ext}$ are vectors of dimension $N_{NMDA_{pCtx}}$, and each element of them corresponds to the input of each neuron of the population.

The currents $I_{pCtx_{GABA}}$ and $I_{Local}$ are also vectors of dimension $N_{{NMDA}_{pCtx}}$.\ The current $I_{{pCtx}_{GABA}}$ models the contribution of inhibitory GABAergic neurons to NMDA-receptor-type neurons within the posterior cortex, and $I_{Local}$ is the recurrent activity of excitatory NMDA-receptor-type neurons in the posterior cortex.\ Recurrent activity of a neural population is crucial, especially for robust maintenance of self-sustained persistent neural activity.\ These synaptic input currents are calculated with the following expressions: 

\begin{equation} \label{eq:recurrent}
\begin{split}
I_{Local}(t) =\sum\limits_{{t_n}} r S_{{NMDA}_{pCtx}}\delta_V (t-t_n), \\
I_{pCtx_{GABA}}(t) =\sum\limits_{{t^*_n}} r S^*_{{pCtx}_{GABA}}\delta^*_V (t-t^*_n).
\end{split}
\end{equation}

The parameter $r$ is a random number which is generated uniformly in the interval $[0,1]$, $\delta_V (t-t_n)$ is a vector of dimension $N_{{NMDA}_{pCtx}} \times 1$ with each element a Dirac delta function $\delta (t-t_n)$ that results in the incremental increase of  $I_{Local}$ at each spike time $t_n$ of a pre-synaptic NMDA-receptor-type neuron in the posterior cortex, and $\delta^*_V (t-t^*_n)$ is a vector of dimension $N_{{GABA}_{pCtx}}\times 1$ with each element a Dirac delta function $\delta (t-t^*_n)$ that results in the incremental increase of $I_{pCtx_{GABA}}$ at each spike time $t^*_{n}$ of a pre-synaptic GABAergic neuron of the posterior cortex.\ The dynamical connections between the pre-synaptic neuron population and the post-synaptic neuron population are represented by the connection matrices $S_{{NMDA}_{pCtx}} \in \mathbb{R}^{N_{{NMDA}_{pCtx}} \times N_{{NMDA}_{pCtx}}}$ and $S^*_{{pCtx}_{GABA}} \in \mathbb{R}^{N_{{NMDA}_{pCtx}} \times N_{{GABA}_{pCtx}}}$.\ We note that, in our model, for the case of an NMDA-type excitatory neuron of the posterior cortex (post-synaptic neuron here), the pre-synaptic neuron that may fire can belong to two populations of neurons, namely: NMDA-type neurons of the posterior cortex and GABAergic inhibitory neurons of the posterior cortex.

Let us consider the computation of the input $I_{Local}$ in equation \eqref{eq:recurrent} for NMDA-type neurons of the posterior cortex.\ In this case, each element $(i,j)$ of the connection matrix  $S_{{NMDA}_{pCtx}}$ is the synaptic strength between the $i$ pre-synaptic NMDA-receptor-type neuron that has fired at time $t_n$ and the $j$ post-synaptic  NMDA-receptor-type neuron.\ The elements $S_{{NMDA}_{pCtx}}(i,j)$, for all $(i, j) \in N_{{NMDA}_{pCtx}} \times  N_{{NMDA}_{pCtx}} $ such that $i=j$, are considered $0$ to indicate that it is not possible to have a connection between a neuron and itself.\ With an abuse of notation each $S_{{NMDA}_{pCtx}}(i,j)$ will be denoted by $s_{{NMDA}_{pCtx}}$ and be referred to as synaptic strength.\ More details on how the connection matrices are built and how they evolve during the simulation are given in Section \ref{dynamical_connections}.\ We consider each synaptic strength $s_{{NMDA}_{pCtx}}$ as a dynamically changing variable,  obtained in the following manner for every time a pre-synaptic neuron (an NMDA-type neuron of the posterior cortex) fires at time $t_n$: 

\begin{itemize}
\item For every $n$, if $t_n \not =t_{{peak}_{{NMDA}_{pCtx}}}$: 
\begin{equation}\label{dyn_connection1}
\frac{ds_{{NMDA}_{pCtx}}(t)}{dt}=-\frac{s_{{NMDA}_{pCtx}}(t)}{\tau_{{NMDA}_{pCtx}}}.
\end{equation}
\item For every $n$, if $t_n =t_{{peak}_{{NMDA}_{pCtx}}}$: 
\begin{equation}\label{dyn_connection2}
\frac{ds_{{NMDA}_{pCtx}}(t)}{dt}=-\frac{s_{{NMDA}_{pCtx}}(t)}{\tau_{{NMDA}_{pCtx}}}+J_{{inc}_{{NMDA}_{pCtx}}} \omega \delta(t-t_n),
\end{equation}
\end{itemize}
 where $t_{{peak}_{{NMDA}_{pCtx}}}$ is the time an $NMDA$-type excitatory neuron of the posterior cortex fires, $\tau_{{NMDA}_{pCtx}}$ is a decaying-effect constant in the dynamical synaptic strength which is given in Table \ref{table:pCtx} as $\tau$. $J_{{inc}_{{NMDA}_{pCtx}}} \in (0,1]$ is a parameter which is the same for each neuron within the same population, and $\omega$ is a parameter with a value between $0$ and $1$ randomly assigned every time the pre-synaptic and the post-synaptic neurons spike simultaneously.\ Thus, the value of $s_{NMDA_{pCtx}}$ is changed with $J_{inc_{NMDA_{pCtx}}}$ and $w$.\ In our dynamical connections, the synaptic strength between two connected neurons decays over time if their firing times do not coincide, but it increases whenever their firing times coincide. These parameters, which define the connection dynamics, are given in Table  \ref{table:connections}.

In an analogous way, we compute each element of the connection matrix  $S^*_{{pCtx}_{GABA}}$. That is,  the synaptic strengths $s^*_{{pCtx}_{GABA}}$ between the pre-synaptic GABAergic neurons of the posterior cortex and the post-synaptic NMDA-type neurons of the posterior cortex are calculated with the following differential equations: 

\begin{itemize}
\item For every $n$, if $t^*_n \not =t_{{peak}_{{NMDA}_{pCtx}}}$: 
\begin{equation}\label{dyn_connection3}
\frac{ds_{{pCtx}_{GABA}}(t)}{dt}=-\frac{s_{{pCtx}_{GABA}}(t)}{\tau_{{NMDA}_{pCtx}}}.
\end{equation}
\item For every $n$, if $t^*_n =t_{{peak}_{{NMDA}_{pCtx}}}$: 
\begin{equation}\label{dyn_connection4}
\frac{ds_{{pCtx}_{GABA}}(t)}{dt}=-\frac{s_{{pCtx}_{GABA}}(t)}{\tau_{{NMDA}_{pCtx}}}+J^{*}_{{inc}_{{NMDA}_{pCtx}}} \omega^{*} \delta(t-t^*_n),
\end{equation}
\end{itemize}

For the computation of $I_{{GABA}_{pCtx}}$, in equation \eqref{eq:IGABApCtx}, the term $I_{pCtx_{NMDA}}$ represents the contribution of $NMDA$-type excitatory neurons of the posterior cortex to GABAergic inhibitory neurons of the posterior cortex; $I^*_{Local}$ and $I_{{back}_{GABA_{pCtx}}}$ are  the recurrent activity and background activity of GABAergic inhibitory neurons of the posterior cortex, respectively.\ $I_{{back}_{GABA_{pCtx}}}$ is obtained from random values generated with a Poisson probability distribution.\  The synaptic input currents $I_{pCtx_{NMDA}}$ and $I^*_{Local}$ are obtained in an analogous way as  \eqref{eq:recurrent}, \eqref{dyn_connection1} and \eqref{dyn_connection2}, substituting $NMDA$ by $GABA$ and considering appropriate dimensions of the involved vectors and matrices depending on the number of neurons of the population of GABAergic neurons and the population of pre-synaptic neurons.

\subsection{Model of the Striatum}

The striatum is the input structure of the  basal ganglia circuit \cite{ref34}.\ Functionally, the striatum is responsible for the coordination of simple body movements and for the processing of information in complex cognitive processes \cite{ref35}.\  The cortical mantle projects  onto the striatum.\ This cortical information is processed within the striatum and passed through the direct and indirect pathways to the output nuclei of the basal ganglia \cite{ref14}.

The proposed model of the striatum consists of three different populations: $D_1$-receptor type neurons, $D_2$-receptor type neurons and fast-spiking interneurons (FSIs).\ $D_1$ and $D_2$-receptor type neurons are medium spiny projection neurons (MSNs).\ MSNs are GABAergic inhibitory cells and account for nearly the 95\% of the total population of striatal neurons \cite{ref36}.\  MSNs are the main target of the neurotransmitter dopamine, most of which is supplied by the substantia nigra pars compacta (SNc) \cite{ref37}.  

Striatal interneurons represent local GABAergic interneurons which are also known as fast-spiking interneurons (FSIs) \cite{ref38}.\ They form the main inhibitory input to the MSNs and provide a winner-takes-all mechanism for MSNs \cite{ref39}.\  When the direct pathway is needed to be dominant over the indirect pathway, FSIs GABAergic input to the $D_2$-receptor-type neurons is scaled up to favour the direct pathway.\ In a similar way, when the indirect pathway is dominant, FSIs inhibit $D_1$-type receptor MSNs activity more than $D_2$-type MSNs.\  By maintaining or releasing their inhibitory output on MSNs, FSIs influence the entire output of the circuit.\ FSIs reflect the MSN-$D_1$-type dynamics on their membrane potentials and exhibit fast-spiking patterns \cite{ref40}.

\begin{figure}[!ht] 
\centering
\includegraphics[width=5.0in]{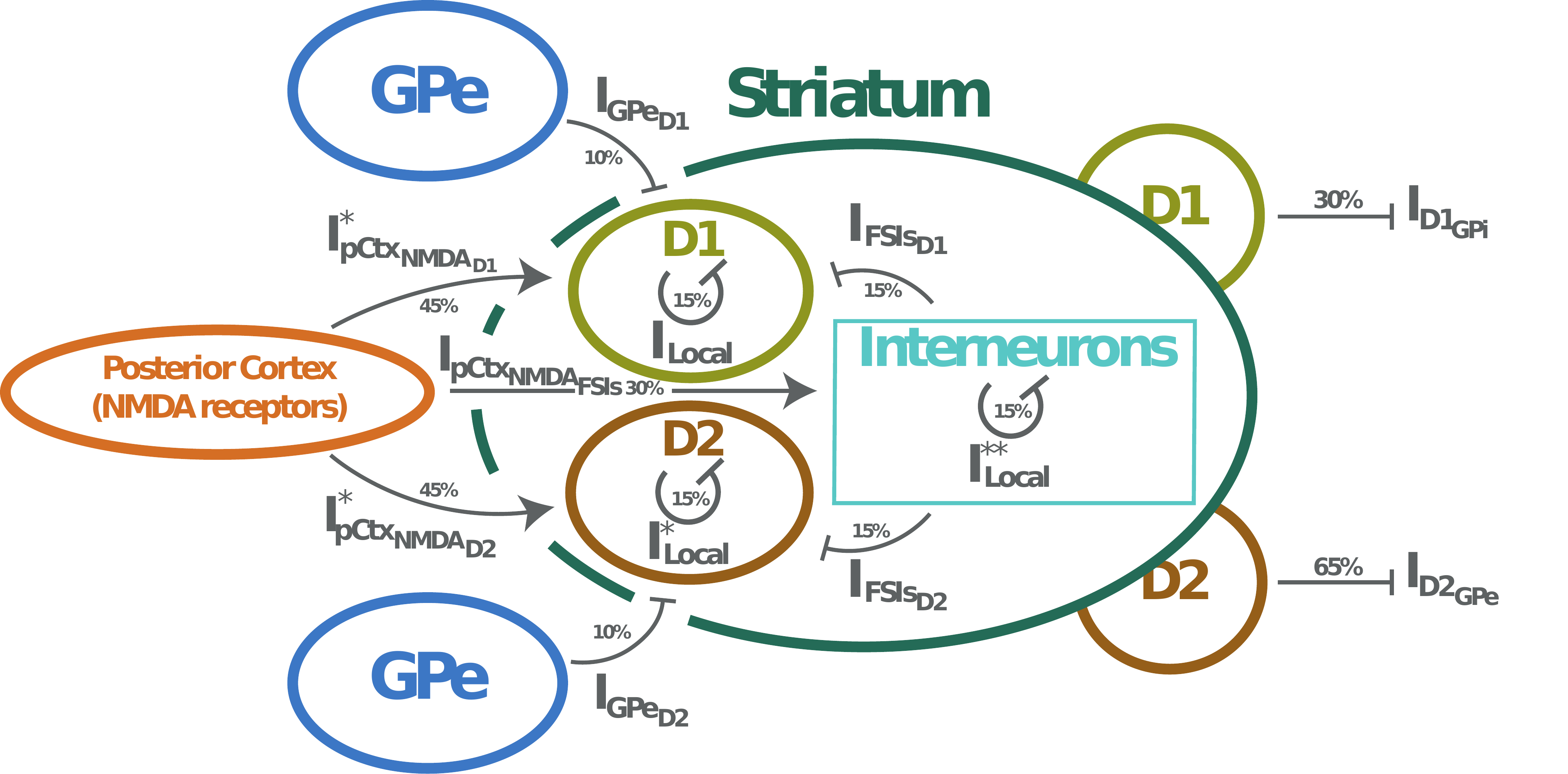}
\caption{The striatum consists of three different neural populations: $D_1$-receptor type neurons, $D_2$-receptor type neurons and fast-spiking interneurons (FSIs).\ $D_1$ and $D_2$-receptor  type neurons are medium spiny neurons (MSNs).\ $D_1$ and $D_2$-receptor type neurons receive cortical inputs from excitatory neurons of the posterior cortex.\ Other inputs are the inhibitory contributions from the GPe, the FSIs and local recurrent connections.\ FSIs have direct input from excitatory neurons of the posterior cortex and from local recurrent connections.\ MSNs have two main targets: outputs from $D_1$-MSNs to the GPi (as part of the direct pathway) and outputs from $D_2$-MSNs to the GPe (as part of the indirect pathway).\ The percentage values indicate connection probabilities of considered connections.\ The arrow $\longrightarrow$ corresponds to excitatory connections, and $\dashv$ corresponds to inhibitory connections.}
\label{fig:str}
\end{figure}

Striatal neurons also receive inhibitory inputs from the GPe.\ $D_1$-receptor type neurons provide inhibitory feedback to the GPi so that the direct pathway is favoured. $D_2$-receptor type neurons project onto the GPe for the indirect pathway to be dominant.\ Other striatal inputs are originated by local recurrent neural activity.\ In Figure \ref{fig:str}, we present the populations of neurons considered in our model of the striatum.

Inspired by \cite{ref27}  and \cite{ref33}, the dynamics for the membrane potential ($v_{D_1}$) and the recovery current ($u_{D_1}$) for  striatal MSNs with $D_1$-type receptors are:

\begin{equation} \label{MSN_D1}
\begin{split}
C\frac{dv_{D_1}(t)}{dt} & = k \bigl[v_{D_1}(t)-v_r\bigr]\bigl[v_{D_1}(t)-v_t\bigr]-u_{D_1}(t)+I_{D_1}(t)+\phi_1 g_{DA}\bigl[v_{D_1}(t)-E_{DA}\bigr],\\
\frac{du_{D_1}(t)}{dt} & = a\biggl\{b\bigl[v_{D_1}(t)-v_r\bigr]-u_{D_1}(t)\biggr\},
\end{split}
\end{equation}
where $v_{D_1}  \in \mathbb{R}^{N_{D_1}}$, $u_{D_1}  \in \mathbb{R}^{N_{D_1}}$ with $N_{D_1}$ the number of MSNs with $D_1$-type receptors within the striatum.\ $C$ is the membrane capacitance, $v_r \in \mathbb{R}^{N_{D_1}}$ the resting membrane potential for each neuron of the population and $v_t \in \mathbb{R}^{N_{D_1}}$ the instantaneous threshold potential for each neuron of the population.\ All the elements of vectors $v_r$ and $v_t$ are the same.\  $I_{D1}(t) \in \mathbb{R}^{N_{D_1}}$ is  the total synaptic current flowing into every neuron at time $t$.\ Parameter $a$ is the recovery time constant; $k$ and $b$ are derived from the single-neuron frequency-current curve ($f-I$ curve) by measuring the instantaneous firing-rate versus the net synaptic current.\ The constant $\phi_1 \in [0,1]$ expresses the proportion of active dopamine $D_1$-type receptors and  is important in our model for the dominance of the the direct and indirect pathways of the basal ganglia.\ Values of $\phi_1$ close to $1$ result in over activation of $D_1$-type receptors by modulating both cortical input (see equation for $I^{*}_{pCtx_{{NMDA}_{D_1}}}(t)$ below) and the membrane potential dynamics.\ For $D_2$-type receptors, we use the constant $\phi_2\in [0,1]$, which appears in the membrane potential dynamics for MSNs with $D_2$-type receptors (equation \eqref{MSN_D2}).\ The higher the value of $\phi_2$ is, the more inhibition in $D_2$ receptors is produced.\ $\phi_1$ and $\phi_2$ can be interpreted as input control parameters.\ In this work, they are considered as constants that are varied to show different dominant states in the network.\ $E_{DA} \in \mathbb{R}^{N_{D_1}}$ and $g_{DA} \in \mathbb{R}$ is the reversal potential for each neuron of the population and the conductance of dopamine regulation, respectively.\ All the elements of vector  $E_{DA}$ are the same.

The spike-generation conditions and reset of every element $i$ of vectors $v_{D_1}$ and $u_{D_1}$ at $t_{{peak}_{D_1}}$:
\begin{align}\label{eq:MSND1_reset}
\text{for all } i, \, \text{if~~} v_{D_1}(i) \geq v_{{peak}_{D_1}} \text{ then} \begin{cases} v_{D_1}(i)\leftarrow c_{D_1}\\ u_{D_1}(i) \leftarrow u_{D_1}(i)+d^i_{D_1}\\ d^i_{D_1} \leftarrow d^i_{D_1}(1-L\phi_1) \end{cases} 
\end{align}
where  $v_{{peak}_{D_1}}$ is the spike cut-off value, $c_{D_1}$ is the voltage reset value -- that is, the value of the membrane potential immediately after the neuron fires, and the initial value of  $d^i_{D_1}, \, \forall i=1,\ldots,N_{D_1}$, is tuned to achieve the desired rate of spiking output.\ The initial value of $d^i_{D_1}$ is the same for every neuron $i$ within the population, but as soon as the simulation starts, $d^i_{D_1}$ may change for each neuron within the population following equation \eqref{eq:MSND1_reset}.\  Finally, parameter $L\in[0,1]$ is a scaling coefficient for the $Ca^{2+}$ current effect.\ $C$, $k$, $a$, $v_r$, $v_t$, $b$, $c$ and $d$ play the same role within the model of each type of neuron, but they have different values for each type of neuron.

For the synaptic currents, we have:
\begin{equation} \label{eq:MSNs D1 afferents1}
\begin{split}
&I_{D_1}(t)=I^{*}_{pCtx_{{NMDA}_{D_1}}}(t)-I_{Local}(t)-(1-\beta_{1}\phi_{1})I_{FSIs_{D_1}}(t)-I_{GPe_{D_1}}(t)+I_{back_{D_1}}(t),\\
&I^{*}_{pCtx_{{NMDA}_{D_1}}}(t)=(1+\beta_{1} \phi_{1})I_{pCtx_{{NMDA}_{D_1}}}(t). 
\end{split}
\end{equation}

 All the currents in \eqref{eq:MSNs D1 afferents1} are vectors of dimension $N_{D_1}$.\ The current $I_{{pCtx}_{{NMDA}_{D_1}}}$ models the excitatory contribution of pyramidal neurons with NMDA-receptors from the posterior cortex, $I_{{FSIs}_{D_1}}$ the inhibitory contribution from FSIs in the striatum, $I_{{GPe}_{D_1}}$ the inhibitory contribution from GPe neurons, and $I_{back_{D_1}}$ the random background input to the $D_1$-MSN population.\ $I_{back_{D_1}}$ is obtained from random values generated with a Poisson probability distribution; in this manner, we can induce random neuronal spike trains in our model.\ $I_{Local}$ models the recurrent activity of MSNs with $D_1$-type receptors.\ $\beta_1 \in (0,1]$ is a scaling coefficient of the dopamine's effect.\ Each synaptic input current $I_{pCtx_{{NMDA}_{D_1}}}$, $I_{FSIs_{D_1}}$, $I_{GPe_{D_1}}$, $I_{Local}$ is  obtained in an analogous way as in \eqref{eq:recurrent}, \eqref{dyn_connection1} and \eqref{dyn_connection2}, substituting $s_{{NMDA}_{pCtx}}$,  $t_{{peak}_{{NMDA}_{pCtx}}}$, $\tau_{{NMDA}_{pCtx}}$ and $J_{{inc}_{{NMDA}_{pCtx}}}$ by $s_{D_1}$, $t_{{peak}_{D_1}}$, $\tau_{D_1}$ and $J_{{inc}_{D_1}}$, respectively, and considering appropriate dimensions of the connection matrices depending on the number of neurons of the pre- and post-synaptic populations.

For  MSNs with $D_2$-type receptors, the dynamics for the membrane potential ($v_{D_2}$) and the recovery current ($u_{D_2}$) are considered as:

\begin{equation} \label{MSN_D2}
\begin{split}
C\frac{dv_{D_2}(t)}{dt} & = k (1-\alpha \phi_2)\bigl[v_{D_2}(t)-v_r\bigr]\bigl[v_{D_2}(t)-v_t\bigr]-u_{D_2}(t)+I_{D_2}(t),\\
\frac{du_{D_2}(t)}{dt} & = a\biggl\{b\bigl[v_{D_2}(t)-v_r\bigr]-u_{D_2}(t)\biggr\},
\end{split}
\end{equation}
where  $v_{D_2}  \in \mathbb{R}^{N_{D_2}}$, $u_{D_2}  \in \mathbb{R}^{N_{D_2}}$ with $N_{D_2}$ the number of MSNs with $D_2$-type receptors within the striatum, $v_r \in \mathbb{R}^{N_{D_2}}$ is the resting membrane potential for each neuron of the population and $v_t \in \mathbb{R}^{N_{D_2}}$ is the instantaneous threshold potential for each neuron of the population.\ All the elements of vectors $v_r$ and $v_t$ are the same.\ Furthermore,  $\alpha \in (0,1]$ is a scaling coefficient of the dopamine's effect on MSN-$D_2$ neurons.

We consider the following spike-generation conditions and reset of every element $i$ of vectors $v_{D_2}$ and $u_{D_2}$ at $t_{{peak}_{D_2}}$:
\begin{align}\label{eq:MSND2_reset}
\text{for all } i, \, \text{if~~} v_{D_2}(i) \geq v_{{peak}_{D_2}} \text{ then} \begin{cases} v_{D_2}(i)\leftarrow c_{D_2}\\ u_{D_2}(i) \leftarrow u_{D_2}(i)+d_{D_2} \end{cases} 
\end{align}
The input current $I_{D_2} \in \mathbb{R}^{N_{D_2}}$ is defined as:

\begin{equation} \label{eq:MSNs D2 afferents}
\begin{split}
&I_{D_2}(t)=I^{*}_{pCtx_{{NMDA}_{D_2}}}(t)-I^*_{Local}(t)-(1+\beta_{2} \phi_{2})I_{FSIs_{D_2}}(t)-I_{GPe_{D_2}}(t)+I_{back_{D_2}}(t),\\
&I^{*}_{pCtx_{{NMDA}_{D_2}}}(t)=I_{pCtx_{{NMDA}_{D_2}}}(t)(1-\beta_{2} \phi_{2}). 
\end{split}
\end{equation}

 All the currents in \eqref{eq:MSNs D2 afferents} are vectors of dimension $N_{D_2}$.\ The current $I_{{pCtx}_{{NMDA}_{D_2}}}$ models the excitatory contribution of pyramidal neurons with NMDA-receptors from the posterior cortex, $I_{{FSIs}_{D_2}}$ models the inhibitory contribution from FSIs in the striatum, $I_{{GPe}_{D_2}}$ represents the inhibitory contribution from GPe neurons, and $I_{back_{D_2}}$ denotes the random excitatory background input to the striatal $D_2$-MSNs.\ $I_{back_{D_2}}$ is obtained from random values generated with a Poisson probability distribution.\ $I^*_{Local}$ models the recurrent activity of MSNs with $D_2$-type receptors.\ $\beta_2 \in (0,1]$ is a scaling coefficient of the dopamine's effect.\ Each synaptic input current  $I_{pCtx_{{NMDA}_{D_2}}}$, $I_{FSIs_{D_2}}$, $I_{GPe_{D_2}}$, $I^*_{Local}$ is  obtained in an analogous way as in equations \eqref{eq:recurrent}, \eqref{dyn_connection1} and \eqref{dyn_connection2}, substituting $s_{{NMDA}_{pCtx}}$,  $t_{{peak}_{{NMDA}_{pCtx}}}$, $\tau_{{NMDA}_{pCtx}}$ and $J_{{inc}_{{NMDA}_{pCtx}}}$ by $s_{D_2}$, $t_{{peak}_{D_2}}$, $\tau_{D_2}$ and $J_{{inc}_{D_2}}$, respectively, and considering appropriate dimensions of the connection matrices depending on the number of neurons of the pre- and post-synaptic populations.

\begin{figure}[!ht]
\centering
\includegraphics[width=4.0in]{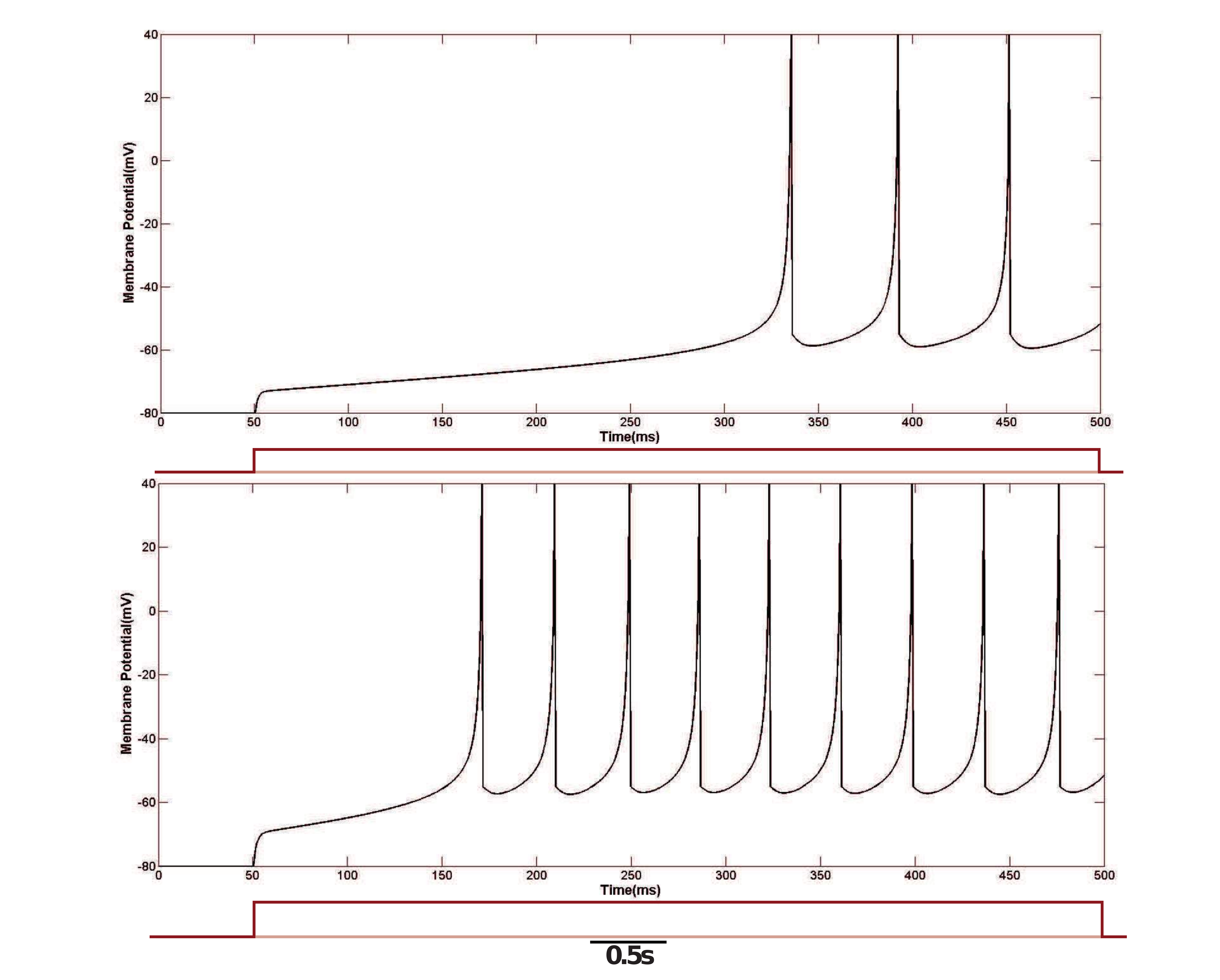}
\caption{Membrane potential of a single MSN in the striatum (for both $D_1$ and $D_2$-receptor-type neurons). Without a stimulation, an MSN stays silent with a hyperpolarised membrane potential. When a low DC current is applied (figure at the top), the MSN shows spike latencies, and a sufficient cortical input is needed to cause a response. In the figure at the bottom, a relatively high current is applied and the MSN reduces the latency time and shows a direct response to input current.}
\label{fig:sMSNs}
\end{figure}

Finally, the dynamics of FSIs are described by:

\begin{equation} \label{FSI}
\begin{split}
C\frac{dv_{FSI}(t)}{dt} & = k \bigl[v_{FSI}(t)-v_r(1-\mu \phi_1)\bigr]\bigl[v_{FSI}(t)-v_t\bigr]-u_{FSI}(t)+I_{FSI}(t),\\
\frac{du_{FSI}(t)}{dt} & = a\biggl\{b\bigl[v_{FSI}(t)-v_r\bigr]-u_{FSI}(t)\biggr\},
\end{split}
\end{equation}

 where  $v_{FSI}  \in \mathbb{R}^{N_{FSI}}$, $u_{FSI}  \in \mathbb{R}^{N_{FSI}}$ with $N_{FSI}$ the number of FSIs within the striatum, $v_r \in \mathbb{R}^{N_{FSI}}$ is the resting membrane potential for each neuron of the population and $v_t \in \mathbb{R}^{N_{FSI}}$ is the instantaneous threshold potential for each neuron of the population.\ All the elements of vectors $v_r$ and $v_t$ are the same.\ $\mu \in (0,1]$ is a scaling coefficient of the dopamine's effect on FSIs.\ The parameter $\phi_1$ is the same as the one for $D_1$-type MSNs.

We consider the following spike-generation conditions and reset of every element $i$ of vectors $v_{FSI}$ and $u_{FSI}$ at $t_{{peak}_{FSI}}$:
\begin{align}\label{eq:FSI_reset}
\text{for all } i, \, \text{if~~} v_{FSI}(i) \geq v_{{peak}_{FSI}} \text{ then} \begin{cases} v_{FSI}(i)\leftarrow c_{FSI}\\ u_{FSI}(i) \leftarrow u_{FSI}(i)+d_{FSI} \end{cases} 
\end{align}
 The input current $I_{FSI} \in \mathbb{R}^{N_{FSI}}$ is defined as:
\begin{equation} \label{eq:FSI afferents}
\begin{split}
&I_{FSI}(t)=I_{pCtx_{{NMDA}_{FSIs}}}(t)-I^{**}_{Local}(t)+I_{back_{FSIs}}(t),\\
&I^{**}_{Local}(t)=I^{FSI}_{Local}(t)(1-\epsilon \phi_{2}). 
\end{split}
\end{equation}

  All the currents in \eqref{eq:FSI afferents} are vectors of dimension $N_{FSI}$.\ The current $I_{{pCtx}_{{NMDA}_{FSIs}}}$ models the excitatory contribution of pyramidal neurons with NMDA-receptors from the posterior cortex, and $I_{back_{FSIs}}$ represents the background activity in the FSIs, which is obtained by using a Poisson probability distribution.\ $I^{FSI}_{Local}$ models the recurrent activity of FSIs.\ $\epsilon \in (0,1]$ is a scaling coefficient of the dopamine's effect.\ Each synaptic input current $I_{pCtx_{{NMDA}_{FSIs}}}$, $I^{FSI}_{Local}$ is  obtained in an analogous way as in equations \eqref{eq:recurrent}, \eqref{dyn_connection1} and \eqref{dyn_connection2}, substituting $s_{{NMDA}_{pCtx}}$,  $t_{{peak}_{{NMDA}_{pCtx}}}$, $\tau_{{NMDA}_{pCtx}}$ and $J_{{inc}_{{NMDA}_{pCtx}}}$ by $s_{FSI}$, $t_{{peak}_{FSI}}$, $\tau_{FSI}$ and $J_{{inc}_{FSI}}$, respectively, and considering appropriate dimensions of the connection matrices depending on the number of neurons of the pre- and post-synaptic populations.

\begin{table}[ht]
\caption{Striatum Parameters}
\centering
\begin{tabular}{c c c c c}
\hline\hline
Parameter & Description & MSN $D_1$ & MSN $D_2$ & FSIs \\ [0.5ex]
\hline
$N$ &  Number of neurons & 100 & 100 & 100\\
$C$ &  Capacitance & 70pF & 70pF & 30pF\\
$v_{r}$ & Reset potential for each neuron of the population & -70mV & -70mV & -55mV\\
$v_{t}$ & Instantaneous threshold potential for each neuron of the population & -40mV & -40mV & -40mV\\
$k$ & Izhikevich's parameter & 0.25 & 0.25 & 1\\
$a$ & '' & 0.02 & 0.02 & 0.45\\
$b$ & '' & 5 & 5 & -2\\
$c$ & '' & -65 & -65 & -55\\
$d$ & '' & 0.1 & 0.1 & 2\\
$v_{peak}$ & Spike threshold & 35mV & 35mV & 25mV\\
$\eta_{background}$ & Background input frequency & 5 & 5 & 5\\
$L$ & Scaling coefficient of $Ca^{2+}$ current effect & 0.731 & - & -\\
$\alpha$ & Sensitivity to injection current & - & 0.932 & -\\
$g_{DA}$ & Conductance of dopamine & 13.7 & - & -\\
$E_{DA}$ & Reversal potential of dopamine for each neuron of the population & -68.4 & - & -\\
$\phi$ (Direct Pathway/Indirect Pathway) & Dopamine receptor occupancy & 0.85/0.3 & 0.7/0.3 & -\\
$\beta$ & Scaling coefficient of dopamine effect & 0.850 & 0.656 & -\\
$\epsilon$ & Scaling coefficient of local input & - & - & 0.525\\
$\mu$ & Scaling coefficient of reset membrane potential & - & - & 0.15\\
$\tau$ & Decaying-effect constant in dynamical connections & 10000 & 10000 & 5000\\[1ex]
\hline
\end{tabular}
\label{table:striatum}
\end{table}

The values for the parameters for striatal medium spiny neurons and fast-spiking interneurons used in our model are given in Table \ref{table:striatum}.

\begin{figure}[!ht]
\centering
\includegraphics[width=5.5in]{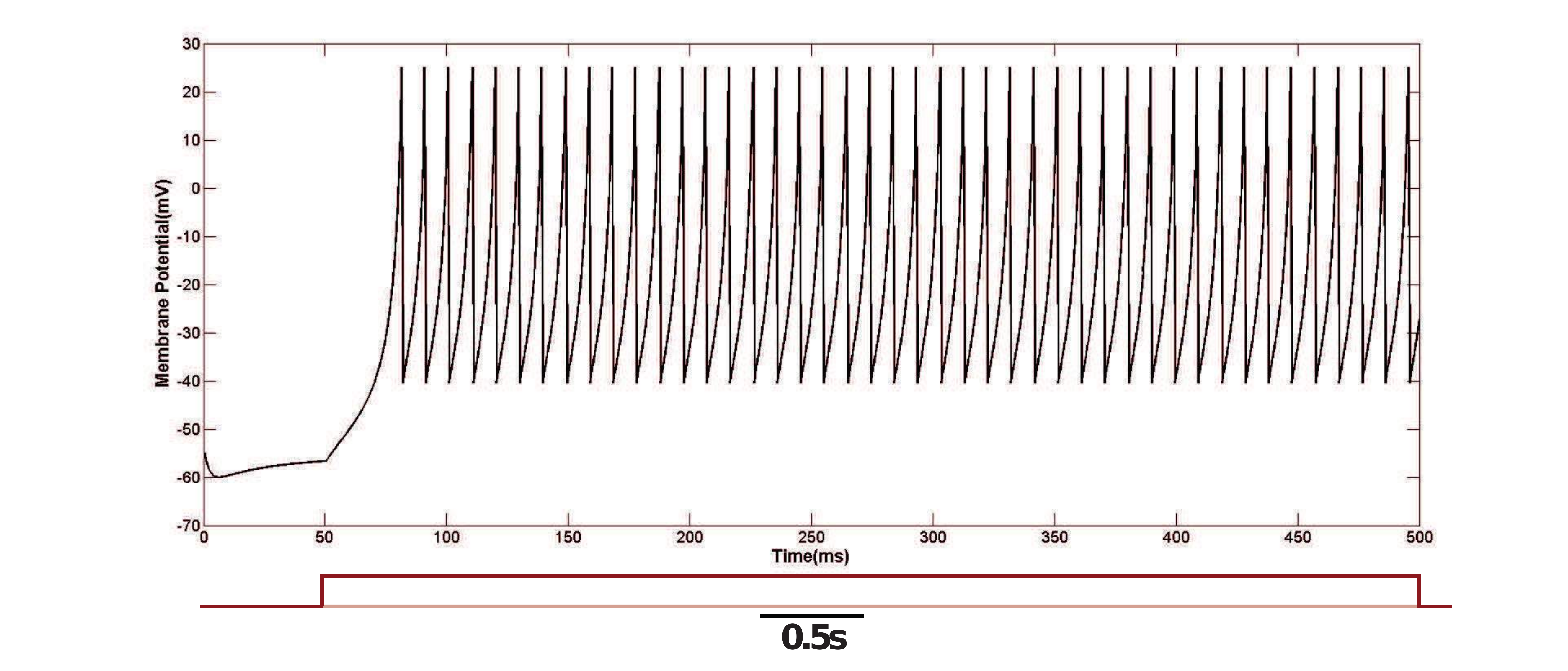}
\caption{Membrane potential of a single FSI in the striatum. FSIs can present fast-spiking activity when stimulated. Their behaviour is distinct from a fast-spiking regime in a LTS neuron. The fast-spiking neuron generates spikes more periodically without frequency-spike adaption.}
\label{fig:sFSIs}
\end{figure}

The dynamical behaviour of the membrane potential for a single MSN and a single FSI obtained by our proposed model for these neurons for parameters of Table \ref{table:striatum} is given in Figures \ref{fig:sMSNs} and \ref{fig:sFSIs}, respectively.

\subsection{Model of the GPe}

The GPe only has connections within the basal ganglia.\ Consequently, the GPe can be considered responsible for the internal regulation of the basal ganglia.\ Although the intrinsic properties of the GPe neurons are not well understood, they are shown to have fast-spiking activity.\ Since most of the afferents of the GPe are inhibitory connections, the GPe should still be capable of having activity within the population.

\begin{figure}[!ht]
\centering
\includegraphics[width=5.0in]{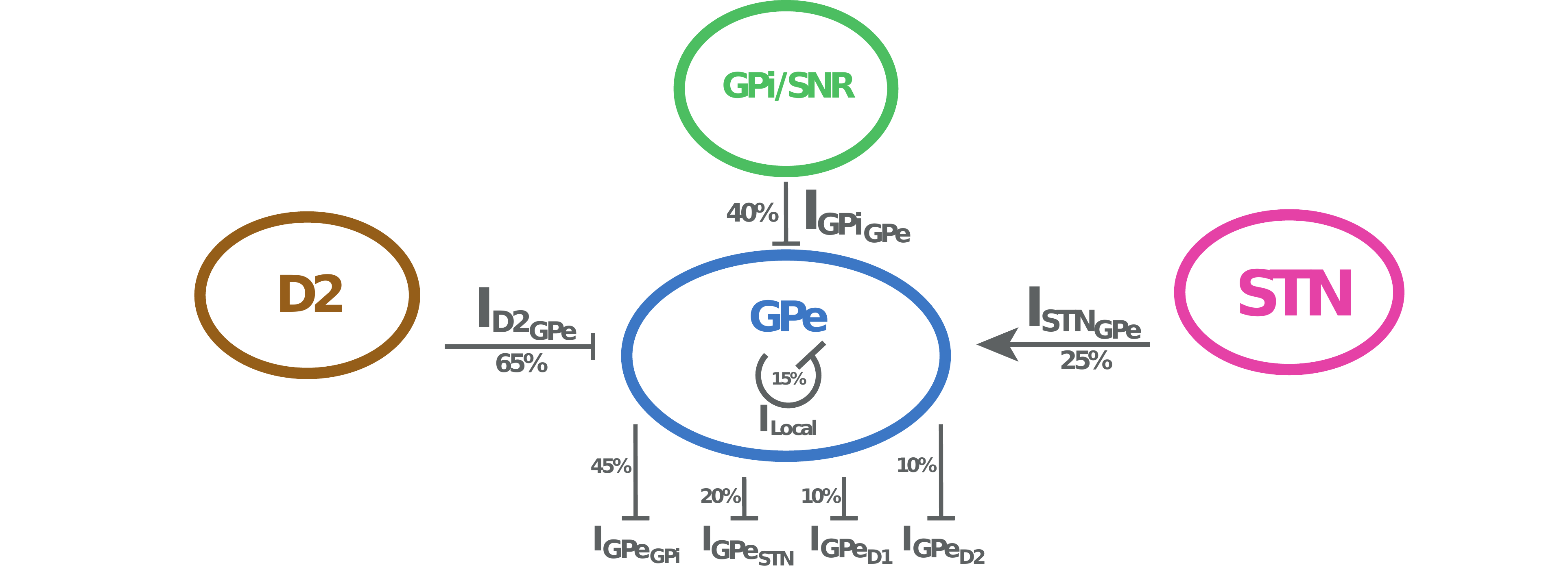}
\caption{The GPe is a key unit for the indirect pathway of the basal ganglia which receives inputs from striatal $D_2$-type MSNs.\ As all the populations, the GPe neurons receive local inputs from its recurrent connections. \ A decrease in dopamine release makes $D_2$-type MSNs to be dominant, which inhibits the GPe.\ This results in the disinhibition of the STN and the inhibition of the thalamus.\ The GPe also receives glutamatergic inputs from the STN.\ In the absence of STN inputs, the activation of these units are expected to be reduced 50\%.\ Reciprocal connections between the STN and the GPe are thought to be important for the regulation of the hyperdirect pathway of the basal ganglia.\ Inhibitory inputs from the GPe inhibits the STN and reduces the activity of the hyperdirect pathway.\ The GPe does not produce any signal that flows out of the basal ganglia.\ This enables the GPe to have an effect on each population of the basal ganglia.\ It projects to the GPi, the STN, and the $D_1$ and $D_2$-type MSNs.\ The probabilities of the considered connections are given with percentages.\ The arrow $\longrightarrow$ corresponds to excitatory connections, and $\dashv$ corresponds to inhibitory connections.}
\label{fig:gpe}
\end{figure}

The GPe has one excitatory connection from the STN ($I_{STN_{GPe}}$) which is related to the hyperdirect pathway regulation. It has also inhibitory inputs from the GPi and $D_2$-type MSNs ($I_{D2_{GPe}}$) as a part of the indirect pathway.\ Moreover, the GPe has recurrent connections.\ Since the GPe has only inhibitory neurons, the recurrent connections ($I_{Local}$) provide inhibitory inputs. In turn, GPe sends information to all the basal ganglia nuclei, providing an internal regulatory mechanism.

The dynamics for neurons of the GPe are described with the same equations as in \eqref{eq:NMDApCtx}-\eqref{eq:NMDApCtx_reset1} substituting $v_{NMDA_{pCtx}}$,
 $u_{NMDA_{pCtx}}$, $I_{NMDA_{pCtx}}$, $v_{peak_{NMDA_{pCtx}}}$, $c_{NMDA_{pCtx}}$ and $d_{NMDA_{pCtx}}$ by $v_{GPe}$, $u_{GPe}$, $I_{GPe}$, $v_{peak_{GPe}}$, $c_{GPe}$ and $d_{GPe}$, respectively.\ Now, $v_{GPe} \in \mathbb{R}^{N_{GPe}}$, $u_{GPe}\in \mathbb{R}^{N_{GPe}}$, $I_{GPe}\in \mathbb{R}^{N_{GPe}}$ with $N_{GPe}$ the number of neurons within the GPe.

The input current $I_{GPe}$ is defined as:

\begin{equation} \label{eq:GPe afferents}
I_{GPe}(t)=I_{{STN}_{GPe}}(t)-I_{{D_2}_{GPe}}(t)-I_{{GPi}_{GPe}}(t)-I_{Local}(t)+I_{back_{GPe}}(t).
\end{equation}

 All the currents in \eqref{eq:GPe afferents} are vectors of dimension $N_{GPe}$.\ The current $I_{{STN}_{GPe}}$ models the excitatory contribution of neurons from the STN, $I_{{D_2}_{GPe}}$ is the inhibitory contribution of $D_2$-type MSNs from the striatum, $I_{{GPi}_{GPe}}$ is the inhibitory contribution from the GPi and $I_{back_{GPe}}$ is the random background activity in the GPe, which is obtained by using a Poisson probability distribution.\ $I_{Local}$ models the recurrent activity of the neurons of the GPe.\ All the connections within the GPe are given in Figure \ref{fig:gpe}.\ Each synaptic input current $z$ (where $z$ is one of ${STN}_{GPe}$, ${D_2}_{GPe}$, ${GPi}_{GPe}$, $Local$) is  obtained in an analogous way as in equations \eqref{eq:recurrent}, \eqref{dyn_connection1} and \eqref{dyn_connection2}, substituting $s_{{NMDA}_{pCtx}}$,  $t_{{peak}_{{NMDA}_{pCtx}}}$, $\tau_{{NMDA}_{pCtx}}$ and $J_{{inc}_{{NMDA}_{pCtx}}}$ by $s_{GPe}$, $t_{{peak}_{GPe}}$, $\tau_{GPe}$ and $J_{{inc}_{GPe}}$, respectively,  and considering appropriate dimensions of the connection matrices depending on the number of neurons of the pre- and post-synaptic populations.

\begin{figure}[!ht]
\centering
\includegraphics[width=5.5in]{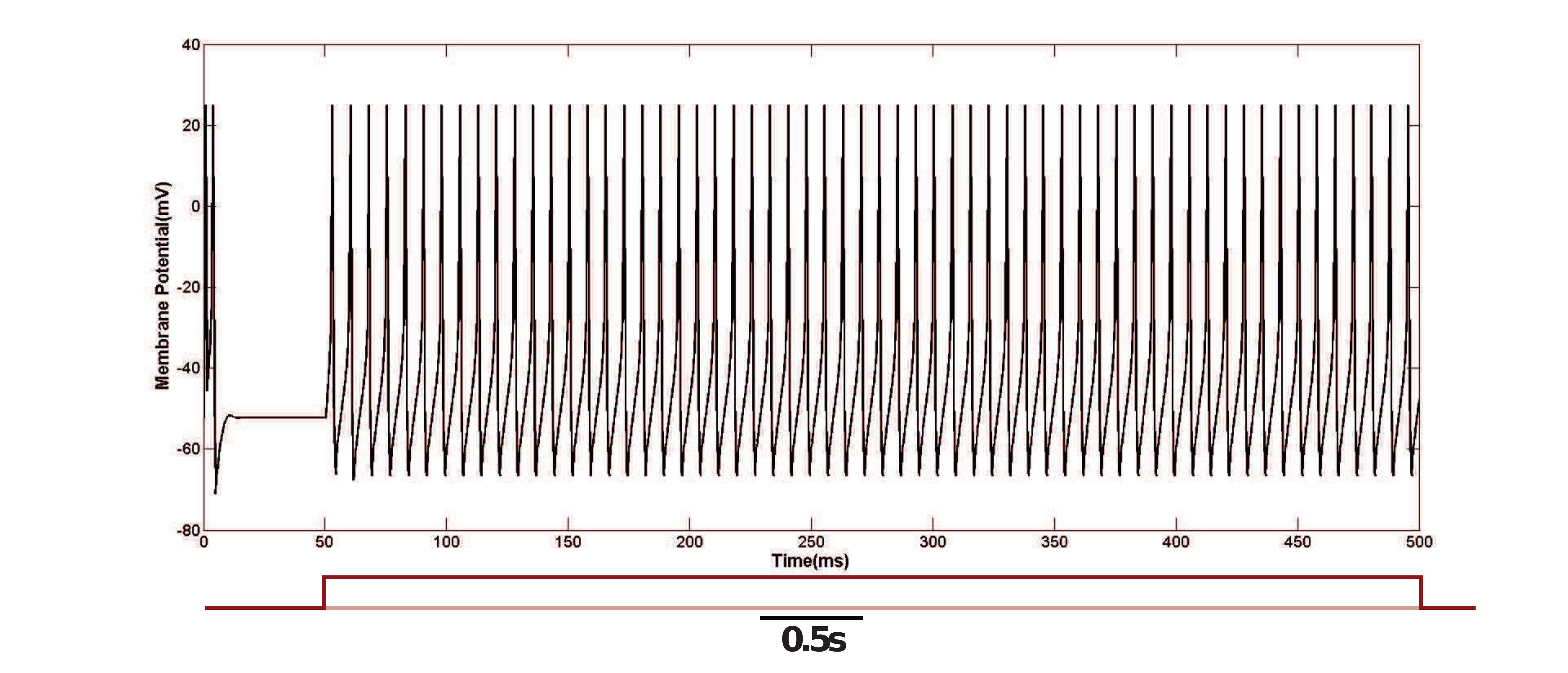}
\caption{Membrane potential of a single neuron of the GPe. As in FSIs, the GPe neurons may show high-frequency activity. The GPe consists of active inhibitory neurons. The single-neuron dynamics are considered in such way so that the GPe neurons can show activity even under the influence of the inhibitory inputs from the GPi and the $D_2$-type MSNs. The model is also able to show transitions between tonic and bursting activity.}
\label{fig:sGPe}
\end{figure}

The membrane potential of a single neuron of the GPe is shown in Figure \ref{fig:sGPe}.\ The values of the parameters used for this simulation are given in Table \ref{table:GPe}.

\begin{table}[ht]
\caption{GP{e} Parameters}
\centering
\begin{tabular}{c c c}
\hline\hline
Parameter & Description & GPe\\ [0.5ex]
\hline 
$N$ &  Number of neurons & 100\\
$C$ &  Capacitance & 5pF\\
$v_{r}$ & Reset potential for each neuron of the population & -55mV\\
$v_{t}$ & Instantaneous threshold potential for each neuron of the population & -45mV\\
$k$ & Izhikevich's parameter & 0.5\\
$a$ & '' & 0.4\\
$b$ & '' & 10\\
$c$ & '' & -50\\
$d$ & '' & 100\\
$v_{peak}$ & Spike threshold & 25mV\\
$\eta_{background}$ & Background input frequency & 5\\
$\tau$ & Decaying-effect constant in dynamical connections & 1000\\[1ex]
\hline
\end{tabular}
\label{table:GPe}
\end{table}

\subsection{Model of the GPi}

The GPi is the output nucleus of the basal ganglia.\ The activity in the GPi is important for our model since it regulates the activity of the thalamus.\ The GPi works as a breaking mechanism for thalamic neurons. Due to the fact that the GPi has only inhibitory neurons, the inhibition of the thalamic cells results in the excitation of the thalamus.

\begin{figure}[!ht]
\centering
\includegraphics[width=5.5in]{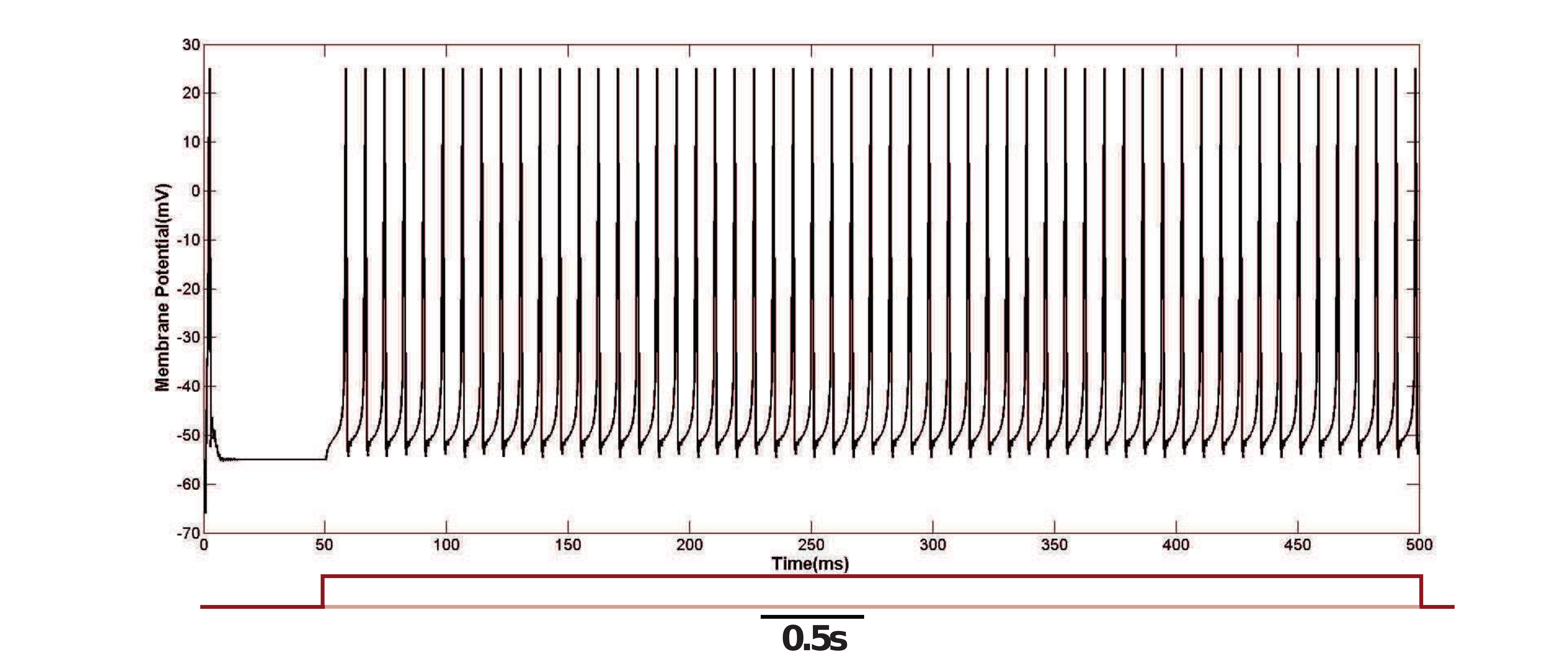}
\caption{Membrane potential of a typical single neuron of the GPi. The neurons in the GPi are highly active in order to provide enough inhibition to thalamic cells. GPi neurons have similar firing properties to GPe neurons, except that they have a higher firing activity.}
\label{fig:sGPi}
\end{figure}

Both the GPe and the GPi are autonomous pacemakers capable of generating fast-spiking activity, even in the absence of excitatory inputs.\ The main difference between the firing patterns of the GPe and the GPi neurons is that the GPi neurons show a higher frequency activity (see Figure \ref{fig:sGPi}).

\begin{figure}[!ht]
\centering
\includegraphics[width=5.0in]{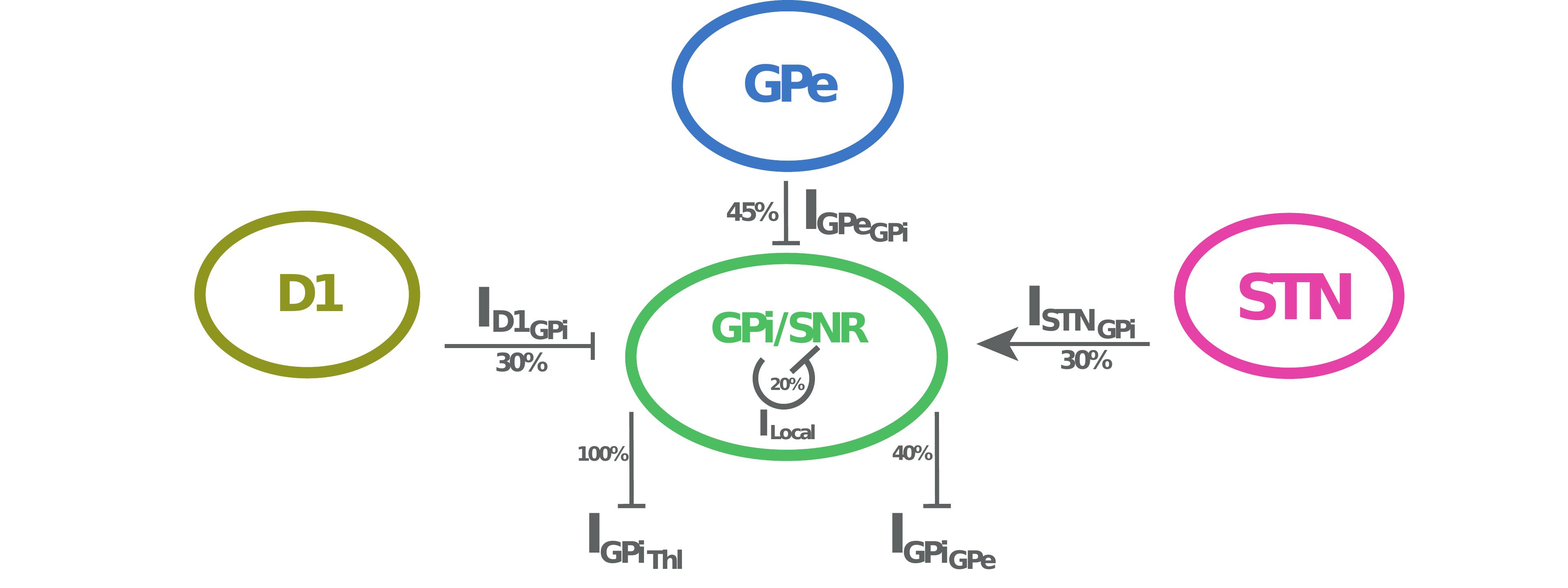}
\caption{The GPi receives inputs from the STN, the GPe and the striatal $D_1$-type MSNs.\ The STN projects in two different ways to the GPi.\ First, when the STN is stimulated by a cortical input, it projects to the GPi with a dominant activity in the hyperdirect pathway.\ Second, the GPi receives an increased input from the STN when the indirect pathway is dominant, providing an extra excitation to the GPi neurons, which results in a reduction of the activity of the thalamus.\ When the direct pathway is dominant, the GPi receives an inhibitory input from striatal $D_1$-type MSNs.\ This input reduces the GPi's activity and results in the excitation of the thalamus.\ The GPi also receives signals from the GPe, which works as a regulatory mechanism for the GPi.\ The input current $I_{Local}$ represents the inhibitory recurrent connections.\ The inhibitory GPi signals affect the thalamic cells and the GPe.\ The probabilities of the connections considered are given by percentages.\ The arrow $\longrightarrow$ corresponds to excitatory connections, and $\dashv$ corresponds to inhibitory connections.}
\label{fig:gpi}
\end{figure}

The dynamics for neurons of the GPi are described with the same equations as in \eqref{eq:NMDApCtx}-\eqref{eq:NMDApCtx_reset1} substituting $v_{NMDA_{pCtx}}$, $u_{NMDA_{pCtx}}$, $I_{NMDA_{pCtx}}$, $v_{peak_{NMDA_{pCtx}}}$, $c_{NMDA_{pCtx}}$ and $d_{NMDA_{pCtx}}$ by $v_{GPi}$, $u_{GPi}$, $I_{GPi}$, $v_{peak_{GPi}}$, $c_{GPi}$ and $d_{GPi}$.\ In this case, $v_{GPi} \in \mathbb{R}^{N_{GPi}}$, $u_{GPi}\in \mathbb{R}^{N_{GPi}}$, $I_{GPi}\in \mathbb{R}^{N_{GPi}}$ with $N_{GPi}$ the number of neurons within the GPi.

The input current $I_{GPi}$ is defined as:

\begin{equation} \label{eq:GPi afferents}
I_{GPi}(t)=I_{{STN}_{GPi}}(t)-I_{{D_1}_{GPi}}(t)-I_{{GPe}_{GPi}}(t)-I_{Local}(t)+I_{back_{GPi}}(t).
\end{equation}

 All the currents in \eqref{eq:GPi afferents} are vectors of dimension $N_{GPi}$.\ The current $I_{{STN}_{GPi}}$ models the excitatory contribution of neurons from the STN, $I_{{D_1}_{GPi}}$ is the inhibitory contribution of $D_1$-type MSNs from the striatum, $I_{{GPe}_{GPi}}$ is the regulatory and inhibitory input from the GPe and $I_{back_{GPi}}$ is the background activity in the GPi, which is obtained with a Poisson probability distribution.\ $I_{Local}$ models the recurrent activity of the neurons of the GPi.\ The main connections of the GPi considered in our model are depicted in Figure \ref{fig:gpi}.\  Each synaptic input current $I_{{STN}_{GPi}}$, $I_{{D_1}_{GPi}}$, $I_{{GPe}_{GPi}}$, $I_{Local}$ is  obtained in an analogous way as in equations \eqref{eq:recurrent}, \eqref{dyn_connection1} and \eqref{dyn_connection2}, substituting $s_{{NMDA}_{pCtx}}$,  $t_{{peak}_{{NMDA}_{pCtx}}}$, $\tau_{{NMDA}_{pCtx}}$ and $J_{{inc}_{{NMDA}_{pCtx}}}$ by $s_{GPi}$, $t_{{peak}_{GPi}}$, $\tau_{GPi}$ and $J_{{inc}_{GPi}}$, respectively, and considering appropriate dimensions of the connection matrices depending on the number of neurons of the pre- and post-synaptic populations.

 Typical values for the parameters for GPi neurons are given in Table \ref{table:GPi}.

\begin{table}[ht]
\caption{GP{i} Parameters}
\centering
\begin{tabular}{c c c}
\hline\hline
Parameter & Description & GPi\\ [0.5ex]
\hline 
$N$ &  Number of neurons & 100\\
$C$ &  Capacitance & 9pF\\
$v_{r}$ & Reset potential for each neuron of the population & -55mV\\
$v_{t}$ & Instantaneous threshold potential for each neuron of the population & -45mV\\
$k$ & Izhikevich's parameter & 1\\
$a$ & '' & 3\\
$b$ & '' & -2\\
$c$ & '' & -55\\
$d$ & '' & 150\\
$v_{peak}$ & Spike threshold & 20mV\\
$\eta_{background}$ & Background input frequency & 4\\
$\tau$ & Decaying-effect constant in dynamical connections & 20000\\[1ex]
\hline
\end{tabular}
\label{table:GPi}
\end{table}

The membrane potential for a single GPi neuron obtained from the model presented above and for the parameters of Table \ref{table:GPi} is shown in Figure \ref{fig:sGPi}.

\subsection{Model of the STN}

The STN plays an important role in movement control \cite{ref41}.\ Most of the projections between the nuclei of the basal ganglia are inhibitory as the projections in the striatum, the GPi and the GPe \cite{ref15}.\ However, the STN projects excitatory glutamatergic outputs to its target nuclei.\ Another key feature of the STN is that it receives direct inputs from the cortex, bypassing the striatum, and is an integral part of the hyperdirect pathway of the basal ganglia \cite{ref40}. 

 In a healthy brain, there are three main different firing patterns of a neuron of the STN.\ In the absence of synaptic stimulation, the cells of the STN fire spontaneously.\ However, by an increased injection, these cells are capable of transiently firing at high frequencies.\ A nonlinearity is also observed in {\em in vivo} recordings; this suggests that when the membrane potential of an STN cell is hyperpolarised below $-65$mV, it can be transiently depolarised \cite{ref45}.\ Furthermore, as observed in patients with Parkinson's disease, the STN neurons generate a burst of spikes \cite{ref45}.

The dynamics for neurons of the STN are described with the same equations as in \eqref{eq:NMDApCtx}-\eqref{eq:NMDApCtx_reset1} substituting $v_{NMDA_{pCtx}}$, $u_{NMDA_{pCtx}}$, $I_{NMDA_{pCtx}}$, $v_{peak_{NMDA_{pCtx}}}$, $c_{NMDA_{pCtx}}$ and $d_{NMDA_{pCtx}}$ by $v_{STN}$, $u_{STN}$, $I_{STN}$, $v_{peak_{STN}}$, $c_{STN}$ and $d_{STN}$.\ In this case, $v_{STN} \in \mathbb{R}^{N_{STN}}$, $u_{STN}\in \mathbb{R}^{N_{STN}}$, $I_{STN}\in \mathbb{R}^{N_{STN}}$ with $N_{STN}$ the number of neurons within the STN.

To allow STN neurons to represent depolarisation when they are hyperpolarised under a certain potential value, we define two different values for parameter $b$ in equation \eqref{eq:NMDApCtx}, which will be different for each neuron $i$ of the STN:

\begin{equation}\label{bSTN}
\begin{cases}
& \text{if } v_{STN}(i) \leq vb_{STN}\text{ then } b=0.485,\\
&\text{if } v_{STN}(i) > vb_{STN}\text{ then } b=0.265,
\end{cases}
\end{equation}
where the constant $vb_{STN}$ is the bursting threshold for STN neurons.

\begin{figure}[!ht]
\centering
\includegraphics[width=6.5in]{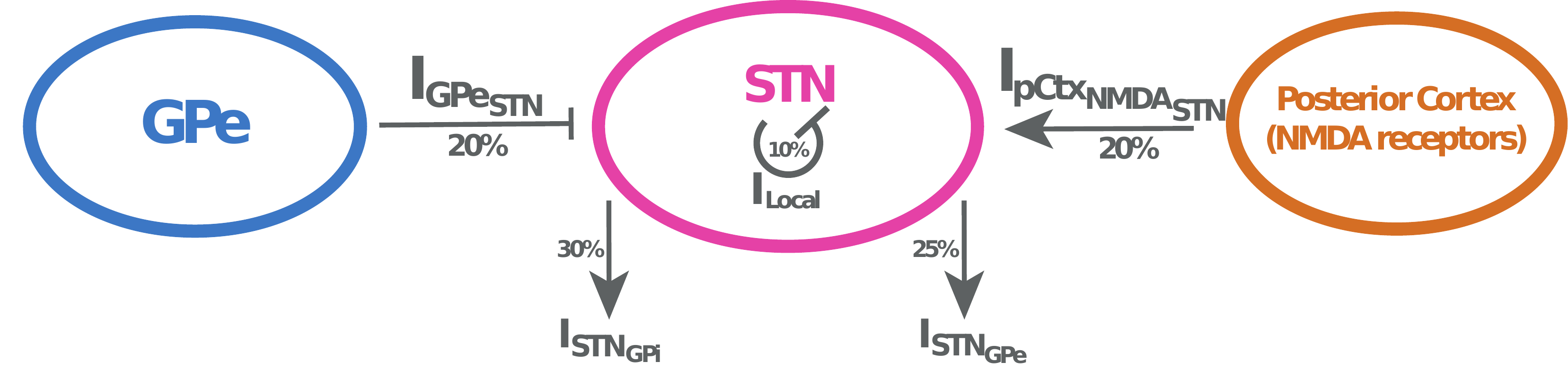}
\caption{The STN receives direct inputs from the posterior cortex and is an integral part of the hyperdirect pathway of the basal ganglia.\ The STN  receives inhibitory afferents from the GPe.\ This projection from the GPe can be interpreted in two different ways.\ On the one hand, the inhibitory signal from the GPe is part of the indirect pathway.\ This inhibitory contribution may be insufficient to suppress the activity of the STN, which results in the excitation of the GPi.\ On the other hand, the STN provides excitatory feedback to the GPe and the excitation of the GPe, in turn, inhibits the STN.\ Consequently, a balance is established between the GPe and the STN.\ Finally, the STN receives excitatory local inputs from recurrent connections.\ The STN excites the GPi and the GPe.\ The probabilities of the considered connections are given by percentages.\ The arrow $\longrightarrow$ corresponds to excitatory connections, and $\dashv$ corresponds to inhibitory connections.}
\label{fig:stn}
\end{figure}

The input current $I_{STN}$ is defined as:
\begin{equation} \label{eq:STN afferents}
I_{STN}(t)=I_{{pCtx}_{{NMDA}_{STN}}}(t)-I_{{GPe}_{STN}}(t)+I_{Local}(t)+I_{back_{STN}}(t).
\end{equation}

 All the currents in \eqref{eq:STN afferents} are vectors of dimension $N_{STN}$.\ The current $I_{{pCtx}_{{NMDA}_{STN}}}$ models the excitatory contribution of neurons from the posterior cortex, $I_{{GPe}_{STN}}$ is the inhibitory contribution of the neurons from the GPe, $I_{Local}$ models the excitatory recurrent activity of the neurons of the STN.\ Moreover, $I_{back_{STN}}$ is the random background activity in the STN, which is obtained with a Poisson probability distribution.\ The main connections of the STN considered in our model are depicted in Figure \ref{fig:stn}.\ Each synaptic input current $I_{{pCtx}_{{NMDA}_{STN}}}$, $I_{{GPe}_{STN}}$, $I_{Local}$ is  obtained in an analogous way as in equations \eqref{eq:recurrent}, \eqref{dyn_connection1} and \eqref{dyn_connection2}, substituting $s_{{NMDA}_{pCtx}}$,  $t_{{peak}_{{NMDA}_{pCtx}}}$, $\tau_{{NMDA}_{pCtx}}$ and $J_{{inc}_{{NMDA}_{pCtx}}}$ by $s_{STN}$, $t_{{peak}_{STN}}$, $\tau_{STN}$ and $J_{{inc}_{STN}}$, respectively, and considering appropriate dimensions of the connection matrices depending on the number of neurons of the pre- and post-synaptic populations.

Typical values for the parameters of STN neurons used in our model are given in Table \ref{table:STN}. 

\begin{table}[ht]
\caption{STN Parameters}
\centering
\begin{tabular}{c c c}
\hline\hline
Parameter & Description & STN\\ [0.5ex]
\hline 
$N$ &  Number of neurons & 100\\
$C$ &  Capacitance & 20pF\\
$v_{r}$ & Reset potential for each neuron of the population & -60\\
$v_{t}$ & Instantaneous threshold potential for each neuron of the population & -45\\
$k$ & Izhikevich's parameter & 1\\
$a$ & '' & 0.005\\
$b$ & '' & 0.265\\
$c$ & '' & -65\\
$d$ & '' & 2\\
$v_{peak}$ & Spike threshold & 30mV\\
$vb$ & Bursting threshold & -65mV\\
$\eta_{background}$ & Background input frequency & 5\\
$\tau$ & Decaying-effect constant in dynamical connections & 8000\\[1ex]
\hline
\end{tabular}
\label{table:STN}
\end{table}

\begin{figure}
\centering
\includegraphics[width=5.5in]{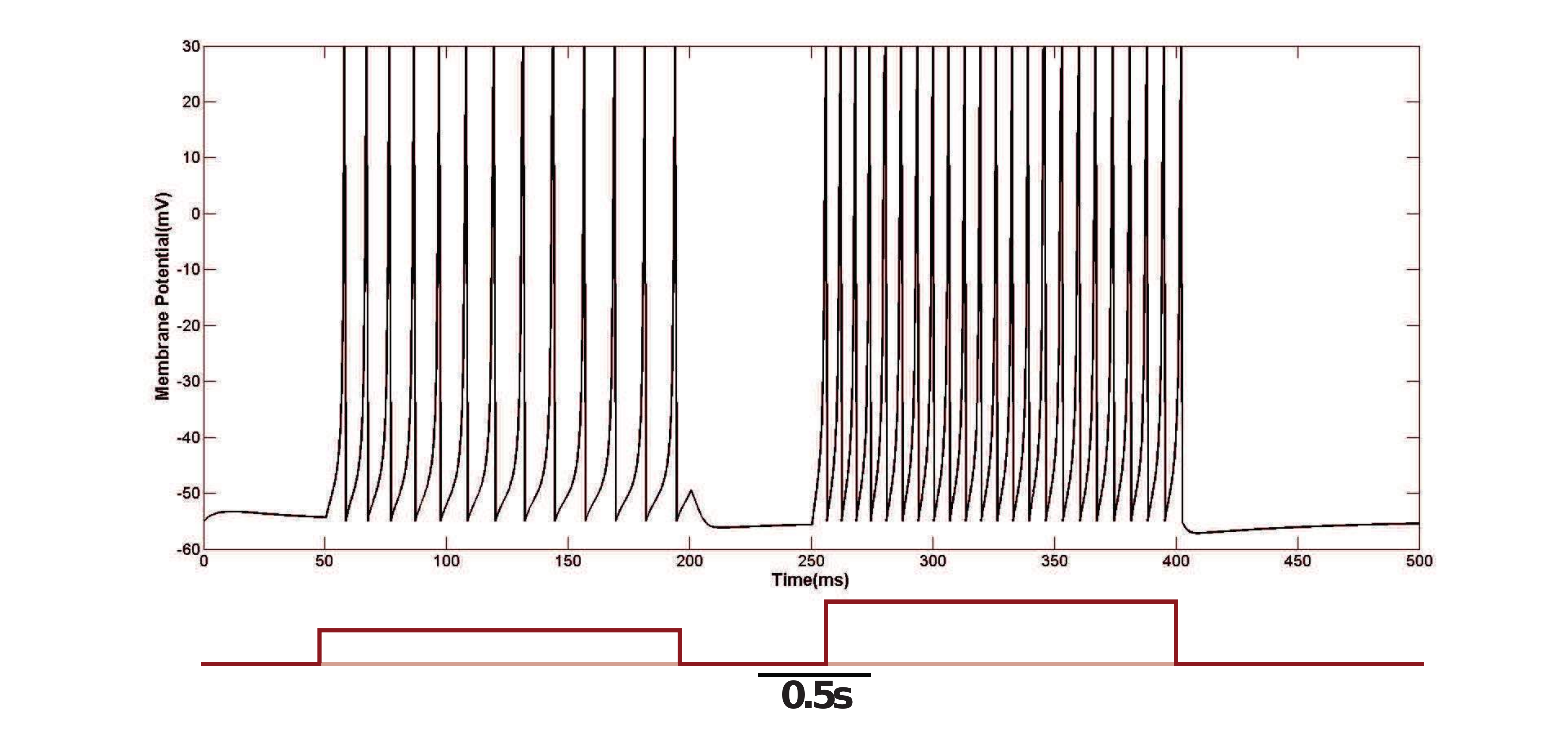}
\caption{Membrane potential for a single neuron of the STN. An STN neuron responds with a high-frequency tonic firing when injected currents get higher. Our model is not capable of showing all the dynamical behaviours of an STN neuron.}
\label{fig:sSTN}
\end{figure}

The behaviour of the membrane potential of a single STN neuron obtained with the proposed model and the parameters of Table \ref{table:STN} is shown in Figure \ref{fig:sSTN}.

\subsection{Model of the Thalamus}

The thalamus is a centrally-located brain structure that controls the flow of information to the cortex.\ In our model, the sensory signals received in the posterior cortex reach the prefrontal cortex by means of the thalamus \cite{ref31}.\ Anatomically, there are three different types of neurons within the thalamus.\ For the sake of simplicity, we consider only two of them: excitatory thalamic neurons and inhibitory reticular nucleus (RTN) neurons.\

While thalamic neurons project to the prefrontal cortex, RTN neurons provide local inhibition for the thalamus.\ The thalamus plays a key role in the proposed working memory network model, since it is the source of background oscillations associated with different working memory processes.

\begin{figure}[!ht]
\centering
\includegraphics[width=4.0in]{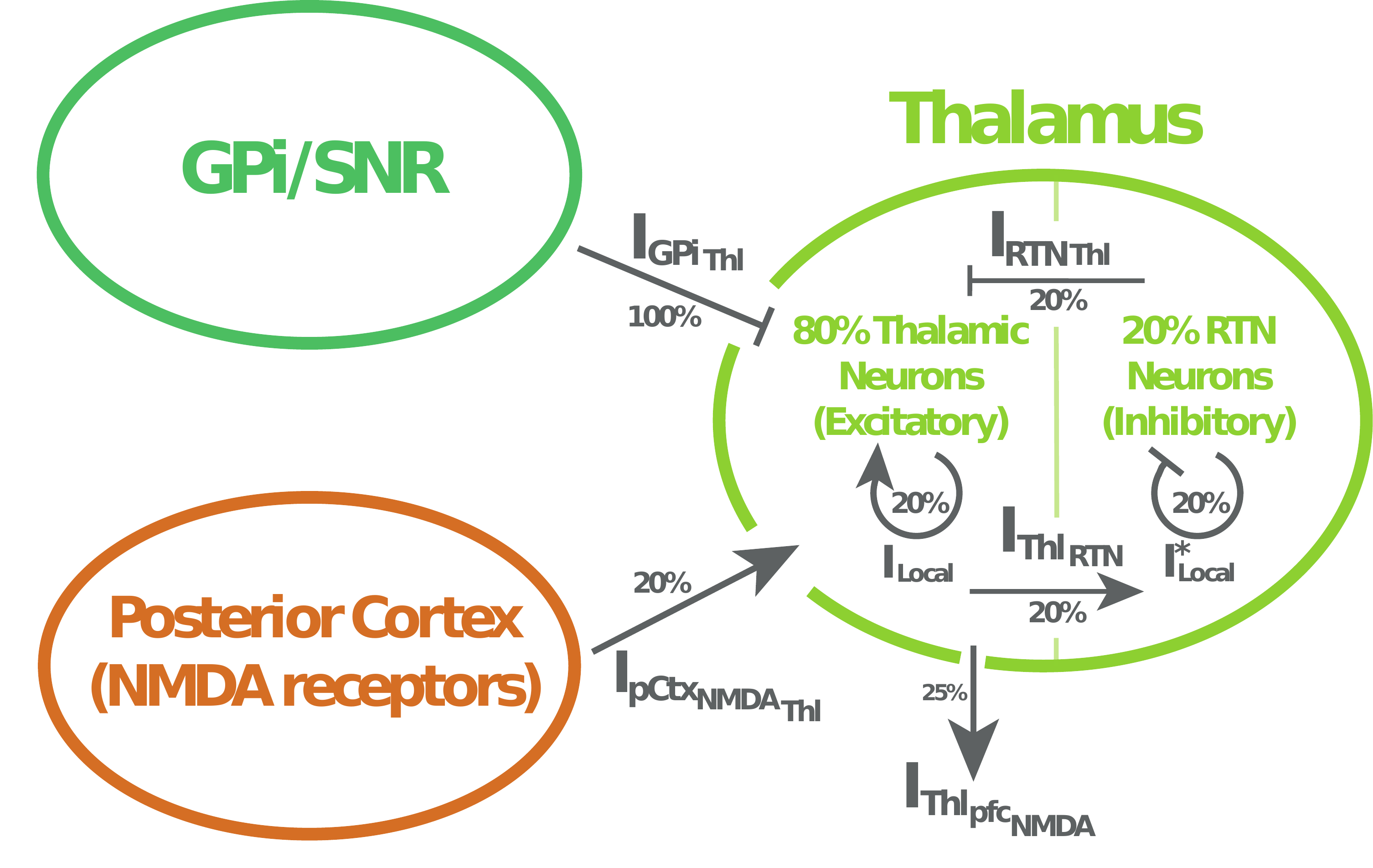}
\caption{Main neuronal populations influencing the thalamus. The GABAergic neurons of the GPi send their axons to specific nuclei of the thalamus.\ This is important since it specifies basal ganglia's message to the prefrontal cortex via the thalamus.\ Inhibition (indirect pathway) or disinhibiton (direct pathway) is a result of the GABAergic input from the GPi to the thalamus.\ Excitatory thalamic neurons also receive local inhibitory input from RTN neurons.\ The excitatory input only comes from NMDA-receptor-type neurons of the posterior cortex, and works as a background input.\ Moreover, thalamic neurons receive excitatory recurrent signals.\ RTN neurons only receive inputs from the thalamic neurons and their recurrent connections.\ Thalamic neurons regulate the activity of the NMDA-receptor-type neurons of the prefrontal cortex.\ The probabilities of the considered connections are given by percentages.\ The arrow $\longrightarrow$ corresponds to excitatory connections, and $\dashv$ corresponds to inhibitory connections.}
\label{fig:thl}
\end{figure}

The dynamics of excitatory thalamic neurons are described by:
\begin{equation} \label{Thalamus}
\begin{split}
C\frac{dv_{Thl}(t)}{dt} & = k \bigl[v_{Thl}(t)-v_r\bigr]\bigl[v_{Thl}(t)-v_t\bigr]-u_{Thl}(t)+I_{Thl}(t),\\
\frac{du_{Thl}(t)}{dt} & = a\biggl\{b\bigl[v_{Thl}(t)-v_r\bigr]-u_{Thl}(t)\biggr\},
\end{split}
\end{equation}
 where  $v_{Thl}  \in \mathbb{R}^{N_{Thl}}$, $u_{Thl}  \in \mathbb{R}^{N_{Thl}}$, $I_{Thl}  \in \mathbb{R}^{N_{Thl}}$ with $N_{Thl}$ the number of excitatory thalamic neurons.\ $v_r \in \mathbb{R}^{N_{Thl}}$ is the resting membrane potential for each neuron of the population and $v_t \in \mathbb{R}^{N_{Thl}}$ is the instantaneous threshold potential for each neuron of the population.\ All the elements of vectors $v_r$ and $v_t$ are the same.

We consider the following spike-generation conditions and reset of every element $i$ of vectors $v_{Thl}$ and $u_{Thl}$ at $t_{{peak}_{Thl}}$, which is a modification of the original reset conditions from \cite{ref33}:
\begin{align}\label{eq:Thl_reset}
\text{for all } i, \, \text{if~~} v_{Thl}(i) \geq v_{{peak}_{Thl}}+0.1 u_{Thl}(i) \text{ then} \begin{cases} v_{Thl}(i)\leftarrow c_{Thl}-0.1u_{Thl}(i)\\ u_{Thl}(i) \leftarrow u_{Thl}(i)+d_{Thl} \end{cases} 
\end{align}

To allow thalamic neurons to have depolarising activity when they are hyperpolarised under a certain potential value, we define two different values for parameter $b$ in equation \eqref{Thalamus}, which will be different for each neuron $i$:

\begin{equation}\label{bThalamus}
\begin{cases}
& \text{if } v_{Thl}(i) \leq vb_{Thl}\text{ then } b=120,\\
&\text{if } v_{Thl}(i) > vb_{Thl}\text{ then } b=2,
\end{cases}
\end{equation}
 where the constant $vb_{Thl}$ is the bursting threshold for thalamic neurons.\ This differentiation is derived from \cite{ref33}.\ However, we take a higher value than the one considered in \cite{ref33}.

The input current $I_{Thl}$ is defined as:
\begin{equation} \label{eq:Thl afferents}
I_{Thl}(t)=I_{{pCtx_{NMDA}}_{Thl}}(t)-I_{{GPi}_{Thl}}(t)-I_{{RTN}_{Thl}}(t)+I_{Local}(t)+I_{back_{Thl}}(t).
\end{equation}

 All the currents in \eqref{eq:Thl afferents} are vectors of dimension $N_{Thl}$.\ The current $I_{{GPi}_{Thl}}$ models the inhibitory contribution of neurons from the GPi, $I_{{RTN}_{Thl}}$ is the inhibitory contribution of the RTN neurons, $I_{{pCtx}_{NMDA_{Thl}}}$ represents the cortical excitatory input and $I_{Local}$ models the excitatory recurrent activity of the thalamic neurons.\ Moreover, $I_{back_{Thl}}$ is the random background activity in the thalamus, which is obtained by using a Poisson probability distribution.\ The main connections of the thalamus considered in our model are depicted in Figure \ref{fig:thl}.\ Each synaptic input current $I_{{GPi}_{Thl}}$, $I_{{RTN}_{Thl}}$, $I_{{pCtx}_{NMDA_{Thl}}}$, $I_{Local}$ is  obtained in an analogous way as in equations \eqref{eq:recurrent}, \eqref{dyn_connection1} and \eqref{dyn_connection2}, substituting $s_{{NMDA}_{pCtx}}$,  $t_{{peak}_{{NMDA}_{pCtx}}}$, $\tau_{{NMDA}_{pCtx}}$ and $J_{{inc}_{{NMDA}_{pCtx}}}$ by $s_{Thl}$, $t_{{peak}_{Thl}}$, $\tau_{Thl}$ and $J_{{inc}_{Thl}}$, respectively, and considering appropriate dimensions of the connection matrices depending on the number of neurons of the pre- and post-synaptic populations.

The dynamics of inhibitory RTN neurons are described by:
\begin{equation} \label{RTN}
\begin{split}
C\frac{dv_{RTN}(t)}{dt} & = k \bigl[v_{RTN}(t)-v_r\bigr]\bigl[v_{RTN}(t)-v_t\bigr]-u_{RTN}(t)+I_{RTN}(t),\\
\frac{du_{RTN}(t)}{dt} & = a\biggl\{b\bigl[v_{RTN}(t)-v_r\bigr]-u_{RTN}(t)\biggr\},
\end{split}
\end{equation}
 where  $v_{RTN}  \in \mathbb{R}^{N_{RTN}}$, $u_{RTN}  \in \mathbb{R}^{N_{RTN}}$, $I_{RTN}  \in \mathbb{R}^{N_{RTN}}$ with $N_{RTN}$ the number of RTN neurons.\ $v_r \in \mathbb{R}^{N_{RTN}}$ is the resting membrane potential for each neuron of the population and $v_t \in \mathbb{R}^{N_{RTN}}$ is the instantaneous threshold potential for each neuron of the population.\ All the elements of vectors $v_r$ and $v_t$ are the same.

We consider the following spike-generation conditions and reset of every element $i$ of vectors $v_{RTN}$ and $u_{RTN}$ at $t_{{peak}_{RTN}}$:
\begin{align}\label{eq:RTN_reset}
\text{for all } i, \, \text{if~~} v_{RTN}(i) \geq v_{{peak}_{RTN}}-0.08 u_{RTN}(i) \text{ then} \begin{cases} v_{RTN}(i)\leftarrow c_{RTN}-0.08u_{RTN}(i)\\ u_{RTN}(i) \leftarrow u_{RTN}(i)+d_{RTN} \end{cases} 
\end{align}

As in the thalamic cell dynamics, we also introduce two different values for parameter $b$ for every neuron $i$ of the population:

\begin{equation}
\begin{cases}
& \text{if } v_{RTN}(i) \leq vb_{RTN}\text{ then } b=100,\\
&\text{if } v_{RTN}(i) > vb_{RTN}\text{ then } b=2,
\end{cases}
\end{equation}
where the constant $vb_{RTN}$ is the bursting threshold for RTN neurons.

The input current $I_{RTN}$ is defined as:
\begin{equation} \label{eq:RTN afferents}
I_{RTN}(t)=I_{{Thl}_{RTN}}(t)-I^*_{Local}(t)+I_{back_{RTN}}(t).
\end{equation}

 All the currents in \eqref{eq:RTN afferents} are vectors of dimension $N_{RTN}$.\ The current $I_{{Thl}_{RTN}}$ models the excitatory contribution of the thalamic neurons, $I^*_{Local}$ models the inhibitory recurrent activity of the RTN neurons.\ Moreover, $I_{back_{RTN}}$ is the random background activity in the RTN neurons, which is obtained by using a Poisson probability distribution.\ Each synaptic input current $I_{{Thl}_{RTN}}$, $I^*_{Local}$ is  obtained in an analogous way as in equations \eqref{eq:recurrent}, \eqref{dyn_connection1} and \eqref{dyn_connection2}, substituting $s_{{NMDA}_{pCtx}}$,  $t_{{peak}_{{NMDA}_{pCtx}}}$, $\tau_{{NMDA}_{pCtx}}$ and $J_{{inc}_{{NMDA}_{pCtx}}}$ by $s_{RTN}$, $t_{{peak}_{RTN}}$, $\tau_{RTN}$ and $J_{{inc}_{RTN}}$, respectively,  and considering appropriate dimensions of the connection matrices depending on the number of neurons of the pre- and post-synaptic populations.

The values of the parameters for both types of neurons considered in our model are given in Table \ref{table:Thalamus}.

\begin{table}[ht]
\caption{Thalamus Parameters}
\centering
\begin{tabular}{c c c c}
\hline\hline
Parameter & Description & Thalamic Cells & RTN \\ [0.5ex]
\hline 
$N$ &  Number of neurons & 80 & 20\\
$C$ &  Capacitance & 10pF & 5pF\\
$v_{r}$ & Reset potential for each neuron of the population & -60mV & -65mV\\
$v_{t}$ & Instantaneous threshold potential for each neuron of the population & -50mV & -45mV\\
$k$ & Izhikevich's parameter & 1 & 0.15\\
$a$ & '' & 0.05 & 0.15\\
$b$ & '' & 2 & 2\\
$c$ & '' & -65 & -55\\
$d$ & '' & 10 & 50\\
$v_{peak}$ & Spike threshold & 35mV & 0mV\\
$vb$ & Bursting threshold & -60mV & -65mV\\
$\eta_{background}$ & Background input frequency & 30 & 5\\
$\tau$ & Decaying-effect constant in dynamical connections & 6000 & 5000\\[1ex]
\hline
\end{tabular}
\label{table:Thalamus}
\end{table}

\begin{figure}[!ht]
\centering
\includegraphics[width=4.5in]{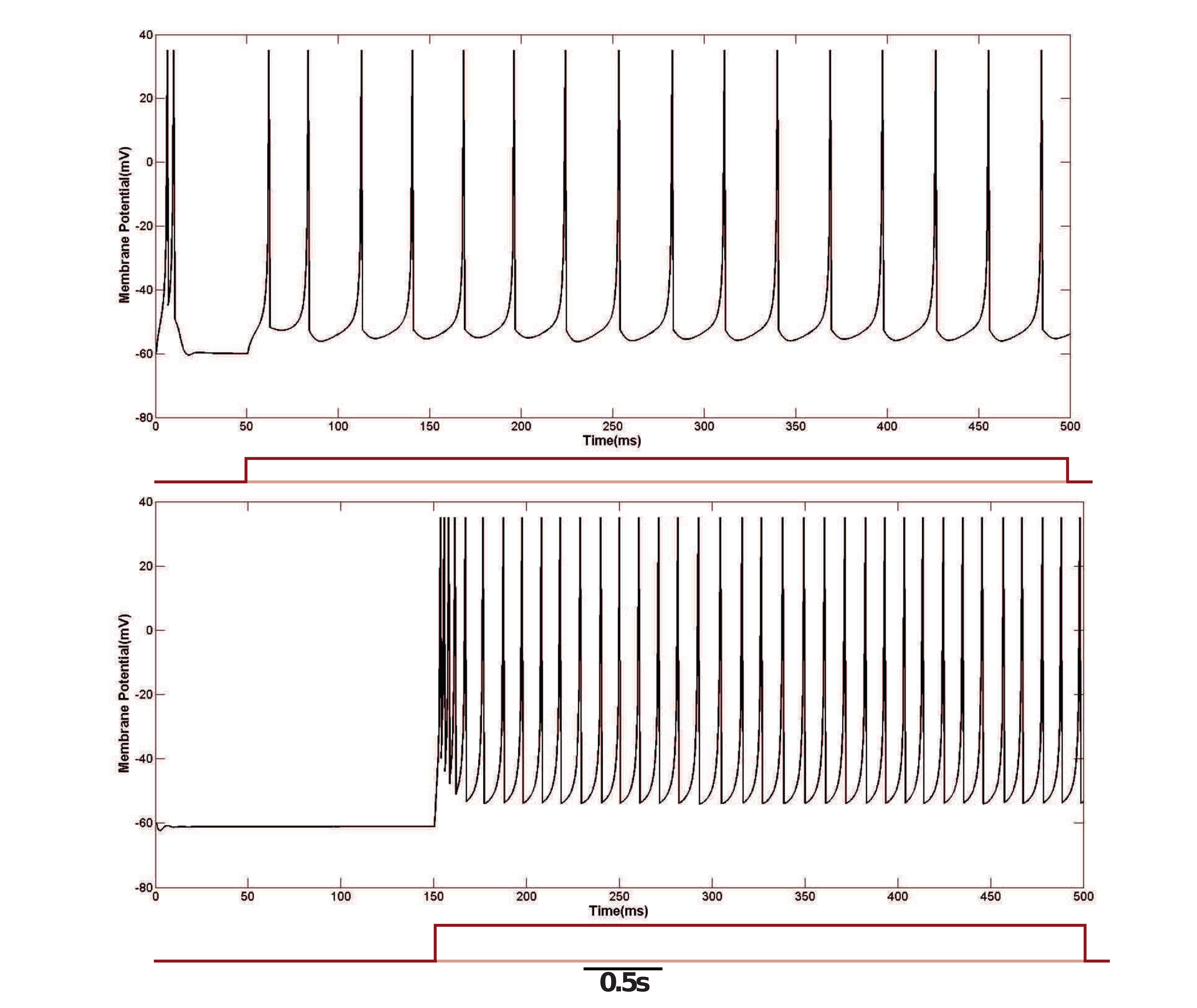}
\caption{Membrane potential of a thalamic cell. Thalamic cells are modelled to show tonic activity while presented with a low-amplitude input and bursting activity in the presence of a high enough input. The figure at the top shows the membrane potential of the neuron when an insufficient, low DC current is applied. In this case, the behaviour of the thalamic neuron is tonic firing. In the figure at the bottom, the thalamic neuron changes its activity to bursting, when a higher current is injected.}
\label{fig:sThalamic}
\end{figure}

\begin{figure}
\centering
\includegraphics[width=4.5in]{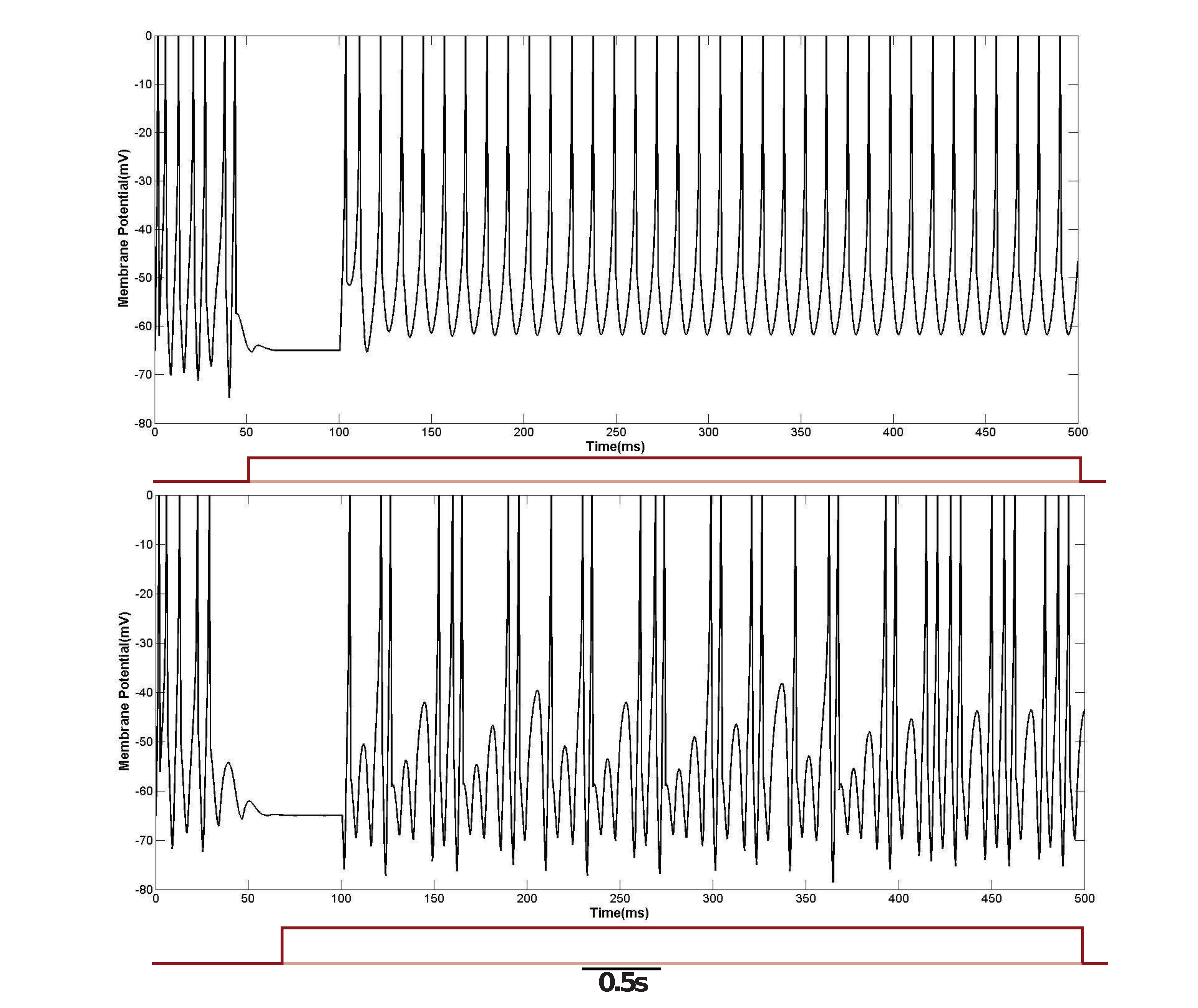}
\caption{Membrane potential of a RTN neuron. RTN neurons show a similar firing pattern to thalamic neurons. The main difference is that RTN neurons present low subthreshold oscillations when stimulated.}
\label{fig:sRTN}
\end{figure}

A simulation for the models presented for a single thalamic neuron and a single RTN neuron for the parameters shown in Table \ref{table:Thalamus} is given in Figure \ref{fig:sThalamic} and Figure \ref{fig:sRTN}, respectively.

\subsection{Model of the Prefrontal Cortex}

The prefrontal cortex (PFC) is the last neuronal structure considered in our model and its activity represents the working memory.\ According to the message from the thalamus, the prefrontal cortex needs to show distinct behaviors, mainly: loading, maintenance or clearance of the information, originated in the posterior cortex.\ In our model, the prefrontal cortex consists of two neuron populations: glutamatergic neurons with NMDA receptors and GABAergic inhibitory neurons, as in the posterior cortex.\ We again choose the ratio of excitatory to inhibitory neurons to be $4$ to $1$.

 The dynamics of NMDA-receptor-type neurons of the prefrontal cortex are described as in \eqref{eq:NMDApCtx}-\eqref{eq:NMDApCtx_reset1} substituting $v_{NMDA_{pCtx}}$, $u_{NMDA_{pCtx}}$, $I_{NMDA_{pCtx}}$, $v_{peak_{NMDA_{pCtx}}}$, $c_{NMDA_{pCtx}}$ and $d_{NMDA_{pCtx}}$ by $v_{NMDA_{pfc}}$, $u_{NMDA_{pfc}}$, $I_{NMDA_{pfc}}$, $v_{peak_{NMDA_{pfc}}}$, $c_{NMDA_{pfc}}$ and $d_{NMDA_{pfc}}$.\ Now, $v_{{NMDA}_{pfc}} \in \mathbb{R}^{N_{{NMDA}_{pfc}}}$, $u_{{NMDA}_{pfc}} \in \mathbb{R}^{N_{{NMDA}_{pfc}}}$, $I_{{NMDA}_{pfc}} \in \mathbb{R}^{N_{{NMDA}_{pfc}}}$ with $N_{{NMDA}_{pfc}}$ the number of NMDA-receptor-type neurons within the prefrontal cortex.

We define:

\begin{equation}\label{eq:INMDApfc}
I_{{NMDA}_{pfc}} =I_{{Thl}_{{pfc}_{NMDA}}}-I_{{pfc}_{GABA}}+I_{Local}+I_{back_{NMDA_{pfc}}}.
\end{equation}

All the currents in \eqref{eq:INMDApfc} are vectors of dimension $N_{{NMDA}_{pfc}}$.\ The current $I_{{Thl}_{{pfc}_{NMDA}}}$ models the excitatory contribution from thalamic neurons, $I_{{pfc}_{GABA}}$ represents the contribution of inhibitory neurons from the prefrontal cortex, and $I_{back_{NMDA_{pfc}}}$ is the random background activity of excitatory neurons in the prefrontal cortex, obtained by using a Poisson probability distribution.\ $I_{Local}$ is the recurrent activity of excitatory neurons in the prefrontal cortex.\ Each synaptic input current $I_{{Thl}_{{pfc}_{NMDA}}}$, $I_{{pfc}_{GABA}}$, $I_{Local}$ is  obtained in an analogous way as in equations \eqref{eq:recurrent}, \eqref{dyn_connection1} and \eqref{dyn_connection2}, substituting $s_{{NMDA}_{pCtx}}$,  $t_{{peak}_{{NMDA}_{pCtx}}}$, $\tau_{{NMDA}_{pCtx}}$ and $J_{{inc}_{{NMDA}_{pCtx}}}$ by $s_{{NMDA}_{pfc}}$, $t_{{NMDA}_{pfc}}$, $\tau_{{NMDA}_{pfc}}$ and $J_{{inc}_{{NMDA}_{pfc}}}$, respectively, and considering appropriate dimensions of the connection matrices depending on the number of neurons of the pre- and post-synaptic populations.

\begin{figure}[!ht]
\centering
\includegraphics[width=5.0in]{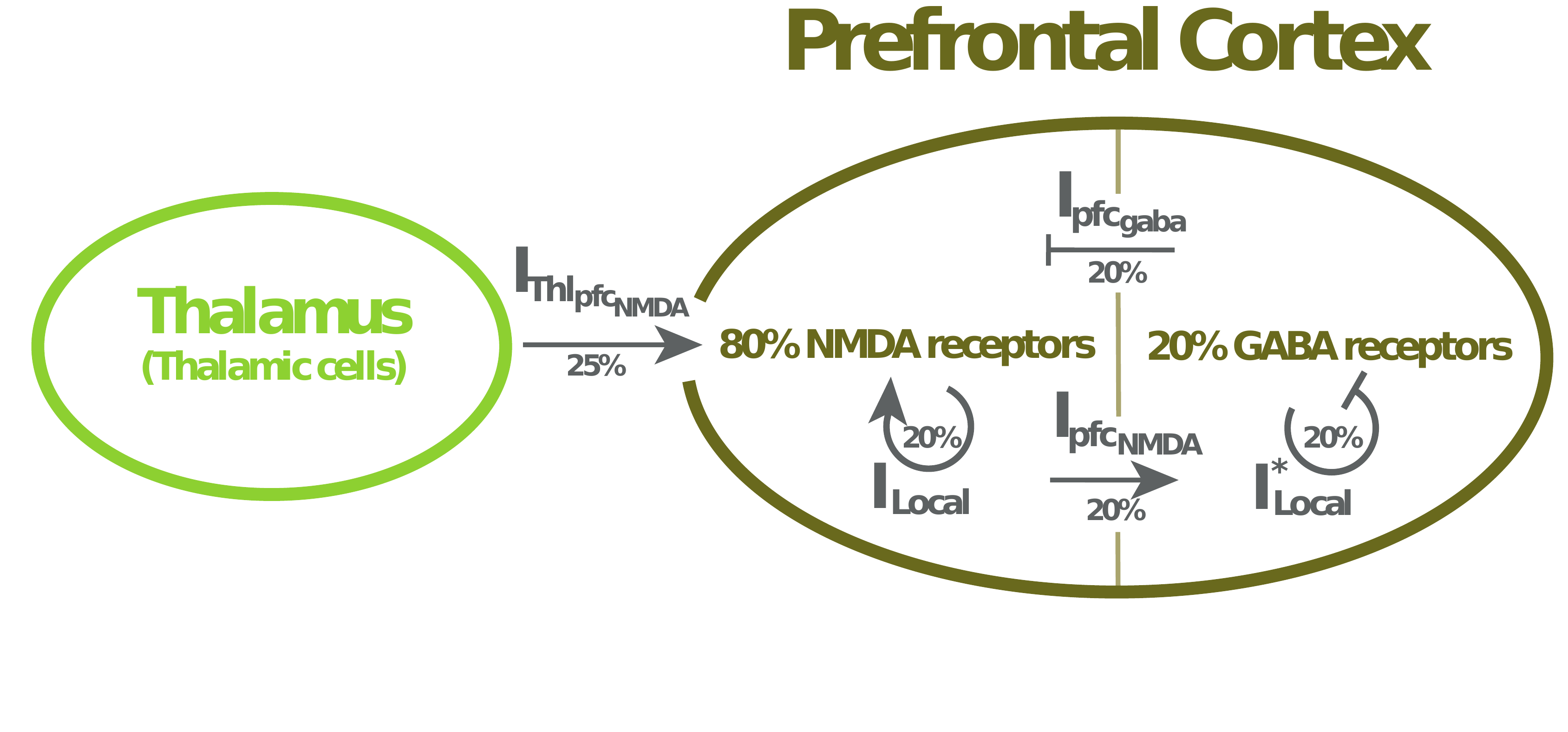}
\caption{The prefrontal cortex is the unit where the circuit considered in our model ends, and its activity represents the working memory.\ The thalamus acts as a translator that translates basal ganglia's message for prefrontal cortex by sending the appropriate signal according to the state of the working memory.\  The  probabilities of the considered connections are given with percentages.\ The arrow $\longrightarrow$ corresponds to excitatory connections, and $\dashv$ corresponds to inhibitory connections.}
\label{fig:pfc}
\end{figure}

The relevant connections and connection probabilities of the prefrontal cortex used in our model are shown in Figure \ref{fig:pfc}.

 The dynamical evolution of the membrane potentials and the recovery currents for GABAergic inhibitory neurons of the prefrontal cortex will follow the same equations as in \eqref{eq:NMDApCtx}-\eqref{eq:NMDApCtx_reset1}, but substituting the following variables and parameters $v_{{NMDA}_{pfc}}$, $u_{{NMDA}_{pfc}}$, $I_{{NMDA}_{pfc}}$, $v_{{peak}_{{NMDA}_{pfc}}}$, $c_{{NMDA}_{pfc}}$ and $d_{{NMDA}_{pfc}}$ by $v_{{GABA}_{pfc}}$, $u_{{GABA}_{pfc}}$, $I_{{GABA}_{pfc}}$, $v_{{peak}_{{GABA}_{pfc}}}$, $c_{{GABA}_{pfc}}$ and $d_{{GABA}_{pfc}}$, respectively.\ Now, $v_{{GABA}_{pfc}} \in \mathbb{R}^{N_{{GABA}_{pfc}}}$, $u_{{GABA}_{pfc}} \in \mathbb{R}^{N_{{GABA}_{pfc}}}$, $I_{{GABA}_{pfc}} \in \mathbb{R}^{N_{{GABA}_{pfc}}}$ with $N_{{GABA}_{pfc}}$ the number of GABAergic neurons within the prefrontal cortex.

 The parameters $C$, $k$, $a$, $v_r$, $v_t$, $b$, $c$ and $d$ are different for inhibitory and excitatory neurons.\ Here, the current  $I_{{GABA}_{pfc}}$ models the total synaptic input current for inhibitory neurons in the prefrontal cortex, and includes the following currents (all vectors of dimension  $N_{{GABA}_{pfc}}$):

\begin{equation} \label{eq:IGABApfc}
I_{{GABA}_{pfc}}=I_{{pfc}_{NMDA}}-I^{*}_{Local}+I_{back_{GABA_{pfc}}},
\end{equation} 
where $I_{{pfc}_{NMDA}}$ models the contribution of $NMDA$-receptor-type excitatory neurons of the prefrontal cortex; and $I^*_{Local}$ and $I_{back_{GABA_{pfc}}}$ models the recurrent activity and background activity of GABAergic inhibitory neurons of the prefrontal cortex, respectively.\ $I_{back_{GABA_{pfc}}}$ is obtained by using a Poisson probability distribution.\  The synaptic input currents $I_{{pfc}_{NMDA}}$ and $I^*_{Local}$ are obtained in an analogous way as described for excitatory neurons, substituting $NMDA$ by $GABA$.

Typical parameter values for both NMDA and GABA-receptor-type neurons used in our model are given in Table \ref{table:PFC}.

\begin{table}[ht]
\caption{Prefrontal Cortex Parameters}
\centering
\begin{tabular}{c c c c}
\hline\hline
Parameter & Description & NMDA Receptor & GABA Receptor \\ [0.5ex]
\hline 
$N$ &  Number of neurons & 80 & 20\\
$C$ &  Capacitance & 50pF & 25pF\\
$v_{r}$ & Reset potential for each neuron of the population & -55mV & -56mV\\
$v_{t}$ & Instantaneous threshold potential for each neuron of the population & -45mV & -42mV\\
$k$ & Izhikevich's parameter & 0.2 & 0.2\\
$a$ & '' & 0.13 & 0.03\\
$b$ & '' & -1.5 & 2\\
$c$ & '' & -50 & -50\\
$d$ & '' & 60 & 2\\
$v_{peak}$ & Spike threshold & 30mV & 40mV\\
$\eta_{background}$ & Background input frequency & 5 & 5\\
$\tau$ & Decaying-effect constant in dynamical connections & 14000 & 5000\\[1ex]
\hline
\end{tabular}
\label{table:PFC}
\end{table}
\subsection{Modelling the Connections between Neurons} \label{dynamical_connections}

In the previous subsections, we have explained the main elements of our working memory network model. The neuron dynamics presented define the evolution of each neuron population's behaviour with time and determine the way neurons behave according to inputs. The connections between neurons and populations of neurons represent the evolution of the relationships between them. This evolution depends on the firing activity and firing times of the neurons. 

In our model, we expect that starting from randomly assigned connections and synaptic strengths, the connections between populations will evolve into a state where the model is capable of showing different behaviours. This is the reason why we have referred to the connections between neurons as dynamical or evolving connections. The behaviours shown by our model are related to different working memory processes, namely: loading, maintainance, ignoring and clearance of information. The output from the basal ganglia generates different activity patterns in the thalamus, depending on the balance between the basal ganglia's pathways. This activity in the thalamus will drive the working memory into a bistable mode or will allow only one possible state to exist. This is accomplished by the interactions of neurons with each other and the evolution of the topology of the different networks of neurons.

In Section \ref{model_pctx}, we gave details on the dynamical connections and synaptic strength evolution between the NMDA-receptor-type neurons within the posterior cortex.\ We will use this example to describe how we model the connections between neurons and populations of neurons. 

The evolving or dynamical connections between neurons are defined by means of what we have called a connection matrix $S$ (see equation \eqref{eq:recurrent}).\ Let us consider that synapses are connecting neurons of two populations, having pre-synaptic and post-synaptic neurons.\ For each pre- and post-synaptic population, we define a different connection matrix $S$.\ The dimension of the matrix $S$ is $N_{post} \times N_{pre}$, with $N_{post}$ the number of neurons in the post-synaptic population and $N_{pre}$ the number of neurons in the pre-synaptic population.\ With this, we are establishing that we have an all-to-all topology between the neurons of two populations.\ However, we will assign a probability to each link between two neurons, and consequently we will finally build random-graph  networks for neurons.\ Each element $(i,j)$ of the connection matrix $S$ contains the synaptic strength ($s$) between the $i$ pre-synaptic neuron that has fired at some time $t_n$ and the $j$ post-synaptic neuron.\ The elements $S(i,j)$, for all $(i, j) \in N_{post} \times  N_{pre} $ such that $i=j$, are considered $0$ to indicate that it is not possible to have a connection between a neuron and itself.

We do not compute the synaptic strengths for every element of the connection matrices $S$.\ The connection probabilities between populations of neurons given in Table \ref{table:connections} determine how sparse the connection matrix will be.\ For example, let us come back to Section \ref{model_pctx} and recall the computation of the elements of matrix $S_{{NMDA}_{pCtx}}$ in order to obtain the recurrent activity of excitatory NMDA-receptor-type neurons in the posterior cortex ($I_{local}$ in \eqref{eq:recurrent}).\ Let us consider that we have $100$ NMDA-receptor-type neurons in the posterior cortex.\ Consequently, the size of $S_{{NMDA}_{pCtx}}$  is $100 \times 100$.\ Since the connection probability between two NMDA-type neurons is $0.8$ (or 80\% as a percentage value as we have indicated in the figures of this paper), there will be $8000$ possible connections between these $100$ neurons.\ We will randomly choose these $8000$ elements within the matrix  $S_{{NMDA}_{pCtx}}$, and we will calculate their values -- that is, the synaptic strengths $s_{{NMDA}_{pCtx}}$ -- in the following manner.

We will first initialise the value for each synaptic strength $s$ between a pre- and post- synaptic neuron with the product of the initial synaptic strength, $J_{s}$, and a uniformly distributed random number between $0$ and $1$, $r$.\ This random scaling parameter $r$ is also used in the computation of some of the synaptic input currents as it is defined in equation \eqref{eq:recurrent}.\ The values used for the initial synaptic strength, $J_{s}$, are given in Table \ref{table:connections}.\ The random value $r$ provides different initial conditions for each simulation.

Once we start simulating our model, the initial values for the synaptic strengths `evolve' according to equations of the same form as equations \eqref{dyn_connection1} and \eqref{dyn_connection2}.\ If a pre-synaptic neuron causes the firing of a post-synaptic neuron, the corresponding synaptic strength $s$ increases as in equation \eqref{dyn_connection2}. How much we increment the connection strength is determined by the product of the increment parameter $J_{inc}$ and a random variable $\omega$ (with a value between $0$ and $1$).\ The values used in our simulations for  parameter $J_{inc}$ are given in Table \ref{table:connections}. However, if the firing times of the two neurons do not coincide, the synaptic strength decays according  to equation \eqref{dyn_connection1}.\ We do not allow synaptic strengths to become negative.\ That is, if $s$ becomes negative according to equations \eqref{dyn_connection1} and \eqref{dyn_connection2}, we make $s=0$, which means that the connection between the corresponding pre-synaptic and post-synaptic neuron has `disappeared'.\ In equations \eqref{dyn_connection1} and \eqref{dyn_connection2}, the parameter $\tau$ is a decaying-effect.\ We use different decaying-effect constants for each population of neurons and the values are given in the parameter tables for the corresponding population.

  In this context, an existing connection between two neurons may disappear (the value of the synaptic strength becomes $0$) or a non-existent connection may be formed (the value of the synaptic strength becomes greater than $0$) depending on the spike times.\ This is what we call {\em evolution} of the connections.\ With this, our model can capture in a manner what is known as {\em structural plasticity} \cite{structural_plasticity}.\ That is, not only is synaptic behaviour modified, but synapses may also be rewired. This concept of structural plasticity can be also associated to changes in the network topology, which is also termed {\em wiring plasticity} \cite{wiring_plasticity}.

\begin{table}[ht]
\caption{Parameters for the dynamical connections between neurons}
\centering
\begin{tabular}{c c c c}
\hline\hline
Connection(From-To) & Connection Probability & Initial Synaptic Strength ($J_{s}$) & Increment of the Synaptic Strength ($J_{inc}$) \\ [0.5ex]
\hline 
$pCtx_{NMDA}\rightarrow pCtx_{NMDA}$ &  0.2 & 15 & 0.4\\
$pCtx_{NMDA}\rightarrow pCtx_{GABA}$ &  0.2 & 5 & 0.3\\
$pCtx_{NMDA}\rightarrow MSN_{D1}$ &  0.45 & 150 & 0.5\\
$pCtx_{NMDA}\rightarrow MSN_{D2}$ &  0.45 & 150 & 0.5\\
$pCtx_{NMDA}\rightarrow FSIs$ &  0.3 & 45 & 0.6\\
$pCtx_{NMDA}\rightarrow STN$ &  0.2 & 10 & 0.3\\
$pCtx_{NMDA}\rightarrow Thalamic Cells$ &  0.2 & 45 & 0.1\\
$pCtx_{GABA}\rightarrow pCtx_{GABA}$ &  0.2 & 5 & 0.4\\
$pCtx_{GABA}\rightarrow pCtx_{NMDA}$ &  0.2 & 5 & 0.6\\
$MSN_{D1}\rightarrow MSN_{D1}$ &  0.15 & 10 & 0.3\\
$MSN_{D1}\rightarrow GPi$ &  0.3 & 60 & 0.5\\
$MSN_{D2}\rightarrow MSN_{D2}$ &  0.15 & 10 & 0.3\\
$MSN_{D2}\rightarrow GPe$ &  0.65 & 175 & 0.3\\
$FSIs\rightarrow FSIs$ &  0.15 & 15 & 0.2\\
$FSIs\rightarrow MSN_{D1}$ &  0.15 & 25 & 0.4\\
$FSIs\rightarrow MSN_{D2}$ &  0.15 & 25 & 0.4\\
$GPi\rightarrow GPi$ &  0.2 & 5 & 0.2\\
$GPi\rightarrow Thalamic Cells$ &  1.0 & 7 & 0.8\\
$GPi\rightarrow GPe$ &  0.4 & 175 & 0.3\\
$GPe\rightarrow GPe$ &  0.15 & 10 & 0.1\\
$GPe\rightarrow STN$ &  0.2 & 5 & 0.2\\
$GPe\rightarrow MSN_{D1}$ &  0.1 & 5 & 0.3\\
$GPe\rightarrow MSN_{D2}$ &  0.1 & 5 & 0.3\\
$GPe\rightarrow GPi$ &  0.45 & 5 & 0.5\\
$STN\rightarrow STN$ &  0.1 & 5 & 0.15\\
$STN\rightarrow GPi$ &  0.3 & 50 & 0.2\\
$STN\rightarrow GPe$ &  0.25 & 200 & 0.2\\
$Thalamic Cells\rightarrow Thalamic Cells$ &  0.20 & 25 & 0.6\\
$Thalamic Cells\rightarrow RTN$ &  0.20 & 55 & 0.3\\
$Thalamic Cells\rightarrow PFC_{NMDA}$ &  0.25 & 45 & 0.2\\
$RTN\rightarrow RTN$ &  0.20 & 5 & 0.2\\
$RTN\rightarrow Thalamic Cells$ &  0.20 & 5 & 0.1\\
$PFC_{NMDA}\rightarrow PFC_{NMDA}$ &  0.2 & 15 & 0.4\\
$PFC_{NMDA}\rightarrow PFC_{GABA}$ &  0.2 & 5 & 0.3\\
$PFC_{GABA}\rightarrow PFC_{GABA}$ &  0.2 & 5 & 0.4\\
$PFC_{GABA}\rightarrow PFC_{NMDA}$ &  0.2 & 5 & 0.6\\[1ex]
\hline
\end{tabular}
\label{table:connections}
\end{table}

\begin{table}
\caption{Population Connections}
\centering
\begin{tabular}{c c c}
\hline\hline
Connection (Pre-synaptic) & Number of Neurons & Targets (Post-synaptic)\\ [0.5ex]
\hline 
$Posterior Cortex NMDA$ & 80 & Local, Posterior Cortex (GABA), $MSNs_{D1}$, $MSNs_{D2}$, FSIs, STN, Thalamic Cells\\
$Posterior Cortex GABA$ &  20 & Local,  Posterior Cortex (NMDA)\\
$MSNs_{D1}$ & 100 & Local, GPi\\
$MSNs_{D2}$ & 100 & Local, GPe\\
$FSIs$ &  100 & Local, $MSNs_{D1}$, $MSNs_{D2}$\\
$GPi$ & 100 & Local, GPe, Thalamic Cells\\
$GPe$ & 100 & Local, GPi, STN, $MSNs_{D1}$, $MSNs_{D2}$\\
$STN$ & 100 & Local, GPi, GPe\\
$Thalamic Cells$ & 80 & Local, RTN, Prefrontal Cortex NMDA\\
$RTN$ & 20 & Local, Thalamic Cells\\
$Prefrontal Cortex NMDA$ & 80 & Local, Prefrontal Cortex GABA\\
$Prefrontal Cortex GABA$ & 20 & Local, Prefrontal Cortex NMDA\\[1ex]
\hline
\end{tabular}
\label{table:PopGraph}
\end{table}

In our model, we have considered that the posterior cortex, the prefrontal cortex and the thalamus consist of two different types of sub-populations: excitatory and inhibitory. For the simulations, we will considered the following set up. The excitatory and inhibitory neurons for these units have the same ratio of 4:1 (80:20 neurons). The inhibitory sub-populations only send/receive signals to/from neurons within the population. The striatum consists of three sub-populations and each of them contains $100$ neurons. The other populations within the basal ganglia have $100$ neurons each. The number of neurons in each pre-synaptic population and the list of the connections are given in Table \ref{table:PopGraph}.

\section{Evaluation of the working memory network model}

We will consider the delayed-match-to-sample (DMS) task to explain how our working memory network model can describe brain activity during different working memory tasks, and how these tasks are associated with the generation of different oscillations in the thalamus.

\begin{figure}[!ht]
\centering
\includegraphics[width=4.5in]{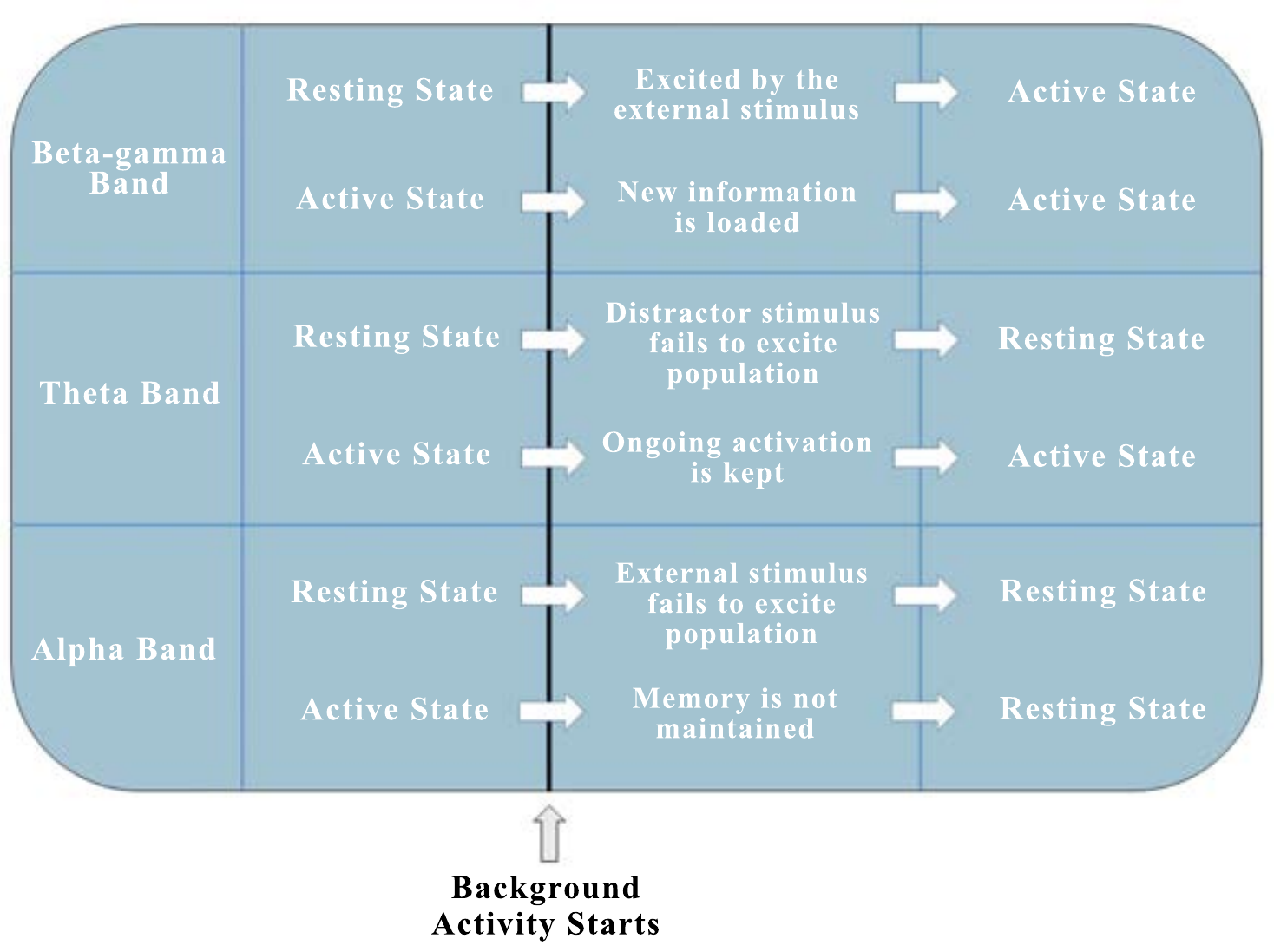}
\caption{A summary of expected behaviours of the working memory from \cite{ref28}. In the presence of beta-gamma-frequency band oscillations, a bistable regime is observed. If there is an ongoing resting state, a transient stimulus can evoke a persistent active response. While there is an ongoing active state, gamma-frequency band activity allows a persistent state. However, new information overwrites the previous one. In the presence of theta-frequency band oscillations, the previous activity is preserved. In this case, no information can reach the working memory, and any interference is avoided. In the presence of alpha-frequency band oscillations, only a resting state is allowed, and the active persistent state is cleared away, and information is no longer available within the working memory.}
\label{fig:bands}
\end{figure}

 In a DMS task, a participant is shown a complex visual pattern named ``the sample'' and, after some short delay, she/he is exposed to other similar visual patterns. The participant must touch the pattern which exactly matches the sample  \cite{ref15}. The external stimulus, which represents the sample in our simulations, is applied between $25-75$ milliseconds. This stimulus excites the population of neurons in the posterior cortex and causes a persistent spiking activity in the thalamus. This corresponds to loading of information and the onset of beta-gamma-frequency band oscillations as described in Section II. In a DMS task during the delay part, participants need to keep the sample visual pattern in their mind while ignoring irrelevant pictures. For our simulation, this means to keep persistent spiking activity alive with theta-frequency band oscillations, which corresponds to maintenance of information as explained in Section II. Theta-frequency band oscillations also provide a mechanism that enables working memory to keep its persistent activity in spite of the distractors (maintenance while ignoring distractors). At the end, when a matching pattern arrives, there is no need to keep the sample information anymore and the clearance of information is carried out. During the clearance of information, alpha-frequency band oscillations are generated and the persistent active state is cleared away, only allowing a resting state. In Figure \ref{fig:bands}, the relationships between the frequency bands and the working memory processes are summarised based on the bistable activation introduced in \cite{ref28}.

As shown in Figure \ref{fig:bands}, for each frequency band, the neural activity can be interpreted as a bistable dynamical system that changes among two possible states: resting state and active state. With the onset of beta-gamma-frequency band activity, there are two possibilities: the system at rest is activated due to an external stimulus or new information is loaded into the system. The first possibility corresponds to the transfer of the system from resting state to the active state, while the second possibility corresponds to keeping the system in the active state, and consequently, allowing new information to be loaded. In terms of the DMS task, the transition from the resting state to the active state means that the first ``sample'' is introduced. The case of having new information loaded means that a new ``sample'' is introduced to the participant. 

Moreover, theta-frequency band oscillations allow the network to stay in a persistent state, and no behavioral change is observed. Hence, the ongoing activity corresponding to the active state linked to the onset of beta-gamma-frequency band activity continues and any stimulus received when theta-frequency band oscillations are present does not change the system activity. In other words, there is no switching from the resting state to the active state, and the maintenance of the ongoing activity is ensured while any distractor is blocked. Since the resting state does not change to active in the presence of a stimulus, the first row of the theta band in Figure \ref{fig:bands} corresponds to blocking of distractor, while the active state continues to be active, which corresponds to the maintenance of the ongoing activity. 

Alpha-frequency band oscillations prevent the prefrontal cortex from being activated, even in the presence of an external stimulus. Consequently, the resting state is kept; and if the system is active, it is switched to the resting state. The change from the active state to the resting state may imply that either new information in the memory is not loaded or the already-loaded information is maintained. 

In the next subsections, we will illustrate some of these working memory processes through simulation results obtained from our model. We will relate these processes to the dominance of the direct pathway and indirect pathway of the basal ganglia and the generation of beta-gamma and theta-frequency-band activity in the thalamus, respectively. The simulations presented in this paper have been obtained with MATLAB. We will confirm the changes of activity in the network by the spiking activity of key populations of neurons, represented in the raster plots given in Figures  \ref{fig:def_raster}, \ref{fig:raster} and \ref{fig:in_raster}. These raster plots correspond to a simulation of $500$ milliseconds of spiking activity with a sampling frequency of $2000$Hz. Furthermore, a frequency analysis of thalamic neurons are given in Figures \ref{fig:def_spectro}, \ref{fig:spectro} and \ref{fig:in_spectro}, to show that the model is capable of generating the expected oscillations observed during working memory tasks.

Before presenting the simulation results corresponding to the dominance of the direct and indirect pathways, we will present the spiking activity of the populations of neurons of our model in the case of having no external stimulus in the posterior cortex. This will be important to clearly understand the changes produced in the network when an external stimulus is applied.

\subsection{Network Activity with no External Stimulus: Default or Resting Network Mode}

By analysing the network activity when no external stimulus is applied to the posterior cortex, we will better understand how the transition from the resting state to any other state is manifested in the changes of the activity in the different neural populations, which depend on the dopamine level in our model. The network neuronal activity with no external stimulus is what we refer to as the default mode or resting state of the network.

\begin{figure}[!ht]
\centering
\includegraphics[width=7.5in]{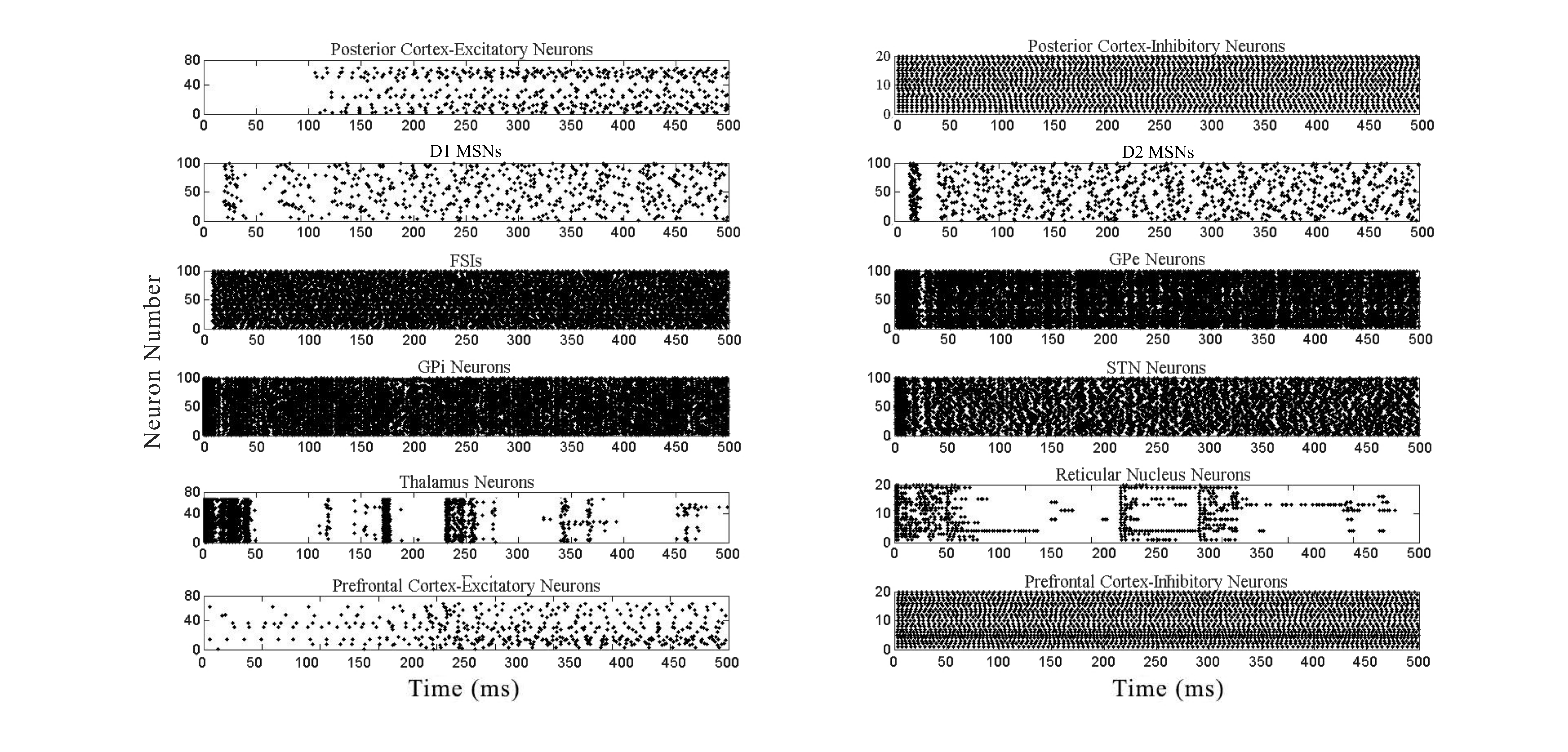}
\caption{Raster plots of the spiking activity of simulated populations of neurons considered in our model when no external stimulus is injected to the posterior cortex. During the simulated case, there is a balance between the direct and indirect pathways, that is, none of the pathways is dominant. We use $\phi_1=0.58$ and $\phi_2=0.55$. Both cortical (posterior and prefrontal) populations stay silent during the simulation since there is no external stimulation for the posterior cortex and no thalamic excitation for the prefrontal cortex. Thalamic neurons also show low activity because of the inhibitory GPi input. Each black dot corresponds to a spike of the corresponding neuron at a given time.}
\label{fig:def_raster}
\end{figure}

The default mode of the network can be interpreted as the neuronal activity within the network when a person is not focused on the environment and she/he is at rest. The default network mode has been also shown to deactivate during goal-oriented tasks such as working memory tasks. Consequently, any neuronal activity during the resting network state becomes irrelevant while the network is performing a task. 

Figure \ref{fig:def_raster} shows the raster plots of the spiking activity of the main populations of neurons considered in our model when there is no stimulation in the posterior cortex. The dopamine levels affecting $D_1$ and $D_2$-MSN populations are considered in such a manner to avoid the dominance of the direct or the indirect pathway of the basal ganglia. We consider $\phi_1=0.58$ and $\phi_2=0.55$. These values were derived from the models of single neurons that provided similar firing patterns for $D_1$ and $D_2$-MSNs under the same input current. These values also allow equivalent distribution of cortical and FSIs input to both MSN populations. 

What we can appreciate in Figure \ref{fig:def_raster} is that the posterior cortex shows a spontaneous and low-frequency activity that is not capable of providing enough excitation to the MSNs striatal populations ($D_1$ and $D_2$-receptor types). The raster plot of the posterior cortex reveals that only a trace of neurons within the posterior cortex can generate spikes. Since there is no incoming information, there is no need for a subcortical decision, which results in the balance between the direct and indirect pathways of the basal ganglia. Furthermore, both MSN populations show a low activity and the uninhibited GPi population generates fast spikes that later provide continuous inhibition over thalamic neurons. Similarly when the indirect pathway is dominant, thalamic neurons must be continuously inhibited since there is no incoming information and the prefrontal cortex should be prevented from being excited. Thus, the excitatory neurons of the prefrontal cortex present spontaneous low activity.

\begin{figure}
\centering
\includegraphics[width=7.0in]{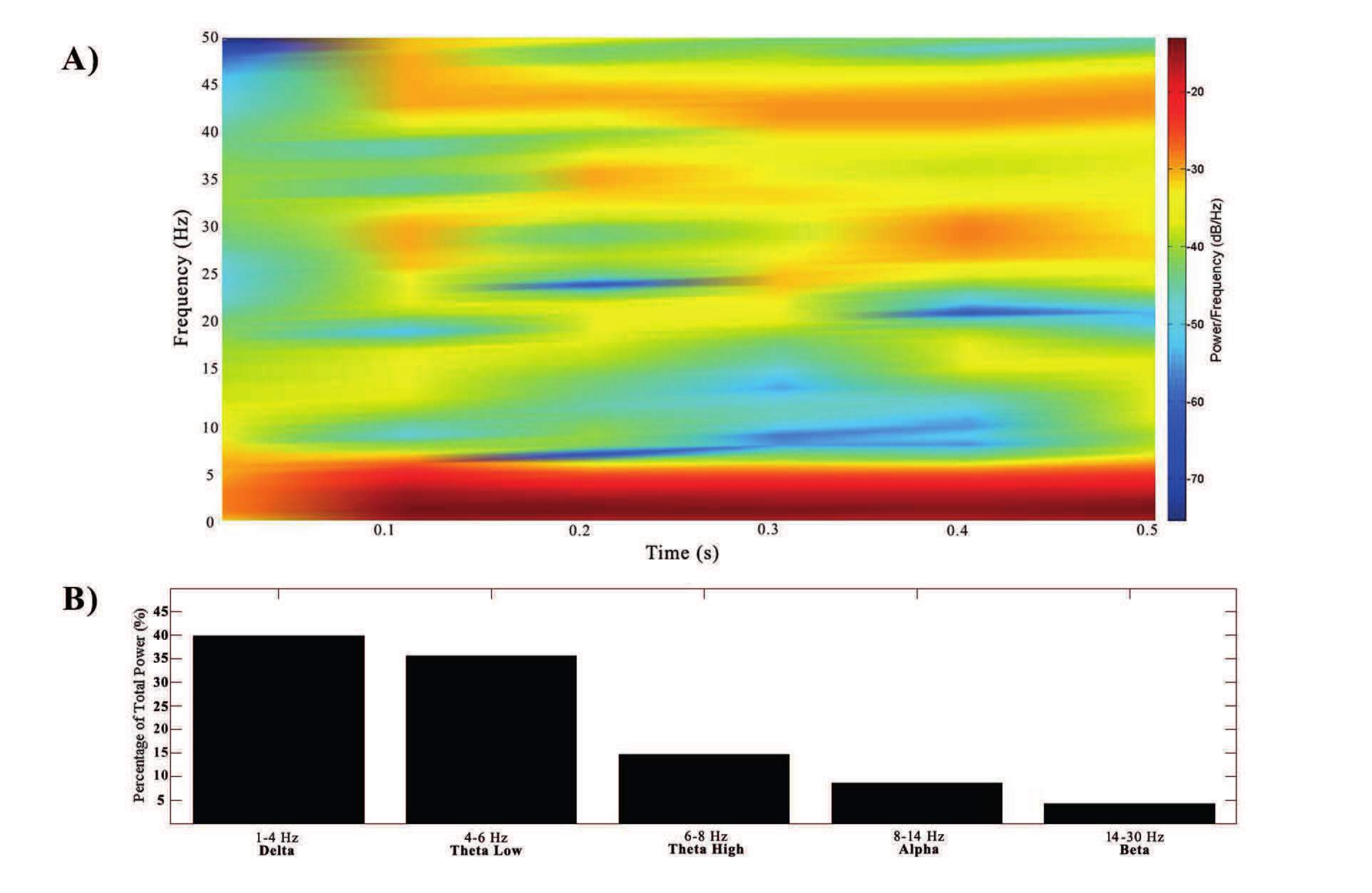}
\caption{Spectrogram analysis of the thalamic activity with no stimulation in the posterior cortex.  A) The spectogram was obtained from local field potentials averaged over $30$ trials. The dominant frequency band is in red. The figure shows how the dominant frequency fluctuates with time: between the delta-frequency band ($1-4$Hz) and a low theta-frequency band ($4-6$Hz). This state can be thought of as the default mode of the network where no goal-directed attentional behaviour is needed. B) Power histogram of the thalamic activity.}
\label{fig:def_spectro}
\end{figure}

Different from the case of having the indirect pathway dominant, when there is no stimulus, the activity within the thalamic neurons presents  even lower frequencies. However, the activity does  not stay within just one frequency-band range: the activity in the thalamus oscillates between the delta-frequency band ($1-4$Hz) and a low theta-frequency-band range ($4-6$Hz), which is consistent with results presented in \cite{ref46_nss}. Frequency aspects of the thalamic neurons are given in Figure \ref{fig:def_spectro}. The thalamic neurons are mostly inhibited by the GPi input, which results in oscillatory activity within the delta and theta-frequency bands.  We note that due to the fact that any neuronal activity during the resting network mode becomes irrelevant while the network is performing a task, the activity of the resting state is subtracted from the spectogram of the direct and indirect pathways. By subtracting the spectrogram we basically refer to removing the default mode neuronal activity from the spectograms of the direct and indirect pathways, and this is what is shown in Figure \ref{fig:def_spectro}.

\subsection{Dominance of the Direct Pathway of the Basal Ganglia}

In this section, we will analyse the scenario when the direct pathway of the basal ganglia is dominant and beta-gamma-frequency band activity is dominant in the thalamus, which is associated to loading of information in the working memory.

An external stimulus is applied to NMDA-receptor type neurons of the posterior cortex between $25-75$ milliseconds. This stimulation was enough to drive these excitatory neurons of the posterior cortex to an active persistent state. These neurons respond to stimulation with a latency at $60$ milliseconds (with a delay of $35$ milliseconds). After this delay, almost all the NMDA-receptor type neurons get activated. The activity of these neurons well represent the external stimulus. Recurrent connections within the neurons of the posterior cortex allow to keep them in an active state throughout the simulation. 

\begin{figure}[!ht]
\centering
\includegraphics[width=7.5in]{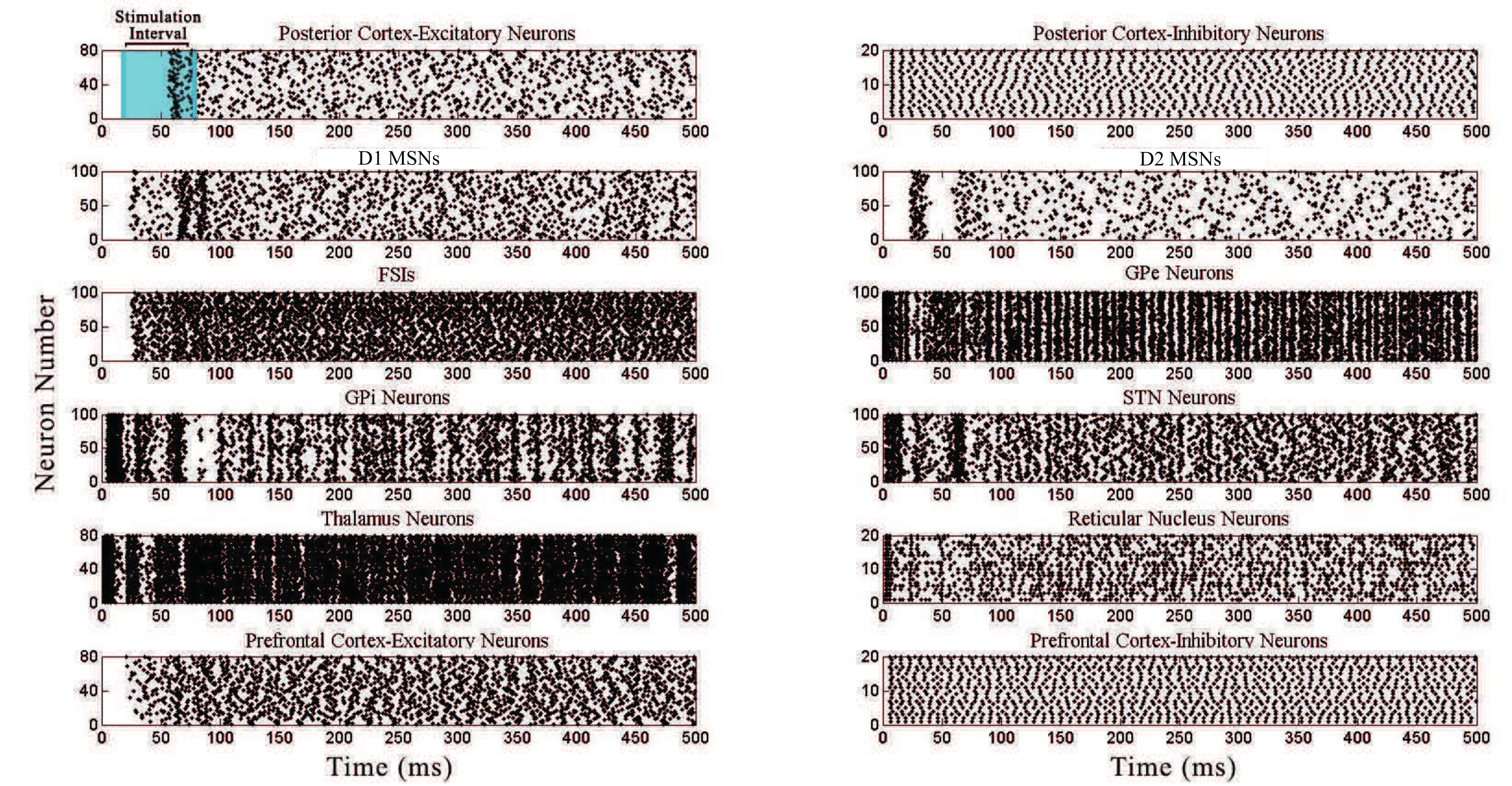}
\caption{Raster plots of the spiking activity of the simulated populations of neurons considered in our model when the direct pathway of the basal ganglia is dominant with ${\phi}_1=0.80$ and ${\phi}_1=0.65$. The populations of the posterior cortex, the prefrontal cortex and the thalamus consist of two different subpopulations. We consider $80$ excitatory neurons and $20$ inhibitory neurons. The inhibitory neurons of the prefrontal and the posterior cortex are modelled to have fast-spiking activity independent from the external stimulation. As a result, the activity in these units starts before the injection of the stimulus. Dopamine release causes depolarisation of $D_1$-MSNs and the hyperpolarisation of $D_2$-MSNs. The inhibition of $D_1$-MSNs in the striatum flows directly to the GPi. In turn, the GPi releases the inhibition over the thalamic neurons. The increased activity within the thalamus is later transmitted to the cortical population. Each black dot corresponds to a spike of the corresponding neuron at a given time.}
\label{fig:raster}
\end{figure}

The posterior cortex sends its output to the striatum, that is, to $D_1$-MSNs, $D_2$-MSNs and FSIs. The striatal sub-circuit makes a decision between rapid updating and robust maintenance. The regulation of the cortical output to the striatum is crucial for the final subcortical decision. Depending on this decision, the stimulus can reach the prefrontal cortex, which will allow the information to flow in or to be ignored. As we explained in Sections I and II, the dopamine release will regulate the selection of the different pathways of the basal ganglia, and in our model, this dopamine release is expressed by the control parameters $\phi_1$ (for $D_1$-MSNs) and $\phi_2$ (for $D_2$-MSNs), both parameters take values between $0$ and $1$. We model the dominance of the direct pathway within the basal ganglia by assigning to $\phi_1$ a big enough value. The dopamine level for the direct pathway is chosen in such a way to allow a single $D_1$-MSN to enhance its response nearly tripling the firing rate to the same stimulus. The values chosen are ${\phi}_1=0.80$ and ${\phi}_1=0.65$ (obtained shift $10$ spike/s to $25$ spike/s to a DC input of $50$pA) \cite{ref47_nss}. The overabundance of dopamine in $D_1$ receptors favours the activation of the direct pathway in the basal ganglia. 



\begin{figure}[!ht]
\centering
\includegraphics[width=7.0in]{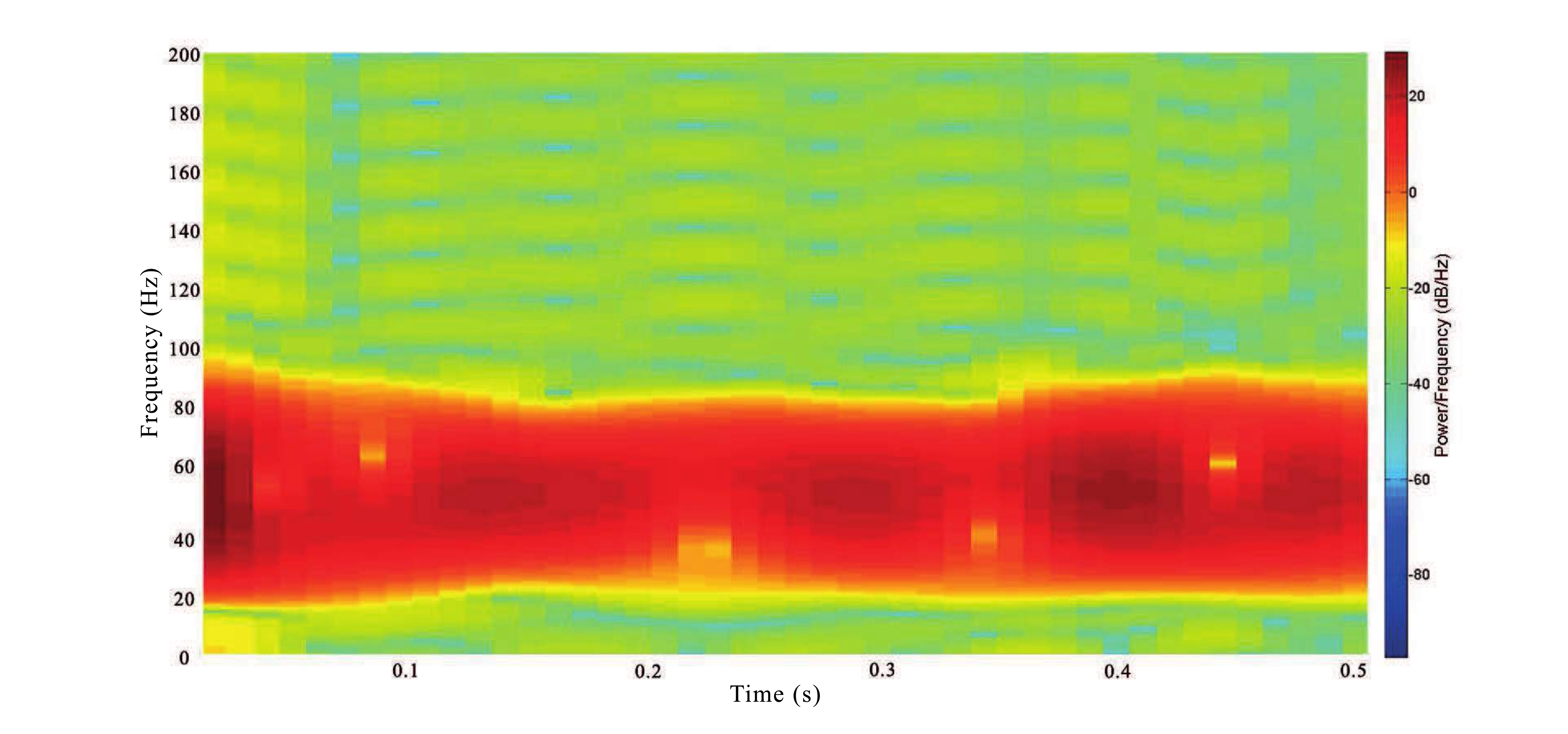}
\caption{Spectrogram of the thalamic activity when the direct pathway of the basal ganglia is dominant. The spectrogram was obtained from local field potentials averaged over $30$ trials. The sprectrogram of the network when there is no stimulus (delta-low-theta-frequency band activity) is extracted from the spectrogram of the local field potentials. Consequently, the sprectrogram presented here represents the increment of the activity from the default/resting network mode, rather than the activity of the network. The dominant frequency band is in red. The figure shows how the dominant frequency changes with time and oscillates between a low gamma-frequency band ($30-50$Hz) and a high gamma-frequency band ($50-100$ Hz) range. These frequency-band ranges are within the expected frequency band.}
\label{fig:spectro}
\end{figure}

We note that the activation of the direct pathway is associated with an overactivity of $D_1$-MSNs and a reduced activity of $D_2$-MSNs in the striatum (as reflected in Figure \ref{fig:raster}). Consequently, the  cortical input cannot provide enough excitation to $D_2$-MSNs.  Moreover, the connections from the FSIs to the $D_1/D_2$-MSNs are arranged in such a way to allow a winner-take-all mechanism by holding the activity within the target cells to promote others. The output of the FSIs to the $D_1$ and $D_2$-MSNs is scaled according to the dopamine level. Since the FSIs are inhibitory neurons, they repress more the activity of the  $D_2$-MSNs than the activity of the $D_1$-MSNs in order to favour the dominance of the direct pathway. At the same time, $D_1$-MSNs reduce the inhibitory outflow from parts of the GPi. The output of the basal ganglia, the GPi, consists of GABAergic and tonically active neurons. Thus, the GPi tonically inhibits the thalamus, which is also a gate for the information flow through cortical targets (prefrontal cortices, motor, supplementary motor) \cite{ref49_nss}. It can be hypothesised that inhibition is a control mechanism which leads to a rapid response in the thalamus and is crucial for encoding and maintaining new information as it occurs. As the result of the circuitry considered, the striatal decision is mapped onto the cortical population via the representation of the output layer, the GPi. The direct pathway allows to map the decision onto a particular action. The increased activity of the $D_1$-MSNs in turn avoids thalamic inhibition. From an indirect way, the reduced activity of the  $D_2$-MSNs decreases its repressive effect on the GPe. The GPe disinhibition holds the glutamatergic STN signal generation, which in turn also inhibits the activity of the GPi. The balance between the two pathways ensures the execution of a given order. In brief, the basal ganglia promotes action about information by disinhibiting the associated target structures. During the disinhibition period, the thalamic neurons start to generate high-frequency bursts of spikes. The behavioural change in the spiking activity is interpreted as a `wake up call' for the cortical targets, and as a result, newly-coming sensory information is allowed to be updated in the prefrontal cortex.

The basal ganglia generate a response to incoming information and send this message to the working memory through the thalamus. Until the basal ganglia make a decision, the GPi has an inhibitory control over the thalamus. However, a `Go' signal allows the thalamic disinhibition, which results in high-frequency activity. Besides the dopaminergic regulation, we also focus on the frequency aspects of thalamic activity. The frequency-time spectrogram of thalamic neurons is given in Figure \ref{fig:spectro}. It is plotted as a function of the local field potential.

\begin{figure}[!ht]
\centering
\includegraphics[width=6.0in]{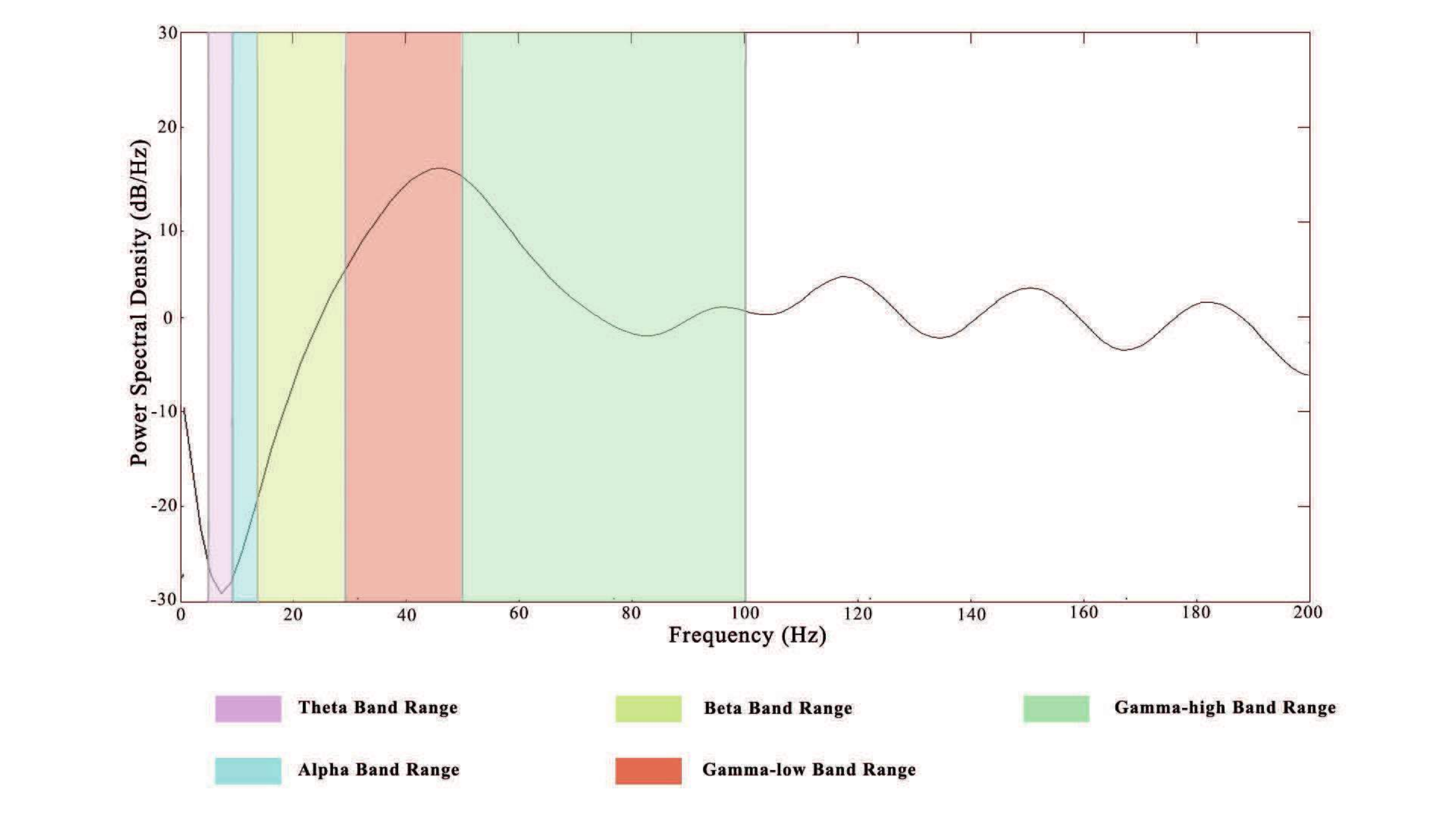}
\caption{Power spectral density of the thalamic neurons when the direct pathway of the basal ganglia is dominant. The spectrum gives the relationship between frequency and relevant power density. The most dominant frequency is calculated as $48$Hz. The figure is an interpretation of the color bar graph within the spectrogram. Power reaches its maximum value at the gamma-frequency band. The power spectral density is computed using Welch's method with $0.01$s hamming window and $50\%$ overlap. The hamming window slides over a $500$ms time interval and the spectrum represents the average of the sliding windows.}
\label{fig:spectrum}
\end{figure}



 Compatible with brain imaging studies and mathematical models, the direct pathway of the basal ganglia allows the thalamic activity to reach a high gamma-frequency band range \cite{ref10,ref28}. According to Fourier decomposition, the thalamic signal contains different frequency components. Frequency fluctuations stay within the low gamma-frequency band and high gamma-frequency band ranges, which provides a high level of depolarising input for the stimulus to flow through the prefrontal cortex. Disinhibition might not be able to result in a synchronised activity within the thalamus. While the thalamic neurons are in an `up' state, they generate high-frequency spikes, followed by a `down' state. During the  `down' state, it is harder, but not impossible, to generate spikes. The oscillating behaviour in the spectrogram might be a result of a difference in the frequency generation between `up' and `down' states of bursting activity. Additionally, low gamma-frequency band activity during `down' states has a high proportion of the total power.  

\begin{figure}[!ht]
\centering
\includegraphics[width=6.0in]{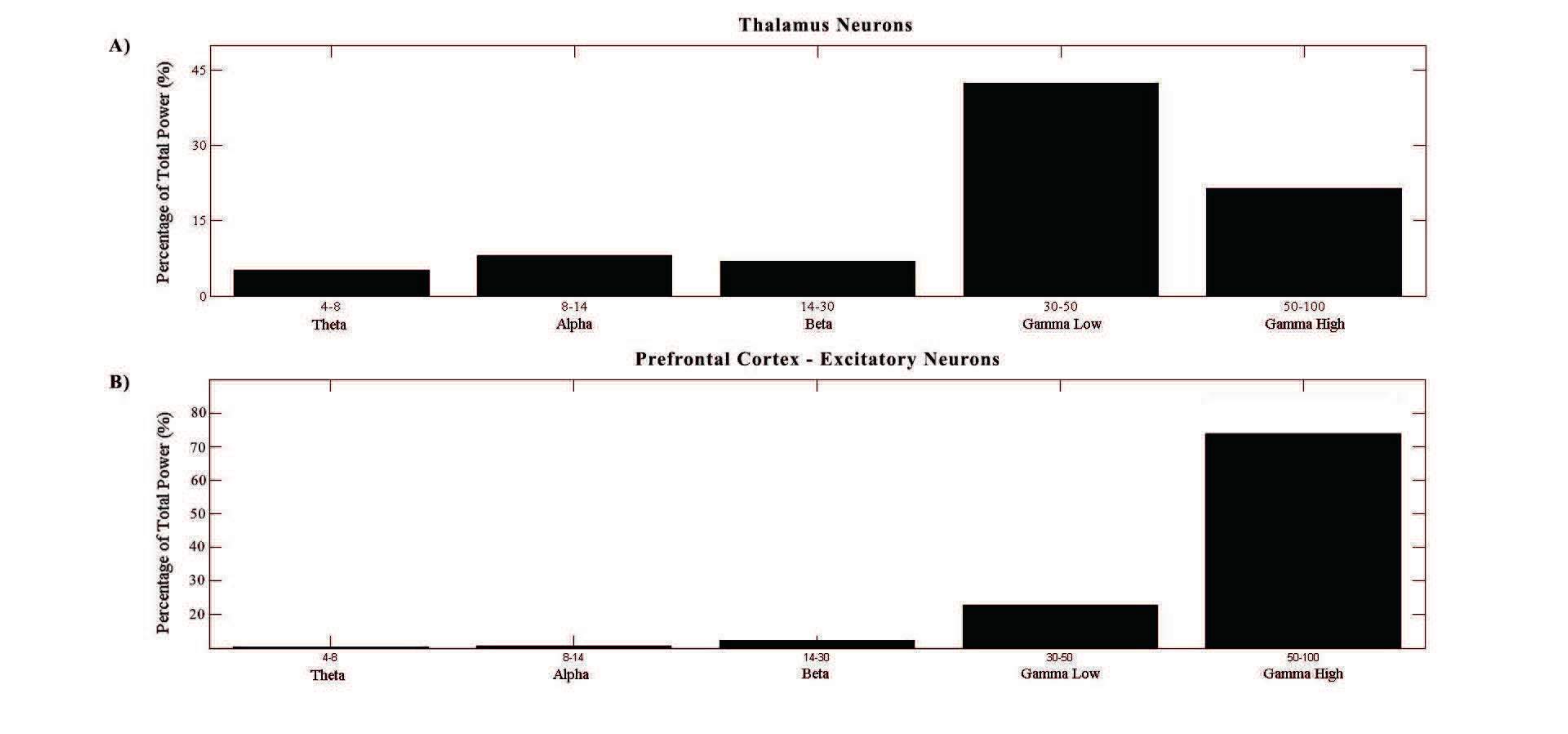}
\caption{Power histogram of the thalamic and the excitatory prefrontal cortical activity when the direct pathway of the basal ganglia is dominant. A) The graph deconstructs frequency components of thalamic activity and represents its ratio with the overall activity. As it can be appreciated from the spectrogram and the power spectrum, the histogram also shows that the dominant activity is within a low gamma-frequency band and high gamma-frequency band ranges. The low gamma-frequency band nearly contains half of the total power. B) The same analysis for cortical neurons, which shows an increment of frequency up to a high gamma-band.}
\label{fig:histogram}
\end{figure}

The power spectral density of the thalamic neurons is given in Figure \ref{fig:spectrum}. This gives a detailed explanation of the frequency distribution. As  it was investigated in the spectrogram, two different frequency values seem to have more power than the rest. There is an overactivity between $30$Hz and $50$Hz, and a gamma-frequency band activity is shown to be the most dominant one when the direct pathway is dominant. This figure also gives clues about the spiking behavioural pattern of the thalamic neurons. The spectrum represents a unimodal distribution which is compatible with results presented in \cite{ref46}. Burst-like activity results in a spread distribution over different orders of magnitude, however, in our case, the spectrum has a peak value which has the maximum power followed by a slightly change through neighbouring frequencies. 

Finally, Figure \ref{fig:histogram} shows a more detailed distribution of power in different frequency ranges. The gamma-frequency band range holds more than the $60\%$ of the total power ($42\%$ power at low gamma-frequency band range and $20\%$ power at high gamma-frequency band range). As mentioned above, gamma-frequency band activity allows the prefrontal cortex to show bistability and switch from a resting state to an active state. This also means that the gamma-frequency band activity keeps the thalamic `gates open', and another incoming sensory stimulus may be overwritten on the previously loaded pattern. Consequently, the gamma-frequency band activity allows the update of information but it does not ensure the maintenance of the information. Figure \ref{fig:histogram} also shows the activity of excitatory neurons of the prefrontal cortex as a consequence of the thalamic activity. Even though low gamma-frequency band activity has the maximum power for thalamic neurons, it was enough to increase the cortical activity to higher levels of the gamma-frequency band range. The high gamma-frequency band entails almost the $75\%$ of the total power.

\subsection{Dominance of the Indirect Pathway of the Basal Ganglia}

In this section, we will analyse the scenario when the indirect pathway of the basal ganglia is dominant and theta-frequency band activity is dominant in the thalamus, which is associated to maintenance of information in the working memory.

\begin{figure}[!ht]
\centering
\includegraphics[width=7.5in]{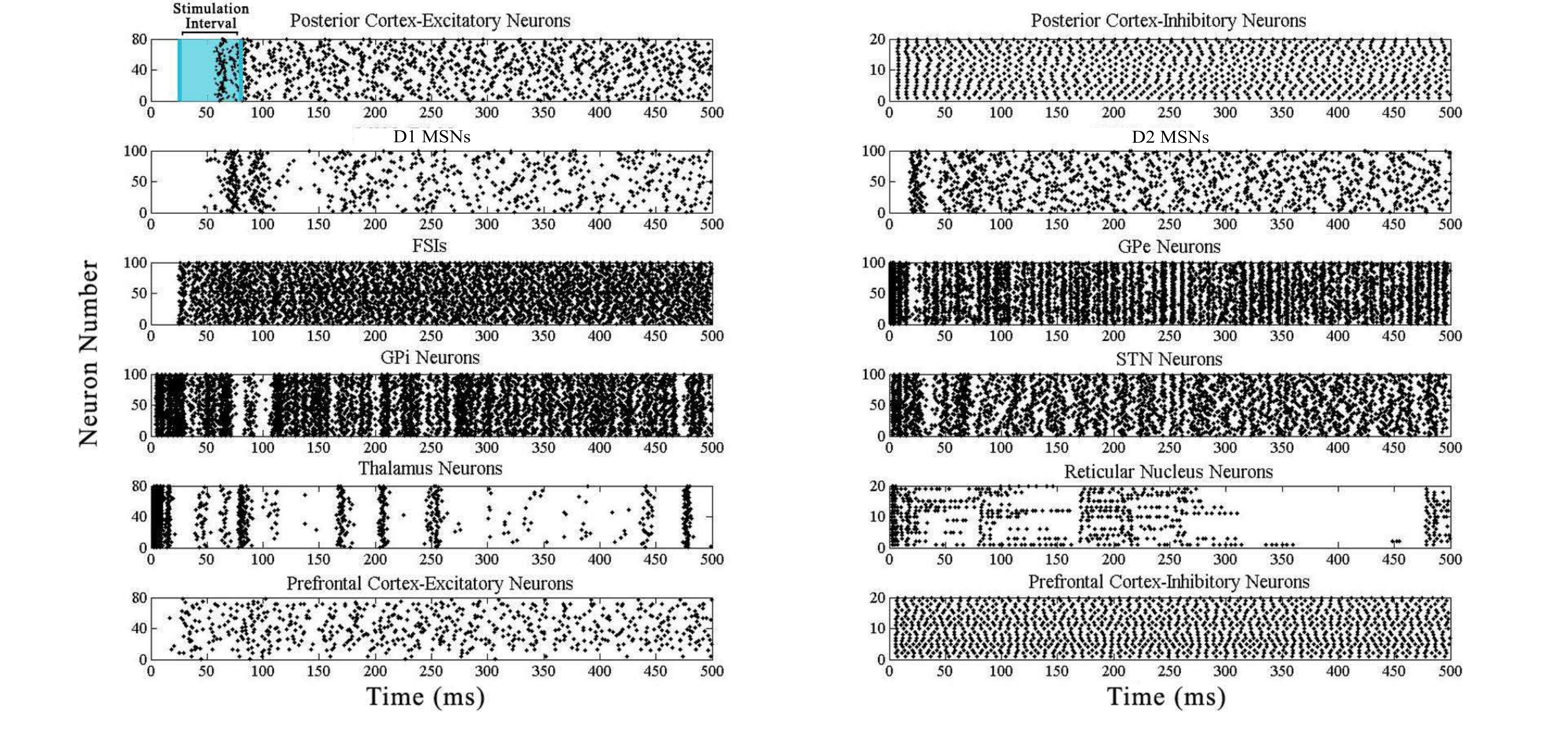}
\caption{Raster plots of the spiking activity of considered populations in the model when the indirect pathway of the basal ganglia is dominant with $\phi_1=0.35$ and $\phi_2=0.35$. The overactivity of $D_2$-MSNs in the striatum supresses the activity of the neurons of the GPe, which in turn stay incapable of reducing the activity of the STN. STN neurons start to send more excitatory inputs to the GPi that later will not allow the thalamic neurons to form a bridge between the external stimulus and the prefrontal cortex. Each black dot corresponds to a spike of the corresponding neuron at a given time.}
\label{fig:in_raster}
\end{figure}

As in previous sections, we  will confirm the changes of activity in the network by the spiking activity of key populations of neurons, represented in the raster plots given in Figure \ref{fig:in_raster}. These raster plots correspond to a simulation of $500$ milliseconds of spiking activity, when an external stimulus is applied to NMDA-receptor type neurons of the posterior cortex between $25-75$ milliseconds. As in the dominance of the direct pathway, these neurons respond to the stimulus with a time delay. Recurrent connections within the neurons of the posterior cortex allow to keep them in an active state throughout the simulation.  In the case study of this section, dopamine release will favour the activation of $D_2$-MSNs neurons over $D_1$-MSNs. For this purpose, we use use $\phi_1=0.35$ and $\phi_2=0.35$.

The neurons in the  posterior cortex send excitatory outputs to both $D_1$-MSNs and $D_2$-MSNs, but they are incapable of causing enough stimulation to $D_1$-MSNs, since cortical neurons' influence to $D_1$-MSNs is scaled down by the level of dopamine ($\phi_1$). However, they stimulate $D_2$-MSNs. Nevertheless, the FSIs produce tonic spike trains even with low levels of dopamine, since they are arranged to show fast-spiking behaviour. The FSIs also have an important effect on MSNs. The inhibitory signals from the FSIs are amplified in the same way as the cortical input amplification to MSNs. Since the FSIs provide inhibitory signals, their post-synaptic signal is scaled up for $D_1$-MSNs to help $D_2$-MSNs become dominant, which favours the indirect pathway over the direct pathway. Moreover, the GPe has inhibitory connections to the neurons of the STN, which in turn has excitatory connections to the neurons of the GPi. Under the influence of inhibitory $D_2$-MSN signals, the GPe remains incapable of suppressing the activity within the STN. This interaction allows the increased STN excitatory signals to be delivered through the GPi. In other words, the output message is a `No-Go' signal, which means that the GPi should have enough inhibitory effect on the thalamus, which in turn will deliver the message to the prefrontal cortex. The overactivation of the GPi does not allow the thalamus to be able to `open the gates' for the information flow. That is, the basal ganglia do not allow the external stimulus to reach the prefrontal cortex and prevents significant activity from appearing. As a consequence, the prefrontal cortex exhibits a low firing activity.

\begin{figure}[!ht]
\centering
\includegraphics[width=6.3in]{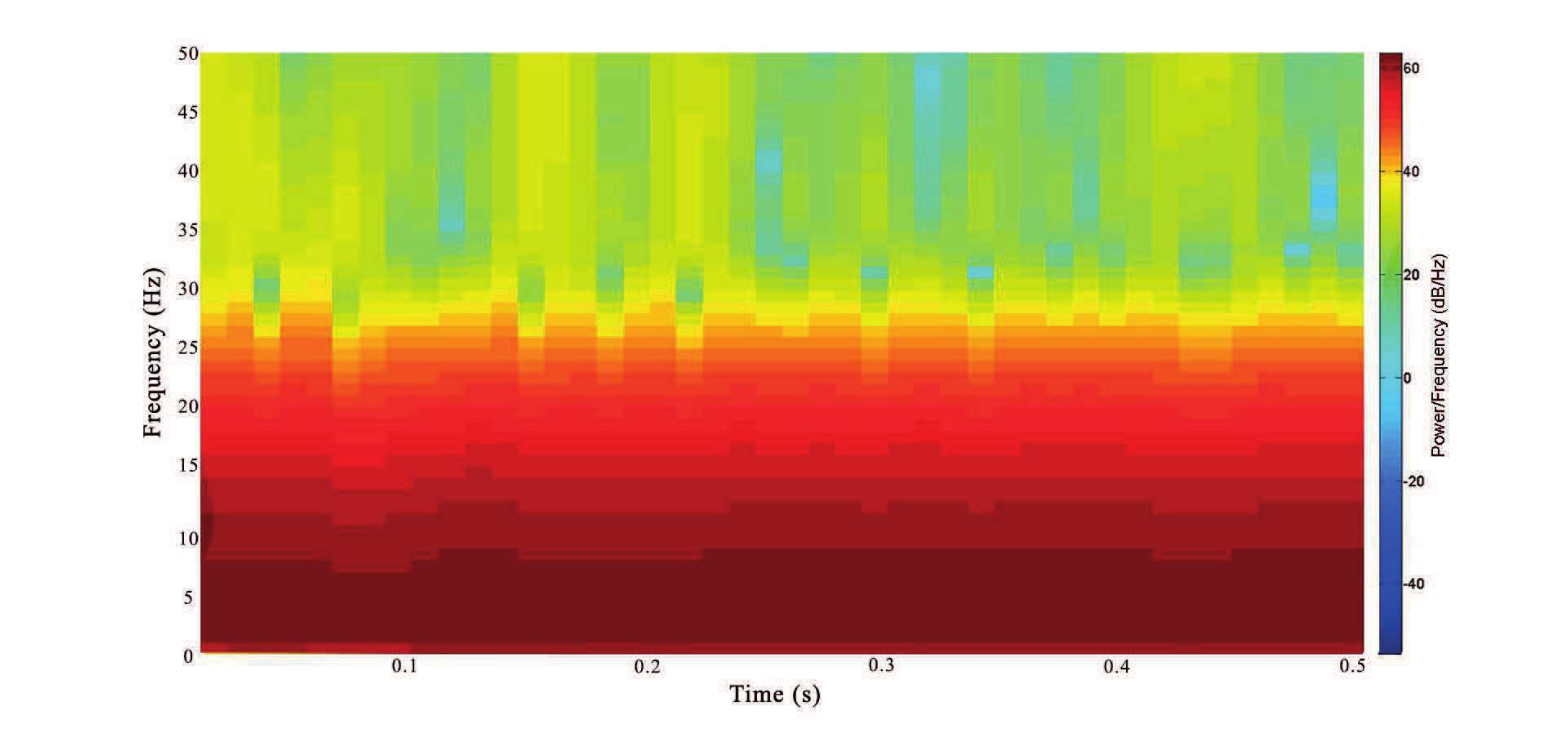}
\caption{Spectrogram of the thalamic activity when the indirect pathway of the basal ganglia is dominant. The spectrogram was obtained from local field potentials averaged over $30$ trials, with each trial lasting $500$ milliseconds. As in the dominance of the direct pathway, the result was filtered with the default values obtained from the no stimulus case. The dominant frequency band is in red. The figure shows that under the influence of an increased activity in the GPi, the thalamic cells reduce their activity to theta-frequency band activity. The dominant frequencies have their peak between $4-8$Hz.}
\label{fig:in_spectro}
\end{figure}

 With regard to the activity of the thalamic neurons, there is a transition from a burst-like activity to a tonic and low-frequency activity.	This is reflected in the frequency-time spectrogram of the thalamic neurons given in Figure \ref{fig:in_spectro}. As imaging studies have revealed, the dominance of the indirect pathway of the basal ganglia results in a reduced thalamic activity and the onset of theta-frequency band activity \cite{ref11}. When the GPi increases its repressive effectiveness, the thalamic network desynchronises especially in the low frequency range, leading to a certain spreading of the spectral intensity. This causes the dominance of the frequency range to spread over the $5-15$Hz interval. However, the power spectral density analysis and the histogram results indicate the dominance of theta-frequency band oscillations. Theta-frequency band oscillations drive the network to a regime so that a persistent active state in the network is prevented from happening, and an incoming signal only results in a transient response. Consequently, these oscillations selectively prevent the  loading of unwanted information from reaching the prefrontal cortex. 

\begin{figure}[!ht]
\centering
\includegraphics[width=6.3in]{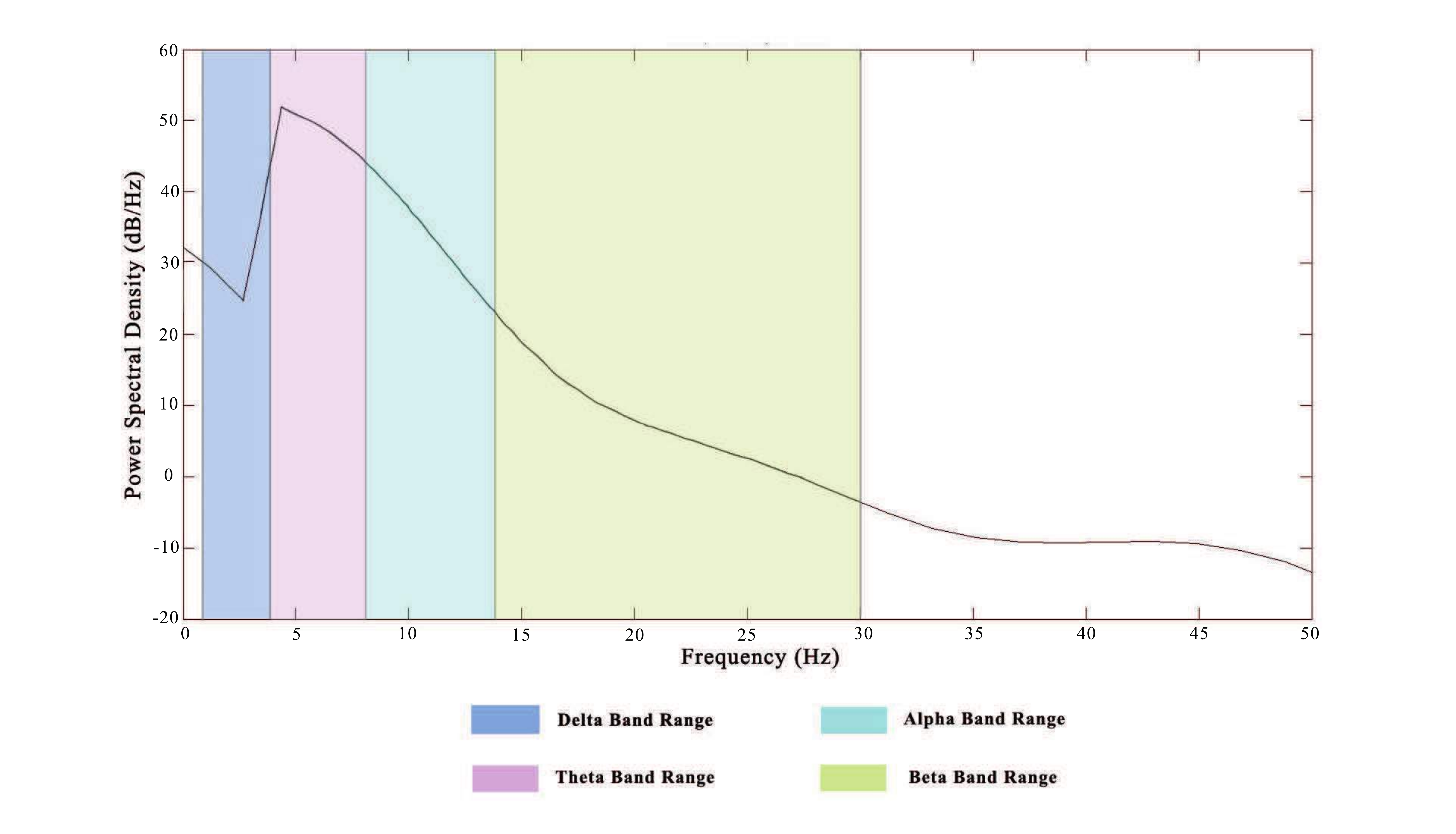}
\caption{Power spectral density of thalamic neurons when the indirect pathway of the basal ganglia is dominant. The spectrum gives the relationship between the frequency values and the related power densities. The most dominant frequency is obtained within the theta-frequency band ($4-8$Hz) which is followed by a slight reduction in power. The figure is an interpretation of the color bar graph within the spectrogram graph.}
\label{fig:in_spectrum}
\end{figure}

\begin{figure}[!ht]
\centering
\includegraphics[width=5.5in]{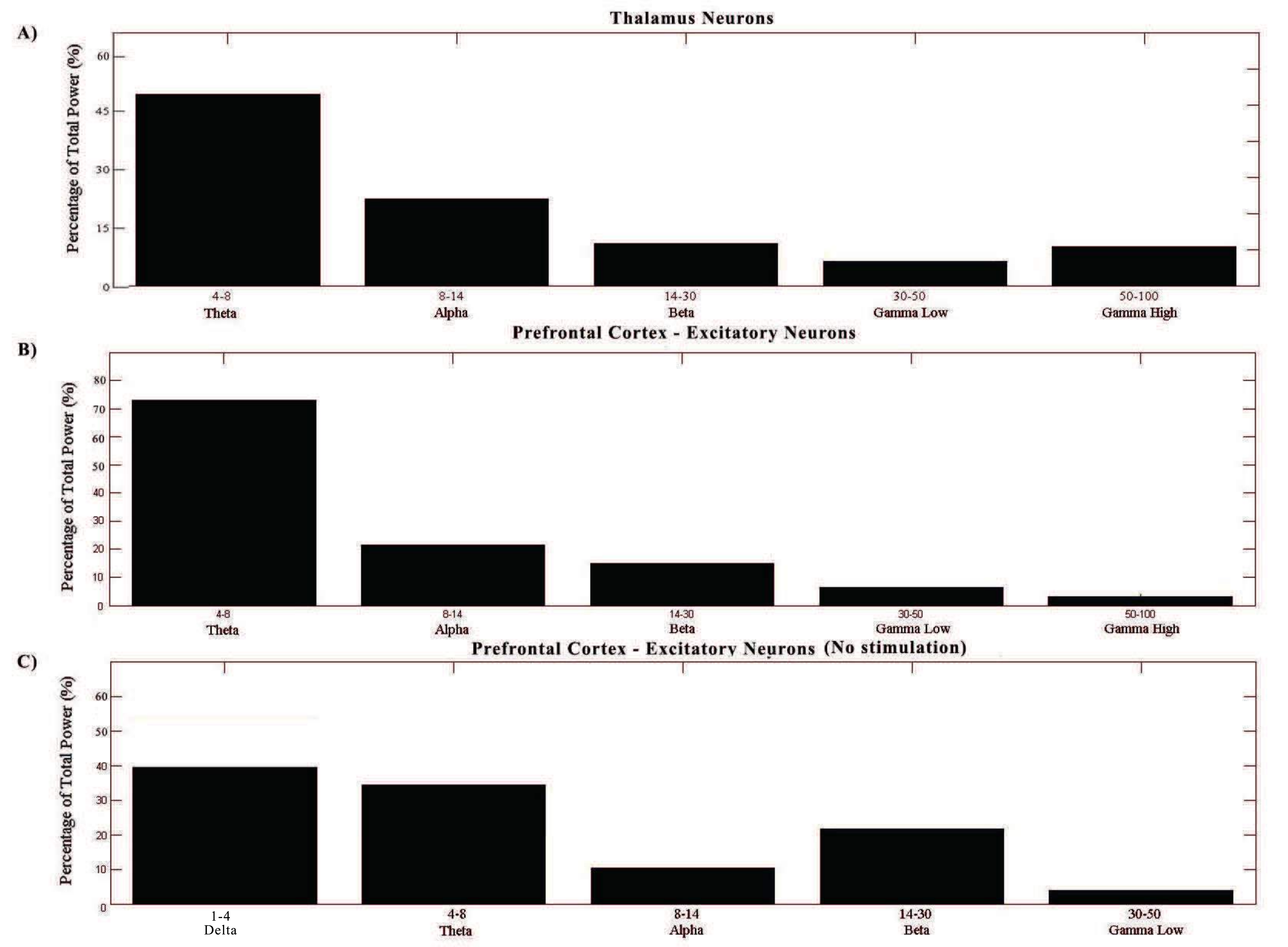}
\caption{Power histogram of thalamic and excitatory prefrontal cortical activity when the indirect pathway of the basal ganglia is dominant.  A) Power distribution of thalamic neurons for different frequency bands. Theta-frequency band activity is dominant and we also obtained a considerable power at neighboring alpha-frequency band.  B) Power distribution of frequency components for the NMDA-receptor-type neurons of the prefrontal cortex. Theta-frequency band oscillations are also observed to be dominant within the NMDA population, which also indicates a resting state activity. When the stimulation decays through the indirect pathway, the external stimulus does not reach the cortex. C) Power distribution of the considered frequency bands for the NMDA-receptor-type neurons of the prefrontal cortex with no external stimulus. NMDA-receptor-type neurons activity fluctuates between delta-frequency and theta-frequency bands. Our results show that when the indirect pathway is dominant, the cortical activity is in a distinct theta-frequency band. However, when there is no stimulus, the oscillatory activity appears to be at lower frequencies within delta-frequency and low-theta-frequency bands.}
\label{fig:in_histogram}
\end{figure}

The power spectral density of the thalamic neurons is given in Figure \ref{fig:in_spectrum}. The graph represents the power spectrum of the thalamic neurons. We obtained the dominance of multiple frequency bands. Thalamic activity fluctuates between the delta-frequency and theta-frequency band ranges, and it reaches its peak value within the theta-frequency band. Additionally, in Figure \ref{fig:in_histogram}, the frequency power histogram reveals an attenuation for frequencies higher than those in the theta-frequency band. Figure  \ref{fig:in_histogram} also presents the frequency-power histograms for NMDA-receptor-type neurons of the prefrontal cortex. With regard to thalamic activity, theta-frequency band oscillations introduce nearly the $50\%$ of the total power that is followed by a low power alpha-frequency band activity. For the rest of the frequency bands, each of them only represents the $10\%$ of the total power. Low activity within the thalamus population also drives the excitatory units of the prefrontal cortex to a resting state within a theta-frequency band range. Theta-frequency band activity covers nearly the $70\%$ of the total power. Moreover, neighbouring alpha-frequency band activity contributes to the power spectrum obtained for the thalamic neurons. The external stimulus is ignored by the striatal decision and consequently, the prefrontal cortex mimics the activity of the thalamic neurons rather than being activated by the stimulation. It is observed that the prefrontal cortex also generates activity at delta-frequency and low theta-frequency band ranges. There seems to be a ground theta-frequency band activity either in the resting state without stimulation or in the state where any persistent neuronal activity is avoided. Our analysis of cortical  activity confirms that the theta-frequency band is successful in preventing activity from appearing within the prefrontal cortex when the indirect pathway of the basal ganglia is dominant.

\section{Conclusions}

We have proposed a novel mathematical-computational model for the basal ganglia-thalamus-cortical network that links different basal ganglia's pathways to different working memory processes. We have made use of the DMS task to help understand the neuronal activity during working memory tasks. Our model well demonstrates how different frequency-band oscillations observed during working memory tasks can emerge due to basal ganglia's activity. The framework proposed here relates the switching behaviour of the basal ganglia to the regulation of the direct and indirect pathways depending on dopamine levels. The dopamine level triggers different pathways to execute appropriate working memory computations, such as: loading, maintainace or  clearance of information. We have also shown the key role of the activity in the thalamus in the generation of oscillations in specific frequency bands, which correlate with each of these working memory computations.

The  model has been validated through simulations for two main cases: when the direct pathway and the indirect pathways of the basal ganglia are dominant.    When the direct pathway is dominant, our model has shown that the thalamus is disinhibited and  gamma-frequency band activity is generated in the thalamic neuronal population. Our model has also shown how increased levels of dopamine influences the balance between $D_1$ and $D_2$-type MSNs, favouring the direct pathway.  Disinhibited thalamic activity allows information to reach the prefrontal cortex. The thalamus translates the basal ganglia output to the working memory by increasing its activity and presenting gamma-frequency band oscillations. Under this scenario, information can be loaded/updated in the working memory. We note that beta-gamma-frequency band oscillations only allow the maintenance of the active state in the prefrontal cortex, but not the maintenance of the updated information. 

Our model is also capable to demonstrate how the dominance of the indirect pathway within the basal ganglia can inhibit the thalamus to block information flow. The maintenance of information in working memory is a result of the generation of theta-frequency band activity in the thalamus. Furthermore, the clearance of information is achieved by the balance between the direct and indirect pathways. The clearance is a result of alpha-frequency band activity within the thalamus. In this case, no persistent activity is allowed, the ongoing activity is blocked and no stimulation can cause any change of activity in the working memory.

Immediate extensions and improvements of our model are the following ones. First, the consideration of the evolution of the network to a self-sustained network. Within this context, a detailed and self-sustained network may provide more information on the communication network of the basal ganglia and the working memory. Second, to introduce synaptic plasticity  -- for example, spike-time dependent plasticity (STDP) -- in the model for the recognition of sensory information. STDP would allow neurons to encode the sensory stimuli by regulating connections. With this, the model could be tested with a complete DMS task. As proposed in \cite{ref22}, this could be achieved by using a stripe-like architecture which would be capable of storing and updating different information individually.


\begin{thebibliography}{99}


\bibitem {ref34}
Albin RL, Young AB, Penney JB (1989) The functional anatomy of basal ganglia disorders. Trends in Neurosciences 12:366-375


\bibitem{ref6}
Alexander GE, DeLong MR, Strick PL (1986) Parallel organization of functionally segregated circuits linking basal ganglia and cortex. Annual Review of Neuroscience 9: 357-381

\bibitem{ref2}
Atkinson RC, Shiffrin RM (1968). Human memory: A proposed system and its control processes. In: Spence KW, Spence JT (eds) The psychology of learning and motivation, Volume 2. Academic Press, New York, pp 81-95


\bibitem{ref1}
Baddeley  AD (1992) Working memory. Science 255:556-559

\bibitem {ref47}
Baddeley AD (2000) The episodic buffer: a new component of working memory?. Trends Cogn Sci 4:417–423

\bibitem{ref4}
Baddeley AD (2003) Working memory: looking back and looking forward. Nature Rev Neuroscience 10:829-839


\bibitem{ref3}
Baddeley AD (2007) Working memory, thought and action. Oxford University Press, Oxford



\bibitem {ref46}
Bair W, Koch C, Newsome W, Britten K (1994) Power spectrum analysis of bursting cells in area MT in the behaving monkey. Journal of Neuroscience 14:2870–2892


\bibitem {ref31}
Benda J, Herz AVM (2003) A universal model for spike-frequency adaptation. Neural Computation 15(11):2523-2564

\bibitem {ref32}
Beatty JA, Sullivan MA, Morikawa H, Wilson CJ (2012) Complex autonomous firing patterns of striatal low-threshold spike interneurons. Journal of Neurophysiology 108(3):77181.

\bibitem {ref48_nss} 
Berns GS, Sejnowski TJ (1998) A computational model of how the basal ganglia produce sequences. J Cogn Neurosci 10(1):108–121

\bibitem {ref14}
Bolam JP, Hanley JJ, Booth PA, Bevan MD (2000) Synaptic organisation of the basal ganglia. J Anatomy 196(4):52742


\bibitem {ref47_nss}
Burkhardt JM, Jin X, Costa RM (2010) Dissociable effects of dopamine on neuronal firing rate and synchrony in the dorsal striatum. Front Int Neurosci 3:28. doi: 10.3389/ neuro.07.028.2009

\bibitem {ref40}
Centonze D, Grande C, Usiello A, Gubellini P, Erbs E, Martin AB, Pisani A, Tognazzi N, Bernardi G, Moratalla R, Borrelli E, Calabresi P (2003) Receptor subtypes involved in the presynaptic and postsynaptic actions of dopamine on striatal interneurons. Journal of  Neuroscience (23):6245-6254


\bibitem {ref26}
Chatham CH, Badre D (2013) Working memory management and predicted utility. Frontiers in Behavioral Neuroscience 7:83



\bibitem {ref25}
Chatham CH, Badre D (2015) Multiple gates on working memory. Current Opinion in Behavioral Sciences 1:23-31. doi:10.1016/j.cobeha.2014.08.001

\bibitem{wiring_plasticity}
Chklovskii DB, Mel BW, Svoboda K (2004) Cortical rewiring and information storage. Nature 431:782–788

\bibitem{ref5} 
DeLong MR, Georgopoulos AP (1981) Motor functions of the basal ganglia. In: Brookhart JM, Mountcastle VB, Brooks VB, Geiger SR (eds) Handbook of physiology: the nervous system: motor control, volume 2. American Physiological Society, pp. 1017-1061

\bibitem {ref16}
DeLong MR (1990) Primate models of movement disorders of basal ganglia origin. Trends in Neurosciences  (13):281-285



\bibitem {ref28}
Dipoppa M, Gutkin BS (2013) Flexible frequency control of cortical oscillations enables computations required for working memory. PNAS 110(31):12828-12833

\bibitem {ref17}
Gerfen CR, Engber TM, Mahan LC, Susel Z, Chase TN, Monsma FJ Jr, Sibley Jr (1990) $D_1$ and $D_2$ dopamine receptor-regulated gene expression of striatonigral and striatopallidal neurons. Science  7 (250):1429-1432


\bibitem {ref22}
Frank MJ, Loughry B, O'Reilly RC (2001) Interactions between frontal cortex and basal ganglia in working memory: a computational model. Cogn Affect Behav Neurosci 1:137-160

\bibitem {ref15}
Frank MJ (2005) Dynamic dopamine modulation in the basal ganglia: a neurocomputational account of cognitive deficits in medicated and nonmedicated Parkinsonism. Journal of Cognitive Neuroscience  17(1):51-77 


\bibitem {ref21}
Frank MJ (2006) Hold your horses: a dynamic computational role for the subthalamic nucleus in decision making. Neural Networks (19):1120-1136 



\bibitem{ref7}
Gerfen CR, Keefe KA, Gauda EB (1995) $D_1$ and $D_2$ dopamine receptor function in the striatum: coactivation of $D_1$- and $D_2$-dopamine receptors on separate populations of neurons results in potentiated immediate early gene response in $D_1$-containing neurons. Journal of Neuroscience (15):8167-8176


\bibitem {ref42}
Hartmann-von Monakow K, Akert K, K\"unzle H (1978) Projections of the precentral motor cortex and other cortical areas of the frontal lobe to the subthalamic nucleus in the monkey. Experimental Brain Research 33(3):395-403



\bibitem{ref10}
Howard MW  et al (2003) Gamma oscillations correlate with working memory load in humans. Cereb Cortex 13(12):1369-1374


\bibitem {ref27}
Humphries M, Lepora N, Wood R, Gurney K (2009) Capturing dopaminergic modulation and bimodal membrane behaviour of striatal medium spiny neurons in accurate, reduced models. Frontiers in Computational Neuroscience 3: article 26. doi: 10.3389/neuro.10.026.2009



\bibitem {ref30}
Izhikevich EM (2003) Simple model of spiking neurons. IEEE Trans Neural Networks 14:1569-1572

\bibitem {ref33}
Izhikevich EM (2007) Dynamical systems in neuroscience: the geometry of excitability and bursting. MIT Press, Cambridge, MA




\bibitem {ref43}
Isoda M, Hikosaka O (2008) Role for subthalamic nucleus neurons in switching from automatic to controlled eye movement. Journal of Neuroscience 28:7209-7218


\bibitem {ref36}
Kemp JM, Powell TPS (1971) The structure of the caudate nucleus of the cat: light and electron microscopy. Philosophical Transactions of the Royal Society of London 262:383-401. doi: 10.1098/rstb.1971.0102


\bibitem {ref12}
Klimesch W, Doppelmayr M, Schwaiger J, Auinger P, Winkler T (1999) Paradoxical alpha synchronization in a memory task. Brain Res Cogn Brain Res 7(4):493-501



\bibitem {ref23}
Klimesch W (2012) Alpha-band oscillations, attention, and controlled access to stored information. Trends Cogn. Sci. 16:606-617

\bibitem {ref38}
Kos T, Tepper JM (1999) Inhibitory control of neostriatal projection neurons by GABAergic interneurons. Nat Neurosci 2:467472

\bibitem{structural_plasticity}
Lamprecht R, LeDoux J (2004) Structural plasticity and memory. Nature Reviews Neurosci 5:45–54

\bibitem {ref37}
Malenka RC, Nestler EJ, Hyman SE (2009) Widely projecting systems: monoamines, acetylcholine, and orexin. In: Sydor A, Brown RY (eds) Molecular neuropharmacology: a foundation for clinical neuroscience, 2nd edn. McGraw-Hill Medical, New York, pp. 147148-154157


\bibitem {ref39}
Mallet N, Le Moine C, Charpier S, Gonon F (2005) Feedforward inhibition of projection neurons by fastspiking GABA interneurons in the rat striatum in vivo. Journal of Neuroscience  25:3857 3869

\bibitem {ref13}
McGeorge AJ, Faull RL (1989) The organization of the projection from the cerebral cortex to the striatum in the rat. Neuroscience 29:503-537



\bibitem {ref18}
Mink JW, Thach WT (1993) Basal ganglia intrinsic circuits and their role in behavior. Current Opinion in Neurobiology 3:950-957



\bibitem {ref46_nss}
Mohajenari MH et al (2013) Spontaneous cortical activity alternates between motifs defined by regional axonal projections. Nature Neuroscience 16:1426-1435




\bibitem {ref19}
Nambu A, Tokuno H, Hamada I, Kita H, Imanishi M, Akazawa T, Hasegawa N (2000) Excitatory cortical inputs to pallidal neurons via the subthalamic nucleus in the monkey. Journal of Neurophysiology 84:289-300

\bibitem{eva:2015}
Navarro-L\'opez EM, Celikok U, Seng\"or NS (2016) Hybrid systems neuroscience. In: El Hady A (ed) Closed-loop neuroscience. Academic Press, in press

\bibitem {ref49_nss} 
Nicola SM, Surmeier J, Malenka RC (2000) Dopaminergic modulation of neuronal excitability in the striatum and nucleus accumbens. Annual Rev Neurosci 23:185–215. doi:10.1146/annurev.neuro.23.1.185


\bibitem {ref29}
Noback CR, Ruggiero DA, Demarest RJ, Strominger NL (2005) The human nervous system: structure and function, 6th edn. Humana Press, Totowa, NJ


\bibitem {ref20}
Plenz D, Kital ST (1999) A basal ganglia pacemaker formed by the subthalamic nucleus and external globus pallidus. Nature 12(400):677-682


\bibitem {ref44}
Postuma RB, Lang AE (2003) Hemiballism: revisiting a classic disorder. Lancet Neurology 2(11):661-668


\bibitem {ref35}
Rolls, ET (1994) Neurophysiology and cognitive functions of the striatum. Rev Neurol(Paris) 150(8-9):648-660



\bibitem {ref24}
Roux F, Wibral M, Singer W, Aru J, Uhlhaas PJ (2013) The phase of thalamic alpha activity modulates cortical gamma-band activity: evidence from resting-state meg recordings. Journal of Neuroscience 33(45):17827-17835

\bibitem{Schmidt:2013}
Schmidt S, Scholz M, Obermayer K, Brandt SA (2013) Patterned brain stimulation, what a framework with rhythmic and noisy components might tell us about recovery maximization. Frontiers in Human Neuroscience 7(325):1-10

\bibitem{ref9}
Spitzer B, Wacker E, Blankenburg F (2010) Oscillatory correlates of vibrotactile frequency processing in human working memory. Journal of Neuroscience 30(12):4496-4502


\bibitem{ref8}
Stocco A, Lebiere C, Anderson JR (2010) Conditional routing of information to the cortex: a model of the basal ganglia's role in cognitive coordination. Psychological Review 117(2):54174


\bibitem{ref11}
Tesche CD, Karhu J (2000) Theta oscillations index human hippocampal activation during a working memory task. PNAS 97(2):919-924



\bibitem{Torres:2014}
Torres JJ, Elices I, Marro J (2014) Efficient transmission of subthreshold signals in complex networks of spiking neurons. arXiv:1410.3992 [physics.bio-ph].



\bibitem {ref45}
Wichmann T, Bergman H, DeLong MR (1994) The primate subthalamic nucleus. I. Functional properties in intact animals. J Neurophysiol 72(2):494-506



\bibitem {ref41}
Yelnik J, Percheron G (1979) Subthalamic neurons in primates: a quantitative and comparative analysis. Neuroscience 4(11):1717-1743








%
%








%
%
%
%
%
%
%
%







\end{thebibliography}
\end{document}